\documentclass[aps, prd, nofootinbib, twocolumn, noeprint]{revtex4-2}
\emergencystretch=\maxdimen
\usepackage[mathlines]{lineno} 
\usepackage{amsmath}    
\usepackage{amssymb}    
\usepackage{bm}         
\usepackage{braket}     
\usepackage{dcolumn}    
\usepackage{diagbox}    
\usepackage{epsfig}     
\usepackage{graphicx}   
\usepackage{hhline}     
\usepackage{mathrsfs}   
\usepackage{txfonts}    
\usepackage{slashed}    
\usepackage{subfigure}  
\usepackage[usenames,dvipsnames]{xcolor}  
\usepackage{mathtools}  
\usepackage{tensor} 
\usepackage{cuted}      
\usepackage{lipsum}     
\usepackage{multirow}  
\usepackage{gensymb}   

\usepackage[pagebackref=false, colorlinks=true]{hyperref}
\definecolor{reddish}{rgb}{0.7,0.2,0.0}  
\definecolor{blueish}{rgb}{0.1,0.1,1}
\hypersetup{
linkcolor=reddish,      
citecolor=blueish,      
filecolor=blue,         
urlcolor=magenta}       

\begin{document}
	\title[]{Exploring Nonlinear Electrodynamics Theories: Shadows of Regular Black Holes and Horizonless Ultra-Compact Objects}
	\author{Rahul Kumar Walia}\email{rahulkumar@arizona.edu}
	\affiliation{ Department of Physics, University of Arizona, 1118 E 4th Street, Tucson, AZ, USA}
	\date{\today}
	
\begin{abstract}
In the Einstein-Maxwell theory with nonlinear electrodynamics (NED) fields, the singularity problem in general relativity is potentially resolved, leading to regular black hole solutions. In NED theories, photons follow null geodesics of an effective geometry that differs from the spacetime geometry itself. This raises an important question: Do NED fields produce unique observational signatures in the electromagnetic spectrum that can test regular black holes and NED theories? We analyze the shadows of two NED-charged regular black holes and their horizonless ultracompact objects (HUCOs) under two accretion models, comparing them with Schwarzschild black holes, focusing on shadow size, central brightness depression, and photon ring characteristics. Our results identify distinctive NED signatures that could be observable by the EHT, providing empirical evidence of NED fields and potentially ruling out previously considered viable candidates for astrophysical black holes models based on shadow measurements. Notably, NED-charged HUCOs generally exhibit only one \textit{unstable} photon ring, thus avoiding the dynamical instability associated with stable photon rings and challenging the idea that objects with photon rings must be black holes.
\end{abstract}
\maketitle

\section{Introduction}\label{sec-1}
The presence of curvature singularities and geodesic incompleteness at the center of black holes remains one of the most intriguing and longstanding problems in general relativity (GR). Resolving this issue has been a primary motivation for various developments in both classical and quantum gravity. 
While reasonable classical modified gravity theories have failed to provide vacuum singularity-free asymptotically flat black hole solutions in four dimensions, various quantum gravity theories have successfully predicted such models, notable examples include loop quantum gravity~\cite{Perez:2017cmj, Rovelli:1997yv}, asymptotic safe gravity~\cite{Bonanno:2000ep, Torres:2014gta}, string theory~\cite{Nicolini:2019irw, Cano:2018aod}, and noncommutative gravity~\cite{Nicolini:2005vd}. However, in this conventional top-down approach to quantum gravity, regular black holes do not usually appear as solutions of the gravitational field equations,  but rather as \textit{ad hoc} models based on quantum gravity arguments. The introduction of the first \emph{model} of a regular black hole by Bardeen in 1968~\cite{Bardeen-RegularBH} prompted the bottom-up approach to derive the regular black hole solutions within the framework of GR, before resorting to the quantum gravity. This approach utilized classical fields to violate certain energy conditions or bypass some other assumptions of singularity theorems in order to obtain regular black hole solutions within GR.

Nonlinear electrodynamics (NED) fields were anticipated as one such possible solution to spacetime singularities \cite{Pellicer:1969cf}. This approach stems from the observation that,  in classical Maxwell electrodynamics theory, the electric field and self-energy of a point-like charged particle diverge at its location, and this infinity could be resolved by nonlinear modifications of Maxwell's theory in strong-field regimes. In the same spirit, one might expect that substituting the Maxwell field with a suitable NED field in Einstein-Maxwell theory could replace the singular Reissner-Nordstr\"om (RN) black hole solution with a regular counterpart.  It turns out that this is indeed the case: NED fields not only resolve field singularities in electrodynamics but also, under certain reasonable conditions \cite{Bronnikov:2000vy,Bronnikov:2000yz}, regularize spacetime singularities in both black holes and early universe cosmology.  Ay\'on-Beato and Garc\'ia~\cite{Ayon-Beato:1998hmi} derived the first regular black hole solution from the Einstein-Maxwell field equations sourced by a NED field from an electric charge, and later showed that the Bardeen regular black hole model can also be explained as an exact solution of the field equations with a NED field from a  magnetic monopole charge~\cite{Ayon-Beato:2000mjt}.

Building upon the foundational works of Born-Infeld~\cite{Born:1934gh} and  Euler-Heisenberg \cite{Heisenberg:1936nmg} on NED field models,  several static and spherically symmetric regular black hole solutions within various well-motivated NED theories are now known~\cite{Fan:2016hvf,Rodrigues:2015ayd,Ayon-Beato:1999kuh,Bronnikov:2017sgg,Burinskii:2002pz,Dymnikova:2015hka}. Some interesting and important no-go theorems for the existences of regular black holes with the NED fields have also been formulated~\cite{Bronnikov:2000vy,Bronnikov:2000yz,Bronnikov:2017tnz,Bokulic:2022cyk,Bokulic:2021xom}. For a recent review and progress on the topic see~\cite{Bronnikov:2022ofk, Sebastiani:2022wbz,Carballo-Rubio:2023mvr, Sorokin:2021tge,Bambi:2023try} and references therein. 

NED theories exhibit a long list of model-independent interesting predictions, including: 
(i) regularization of electromagnetic fields; 
(ii) vacuum polarization of virtual $e^{-}e^{+}$ pair; (ii) pair production;  (iii) photon-photon scattering; (iii) birefringence, wherein an electromagnetic wave splits into two normal modes with mutually orthogonal linear polarization propagating at distinct velocities (Born-Infeld NED model is an exception for the birefringence \cite{Kim:2022fkt}); (iv) photons propagating along the null geodesics of an ``effective metric" modified by the NED field instead of the background metric; (v) significant enhancement of light polarization;  (vi) bending of light purely due to the NED field in vacuum. Since these effects are absent in the Einstein-Maxwell theory, they offer exciting opportunities to explore distinct observational signatures of NED charged regular black holes.  

Of course, it is still unclear whether ordinary astrophysical conditions allow for the existence of such black holes. Electrically charged black holes can, in principle, be produced in gravitational collapse \cite{Nathanail:2017wly}, but they are normally expected to be rapidly neutralized by the abundance of free charges in astrophysical plasma. Formation scenarios for black holes with magnetic monopole charges are not available at present and are objectively hard to justify. However, hereafter, we will simply assume that such a black hole has somehow been produced (eternal or primordial black hole sourced by GUT predicted monopole), and we will explore the consequences of this assumption.  In particular, we examine photon geodesics influenced by the effective metric of these black holes, analyze the resulting shadows, and propose further tests of NED fields in astrophysical contexts.  Specifically, we address the following key questions: 
(i) How do the shadows of NED charged regular black holes, as determined from the effective metric, differ from those dictated by the background metric? 
(ii) What are the similarities and differences between the shadow of a NED charged regular black hole and that of a singular black hole, such as a Schwarzschild black hole? 
(iii) Can a regular horizonless ultracompact object (HUCO) mimic a regular black hole as far as its shadow size is concerned  within the current Event Horizon Telescope (EHT) angular resolutions? 
(iv) Do the EHT's shadow size measurements of M87* and Sgr A* support the predictions of regular black holes from the effective metric or the background metric?
(v) Could the EHT observations of M87* and Sgr A* black holes be attributed to regular HUCO rather than black holes? 
(vi) How does the interior geometry of regular black holes and HUCOs influence their shadows?

These questions are important from both the perspective of EHT shadow observations and for understanding the NED field effects in curved spacetimes. Our approach to address these questions is rather simple. We consider two static and spherically symmetric NED charged regular spacetimes: Bardeen \cite{Bardeen-RegularBH} and the Ghosh-Culetu (GC) \cite{Ghosh:2014pba,Culetu:2014lca} models. Both metrics smoothly interpolate between regular black holes ($k\leq k_{\rm E}$) and regular HUCO spacetimes ($k> k_{\rm E}$) depending on the value of the regularization parameter $k$ that enters the metric; values of $k_{\rm E}$ are given in Secs.~\ref{sec-5a} and \ref{sec-5b}.To distinguish two cases, we use Bardeen-BH and GC-BH to represent black hole spacetimes, and Bardeen-HUCO and GC-HUCO for regular HUCO spacetimes. While both Bardeen and GC spacetimes are globally regular and can be attributed to distinct NED fields originating from magnetic monopole charges $k$, they are not particularly unique among other known regular black hole models. However, they exhibit significant differences in interior geometry and in the weak-field limits, representing two distinct classes of regular black holes, thus making them suitable candidates for addressing previously raised inquiries. Additionally, we consider two simple astrophysical scenarios: a spherically symmetric and radially infalling Bondi-Michel flow, and a Novikov-Thorne thin disk. Note this sets up an interesting challenge for developing a numerical ray-tracing technique involving two spacetime metrics: (i) the accreting matter follows timelike geodesics of the background metric; (ii) the emitted photons follow null geodesics of the effective metric; (iii) the observations are conducted within the background metric.  Comparing the shadows of two black holes and their regular HUCO counterparts, both internally and against that of the Schwarzschild black hole under identical astrophysical conditions, we address the posed questions.

The paper is structured as follows: We begin by providing an executive summary of the main results in Sec.~\ref{sec-1-0}, and addressing the motivation for NED fields for regular black holes, along with their current experimental and observational evidence in Sec.~\ref{sec-2}. Because the EHT shadow size bounds ruled out the Bardeen-BH and Bardeen-HUCO based on the shadows deduced from the effective metric, in the rest of the paper we focus on the GC spacetimes shadows. In Sec.~\ref{sec-3}, we derive the effective metric for photon geodesics. Section~\ref{sec-4} covers the setup for the black hole shadow and two accretion models. We present the shadows of GC-BH and GC-HUCO, using the two accretion scenarios in Sec.~\ref{sec-5}.  Some general results for the NED fields in regular black hole spacetime are discussed in Sec.~\ref{sec-6}. Finally, Sec.~\ref{sec-7} summarizes our main results. We present the shadows of the GC-BH and GC-HUCO deduced from the background metric in the Appendix~\ref{sec-8A} and those of Schwarzschild black hole under identical accretion models in the Appendix~\ref{sec-8B}.

\section{Executive Summary}\label{sec-1-0}
Here, we highlight the important findings of this paper. These results have significant implications for both EHT theory and observation, offering novel perspectives on NED-charged regular black holes.
\begin{enumerate}
	\item The analytical expression for the shadow radius of a general static and spherically symmetric magnetically charged black hole in NED theory is presented in Eq.~(\ref{shadowsize}).
	\item Regular HUCOs exhibit null circular orbits from the background metric \textit{only} within a limited parameter range of $k$, whereas those from the effective metric exist for arbitrarily large values of $k$ (cf. Figs.~\ref{fig:BarRadii} and \ref{fig:NSRadii}).
	\item While the background metric is singularity-free and geodesically complete, the effective metrics for Bardeen and GC spacetimes exhibit curvature singularities that are observable only by photons (see Secs.~\ref{sec-4a} and \ref{sec-4b}). These singularities become particularly relevant for HUCOs, where they influence the resulting shadows. Refer to Sec.~\ref{sec-6} for detail discussion on general magnetically charged NED spacetimes.
	\item HUCOs cast shadows with a central brightness depression and shadow size comparable to that of a black hole (see Sec.~\ref{sec-4b}).
	\item The effects of magnetic monopole charge are more evident in the shadows derived from the effective metric than in those predicted by the background metric. Bardeen black holes can be ruled out by the EHT shadow size measurements of the Sgr A* and M87* black holes (see Sec.~\ref{sec-2}).
	\item  Absence of \textit{stable} photon rings renders NED charged regular HUCOs as viable alternatives to astrophysical black holes.
\end{enumerate}

\section{Experimental and observational interests for NED}\label{sec-2}
Although initially purely theoretical, interest in NED models has grown over time, as evidenced by reports \cite{Ruffini:2009hg,Sorokin:2021tge}. The main obstacle in obtaining direct evidence of the NED field is its exceedingly high Schwinger limit,  $E_{\text{cric}}\sim 10^{18}$ V/m and $B_{\text{cric}}\sim 10^{13}$G. Despite this challenge, recent advancement in experimental facilities offer promising opportunities to reach the Schwinger limit.  
For instance, laser facilities like Extreme Light Infrastructure (ELI) \cite{ELI} are actively searching for evidence of photon-photon scattering and $e^+e^-$ pair creation processes. The X-ray Free Electron Lasers (XFEL) \cite{XFEL} facility is dedicated to investigating the nonlinear response of matter to extremely high-intensity electromagnetic fields and has recently observed multi-photon nonlinear Compton scattering and second harmonic generation at x-ray wavelengths for the first time. The pair production process, such as the Breit-Wheeler process $2\gamma=e^+ + e^-$, conceptually modifies the linearity of Maxwell's theory and naturally necessitates a transition to a NED theory. Thus, observing the Breit-Wheeler process would experimentally confirm the NED theory. The LUXE experiment \cite{LUXE} proposed at DESY aims to investigate nonperturbative effects of QED and measure the pair production rate from the QED vacuum through collisions between a 16.5 GeV electron beam and a 40 TW optical laser. Relativistic electrons in LUXE experience electric fields approaching or exceeding the Schwinger limit in their frame of reference compared to the actual laser electric field in the laboratory frame of reference. Another approach to probe the NED scale is through relativistic heavy-ion collisions $Z\geq \frac{1}{\alpha}$, where ions scatter quasielastically with an impact parameter larger than the sum of their radii. The recent direct experimental detection of  $\gamma\gamma\to \gamma\gamma$ scattering in ultrarelativistic Lead ions collisions in the ATLAS detector at the LHC stands as one of the most cleanest and strongest confirmations of the nonlinear extension of the Maxwell's electrodynamics  theory \cite{ATLAS:2017fur,ATLAS:2019azn}. Electromagnetic vacuum birefringence and dichroism have been searched for in a number of experiments, one of which is PVLAS \cite{Ejlli:2020yhk}. Although the experimentally measured value did not match the predicted value from QED, constraints could be placed on the Euler-Heisenberg NED parameter. At $B=2.5 T \sim 10^{-10} B_{\text{cric}}$, the difference in the refractive index for two different light polarization mode is tiny, $\Delta n\sim \alpha B^2/B_{\rm c}^2 \sim 10^{-23}$, which can be significantly increased by the strong background magnetic field \cite{Ejlli:2020yhk}.

As experimental facilities continue to improve to achieve the Schwinger limit, some astrophysical environments such as neutron stars, magnetars and black holes naturally offer electromagnetic fields exceeding this limit. Magnetars, in particular, have been spotted with magnetic fields stronger than $10^{13}$ G. The Imaging X-ray Polarimetry Explorer (IXPE)~\cite{IXPE} is one of the space observational facilities designed to measure light polarization from black holes, neutron stars, pulsars, and galactic centers. In May 2022, IXPE detected linearly polarized x-ray emission from the pulsar 4U 0142+61, with an estimated dipole magnetic field of $10^{14}$G \cite{Taverna:2022jgl}, marking the first-ever detection of x-rays polarization in a magnetar source. The observed polarization values in two competing polarization modes lend support for the vacuum birefringence and, consequently, the NED theories. Additionally, vacuum birefringence has been confirmed through optical-polarimetry measurements of the isolated neutron star RX J1856.5-3754, showing a polarization degree of 16.43$\pm$5.26 \%  \cite{Mignani:2016fwz}. A  recent addition to the light polarization space observational facility is X-ray Polarimeter Satellite (XPoSat), launched in  2024, by ISRO. 

However, these existing data does not favor any specific NED model, necessitating to gather diverse observational signatures of the NED field. Of particular importance is the calculation of potential observables within black hole spacetimes, given the theoretical significance of NED fields in mitigating curvature singularities.

NED charged regular black holes, considering their effects solely within the background metric, have already been tested through various observations including EHT and LIGO/Virgo observations~\cite{KumarWalia:2022aop,Kumar:2020yem,Vagnozzi:2022moj,Ghosh:2020spb,Kumar:2020yem,Kumar:2018ple,Kar:2023dko,Uniyal:2022vdu}, and with X-ray data from the Cygnus X-1 \cite{Bambi:2014nta}. However, only recently have the effects of the effective geometry on observational features been considered ~\cite{Schee:2019,Schee:2019gki,Rayimbaev:2020hjs,Toshmatov:2021fgm,dePaula:2023ozi}, led by Stuchl\'ik and Schee~\cite{Stuchlik:2019uvf}. Given that EHT observations directly involve the observation of light photons from synchrotron emission, accounting for the effective metric $\tilde{g}_{\mu\nu}$ becomes imperative to calculate accurate predictions for the shadows.  

The EHT, with an observing frequency of 230 GHz, captured total-intensity and polarized images of the horizon-scaled radio emission regions of the supermassive black holes, Sgr A*~\cite{EventHorizonTelescope:2022wkp,EventHorizonTelescope:2022urf,EventHorizonTelescope:2022xqj,Collaboration:2024unf} and M87*~\cite{EventHorizonTelescope:2019dse, EventHorizonTelescope:2019pgp,EHT:2023ujh}. For testing the observational predictions of NED theories and our black hole models,  and to address the queries raised in the introduction section, our focus lies solely on the observed angular sizes of Sgr A* and M87* shadows. The bounds, within $1\sigma$ uncertainty region, on the shadow angular diameters of the M87* and the Sgr A* black holes, respectively, are translated to bounds on the shadow radius as follows~\cite{EventHorizonTelescope:2021dqv,EventHorizonTelescope:2022wkp}
\begin{eqnarray}
	4.31 M\leq \; &R_{\text{sh}}& \leq \; 6.08M\\
	4.54 M \leq \; & R_{\text{sh}}&\leq \; 5.22M.\label{EHTbound}
\end{eqnarray}

\begin{figure}[h!]
	\includegraphics[scale=0.87]{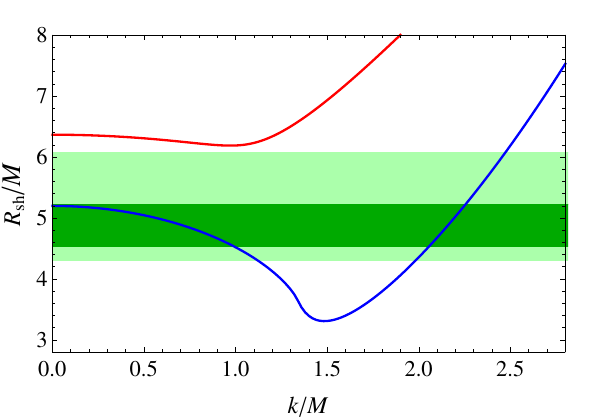}
	\caption{The shadow radii of Bardeen spacetime (red curve) and GC spacetime (blue curve) predicted by the effective metric are shown. The darker green and lighter green regions indicate the EHT bounds for the Sgr A* and M87* black hole shadow radii, respectively. Bardeen black hole and HUCO can be ruled out solely from the EHT shadow-size measurement. }
	\label{fig:Shadow-Eff}
\end{figure}

Figure \ref{fig:Shadow-Eff} illustrates the sizes of black hole shadows for Bardeen and GC spacetimes as a function of NED charge $k$, inferred from photons following the null geodesics of their effective metrics. Few comments are in order. Contrary to expectations, in Bardeen spacetime, $R_{\text{sh}}$ does not converge to the Schwarzschild value even in the vanishing charge limit as $k \to 0$. A relevant question arises: Is this discontinuous limit at $k\to 0$ specific to only the Bardeen spacetime?  No, this holds true for a class of magnetically charged NED black holes, as we shall see in the following Section~\ref{sec-6}. GC spacetime is one such example, where the shadow size predicted by the effective metric has a smooth Schwarzschild limit as $k\to 0$. The shadow sizes $R_{\text{sh}}$ for both Bardeen and GC spacetimes decreases slowly with $k$, reaches a minimum value, and thereafter monotonically increases with $k$, indicating that now HUCOs also cast shadows. The smallest shadow appears for a HUCO with $k=0.97M$ and $R_{\text{sh}}=6.18519M$ for Bardeen spacetime, and $k=1.4856M$ and $R_{\text{sh}}=3.3061M$ for GC spacetime. 
Shadow radii $R_{\text{sh}}$ of Bardeen-BHs and Bardeen-HUCOs fall outside of the EHT bounds (cf. Fig.~\ref{fig:Shadow-Eff}): Shadows of Bardeen-BHs and Bardeen-HUCOs, as predicted by the effective metric, fail to meet the shadow size bounds set by Sgr A* and M87*.  On the other hand, the shadow sizes of GC-BHs within $0\leq k\leq 0.986M$ and $0\leq k\leq 1.119M$ are consistent, respectively, with the Sgr A* and M87*  shadow sizes within $1\sigma$ bounds. Additionally, GC-HUCOs within $2.059M\leq k\leq 2.257M$ and $1.987M\leq k\leq 2.476M$ are also consistent, respectively, with the observed shadows of Sgr A* and M87*. This results in a degeneracy between GC-BHs and GC-HUCOs, where a shadow of a given size could correspond to either a black hole or a HUCO. Furthermore, the observed shadows of M87* and Sgr A* can be explained not only by a GC-BH but also by a GC-HUCO. Thus, further observables are required to distinguish a black hole from a HUCO, such as photon rings as discussed in this paper.  In summary, as predicted by the null geodesics of the effective metric, unlike the Bardeen model, where both black holes and the HUCOs were ruled out by the EHT shadow size observations, GC-BHs and GC-HUCOs satisfy the observed shadow size. This finding is one of the important results of this work.

In light of the conclusive evidence against Bardeen spacetime provided by the EHT shadow size measurement, our focus in the subsequent sections of this paper will shift exclusively to the examination of the GC spacetime shadows as derived from the effective metric.

\section{Effective Spacetime Geometry for Photon Propagation in NED Fields}\label{sec-3}
The main object of our interest in this paper is the effective metric, so we briefly outline the formulation of the effective metric $\tilde{g}_{\mu\nu}$ for photon propagation in the NED charged regular black hole spacetime. We start with the Einstein-Hilbert action minimally coupled with a generic NED field featuring a  smoothly varying Lagrangian density $\mathcal{L(F)}$ \cite{Ayon-Beato:2000mjt}:
\begin{equation}
	S=\frac{c^4}{16\pi G}\int d^4x\sqrt{-g}\left(R-\mathcal{L(F)}\right)\label{action},
\end{equation}	
where $R$ is the Ricci scalar. The choice of $\mathcal{L(F)}$ is guided by the specific NED model phenomenology under investigation. One might expect that under weak electromagnetic fields, the action would reduce to that of Einstein-Maxwell theory. However, as we will see later, there exist interesting NED models that defy these assumptions. Likewise Maxwell field,  the NED field $\mathcal{L(F)}$ respects $U(1)$ gauge invariance and Lorentz symmetry group, making it a function of the invariant   $\mathcal{F}=F_{\mu\nu}F^{\mu\nu}$, where $F_{\mu\nu}$ is the Faraday tensor associated with the four-potential $A_{\mu}$ through $F_{\mu\nu}=\partial_{[\mu}A_{\nu]}$.  $\mathcal{L(F)}$ is a nonlinear function of the single Lorentz  invariant $\mathcal{F}=2(\vec{E}^2-\vec{B}^2)$; $\vec{E}$ and $\vec{B}$ are the radial electric and magnetic fields. On varying the action (\ref{action}) with the metric field tensor $g_{\mu\nu}$ and gauge field $A_{\mu}$, the Einstein-Maxwell field equations read  \cite{Ayon-Beato:2000mjt}
\begin{eqnarray}
	&&G_{\mu\nu}=T_{\mu\nu}\equiv 2\Big(\mathcal{L'}\tensor{F}{_\mu^\alpha}\tensor{F}{_{\nu\alpha}}-\frac{1}{4}g_{\mu\nu}\mathcal{L}\Big),\label{EfE}\\
	&&\nabla_{\mu}\Big(\mathcal{L'}F^{\mu\nu}\Big)=0,\label{EfE1}\\
	&&\nabla_{\mu}\Big(  {}^*F^{\mu\nu}\Big)=0, \label{EfE2}
\end{eqnarray}	
where $\mathcal{L'}:=\partial\mathcal{L(F)}/\partial\mathcal{F},\; \mathcal{L''}:=\partial^2\mathcal{L(F)}/\partial\mathcal{F}^2$ and $^*F^{\mu\nu}$ is the Faraday dual tensor. 

Contrary to the Einstein-Maxwell theory, where the action is quadratic and the field equations are \textit{linear} in $F_{\mu\nu}$, NED theories exhibit nonlinear behavior, as both the action (\ref{action}) and the field equations (\ref{EfE}) and (\ref{EfE2}) are \textit{nonlinear} functions of $F_{\mu\nu}$. The NED field stress tensor is diagonal but not trace-free and violates the strong energy condition
\begin{eqnarray}
	\tensor{T}{^\mu_\nu}=-\text{diag}\Big[\frac{1}{2}\mathcal{L},\, \frac{1}{2}\mathcal{L},\,  \frac{1}{2}\mathcal{L}-\mathcal{F}\mathcal{L'},\, \frac{1}{2}\mathcal{L}-\mathcal{F}\mathcal{L'}\Big].\label{EMT}
\end{eqnarray}
Equation~(\ref{EfE1}) characterizes the nonlinear interaction between electromagnetic fields and Eq.~(\ref{EfE2}) is the consequence of the Bianchi identity. 

Because we are interested in static and spherically symmetric black hole spacetimes, we opt for a general metric ansatz in Schwarzschild coordinates $\{x^{\mu}\}= (t, r, \theta, \phi$)
\begin{equation}
	ds^2=g_{\mu\nu}dx^{\mu}dx^{\nu}=-A(r)dt^2+B(r)\,dr^2+C(r)\,d\Omega_2^2,\label{metric}
\end{equation}
where $d\Omega_2^2=(d\theta^2+\sin^2\theta\, d\phi^2)$ is metric on unit 2-sphere, the Faraday tensor is $F_{\mu\nu}=2\delta^{\theta}_{[\mu}\delta^{\phi}_{\nu]}k\sin\theta$, $\mathcal{F}=\frac{2k}{C^2(r)}$, and $k$ is the magnetic-monopole charge. In spherically symmetric spacetime, electromagnetic fields exhibit two primary configurations: radial electric fields and radial magnetic fields arising from charged monopoles.  Here, we focus only to the magnetic monopole fields. Making a suitable choice of NED Lagrangian density $\mathcal{L(F)}$, one can solve the field equations above to get the metric functions in (\ref{metric}). Bronnikov~\cite{Bronnikov:2000vy} demonstrated that simply requiring finite values for the electromagnetic fields ($\vec{E}$, $\vec{B}$) and the self-energy of a charged point particle in NED models is not enough to produce regular black hole solutions from the gravitational field equations; and that additional conditions are required. For instance, although the Born-Infeld NED model \cite{Born:1934gh} regularizes the $\vec{E}$ field and self-energy of a test charged particle, it fails to produce regular black hole solutions when coupled with gravity. Intriguingly, regular black holes naturally result from Eqs.~(\ref{EfE}) and (\ref{EfE2}), if in the weak-field limit, $\mathcal{L(F)}\to 0$ as $\mathcal{F} \to 0$, and in  the strong-field limit, $\mathcal{L(F)}\to \text{finite}$ as $\mathcal{F} \to \infty$ \cite{Bronnikov:2000vy,Bronnikov:2000yz,Bokulic:2022cyk}.

Using Hadamard's approach of light propagation \cite{Hadamard} -- where the electromagnetic field is continuous, but its derivatives are discontinuous across the light wavefront -- Novello \textit{ et al.}~\cite{Novello:1999pg,Novello:2000km} have shown that  photons no longer propagate along the null geodesics of a fixed background geometry $g_{\mu\nu}$; instead, they follow the null geodesics of an ``effective" Riemannian geometry $\tilde{g}_{\mu\nu}$ modified by the NED effects. This results in the fact that the photon four-momentum, $p_{\mu}$, is a geodesic vector satisfying the following equations \cite{DeLorenci:2001gf,Novello:1999pg,Novello:2000km,Novello:2001gk,Novello:2000xw} 
\begin{eqnarray}
	&&p^{\mu}\nabla_{\mu}p^{\nu}=0,\\
	&&g_{\mu\nu}p^{\mu}p^{\nu} \neq 0,\qquad \text{rather}\qquad \tilde{g}_{\mu\nu}p^{\mu}p^{\nu}=0.
\end{eqnarray}
The effective metric $\tilde{g}_{\mu\nu}$ for photons depends on the background metric $g_{\mu\nu}$ and the NED field $\mathcal{L(F)}$, such that\footnotemark~
\begin{equation}
	\tilde{g}^{\mu\nu}= \mathcal{L'} g^{\mu\nu} - 4 \mathcal{L''} F^{\mu\alpha} \tensor{F}{_\alpha^\nu},\label{EffMetric}
\end{equation}
where
\begin{equation}
	\tilde{g}_{\mu\nu} \tilde{g}^{\mu\sigma}= \delta^{\sigma}_{\nu}.
\end{equation}
\footnotetext[1]{A similar effective metric can be derived with an electric charge induced NED field \cite{Novello:1999pg,Novello:2001gk}.} This effective metric for photon propagation in NED backgrounds is an example of nonlinear medium induced modification on light propagation which is generic for any nonlinear field theory and known for a quite long time now~\cite{Plebanski,Plebanski:1970zz,Novello:2003je,Novello:1999pg,Gibbons:2001gy}. 

For the background geometry $g_{\mu\nu}$ given in Eq.~(\ref{metric}), the effective metric $\tilde{g}_{\mu\nu}$  reads~\cite{Novello:1999pg,Novello:2001gk}
\begin{eqnarray}
	\tilde{\rm ds}^{2}&=&\tilde{g}_{\mu\nu}dx^{\nu}dx^{\nu}\,\nonumber\\
	&=&-\frac{A(r)}{\mathcal{L'}}dt^2+\frac{B(r) }{\mathcal{L'}}dr^2+\frac{C(r)}{\Phi}\, d\Omega_2^2\label{effmetric},		
\end{eqnarray}
with
\begin{equation}
	\Phi:=\mathcal{L'}+2 \mathcal{F} \mathcal{L''}.\label{phi}
\end{equation}
The effective metric (\ref{effmetric}) mirrors the static and spherically symmetric nature of the background metric (\ref{metric}). Note that the effective metric $\tilde{g}_{\mu\nu}$ is relevant \textit{only} for photon motion; other uncharged, massless (or massive) particles remain unaffected by the nonlinearity of the NED field and adhere to the null (or timelike) geodesics of the background metric $g_{\mu\nu}$. Additionally, under a linear Maxwell's field ($\mathcal{L(F)}\sim \mathcal{F}$), photons also traverse along the null geodesics of the background metric, indicating that the conformal modification outlined in Eq.~(\ref{effmetric}) does not alter their trajectory. Furthermore, contrary to expectations, the metric $\tilde{g}_{\mu\nu}$ does not necessarily match the $g_{\mu\nu}$ metric in the large-$r$ limit. This will be emphasized in Section (\ref{sec-5}), where we discuss two black hole models with different weak-field limits of $\tilde{g}_{\mu\nu}$. 

In summary, in the presence of NED fields, the spacetime geometry seen by the photons is more complex than the geometry seen by other relativistic particles yielding some unexpected and wide-ranging consequences. 

Notably, Bardeen metric is also treated as quantum-corrected Schwarzschild spacetime \cite{Maluf:2018ksj,Konoplya:2023ahd}, with parameter $k$ controlling the quantum correction. Now,  instead of a NED field, stringy effects create a de Sitter core at the origin, stabilizing the matter configuration against collapse. This offers an alternative explanation for the most of regular black holes, wherein photons continue to follow the null geodesics of the background metric. However, due to lack of a well-defined action principle, these alternative explanations are not widely accepted. 
Nevertheless, considering the Einstein-NED theory resulting in regular black holes, our focus is to determine the qualitative effects of the NED fields on black hole's observational features . For this, in the rest of the paper, we compare the shadows predicted by the null geodesics of the effective metric $\tilde{g}_{\mu\nu}$, presented in section~\ref{sec-5b}, with those predicted by the null geodesics of the background metric $g_{\mu\nu}$, presented in appendix~\ref{sec-8A}. This comparative analysis serves as the focal point of our investigation, shedding light on the distinct outcomes stemming from the inclusion of NED fields within the black hole spacetimes.

\section{Black Hole Shadow}\label{sec-4}
\begin{figure}
	\includegraphics[scale=0.7]{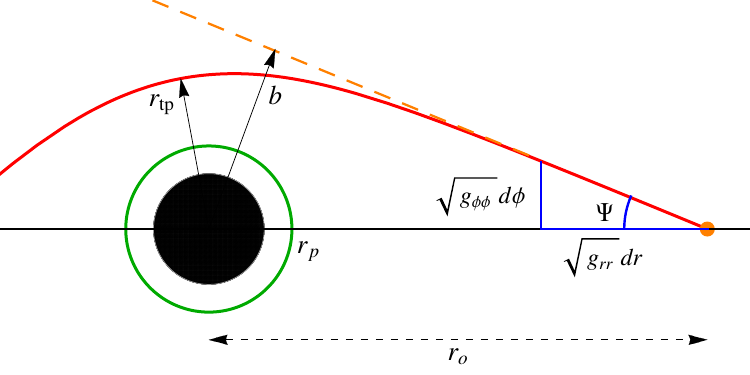}
	\caption{A schematic diagram for the black hole shadow. The observer, represented by an orange dot, is situated at a distance  $r_o$ from the black hole. A light ray with an impact parameter $b$ experiences a turning point at radial distance $r_{\text{tp}}$ and makes an angle $\Psi$ on the observer's screen. See text in Section~\ref{sec-4} for more details.}\label{fig:schematic}
\end{figure}

Isometries of the effective metric (\ref{effmetric}) enable the formulation of the photon geodesic equations; the four-velocity of photon reads as follows:
\begin{eqnarray}
	\{\dot{x}^{\mu} \}&=&\Big\{ \frac{E \mathcal{L'}}{A(r)},\; \pm \frac{\mathcal{L'}}{B(r)} \sqrt{ \frac{B(r)}{A(r)}E^2  -(\mathcal{K}+L^2) \frac{B(r)\Phi}{C(r) \mathcal{L'}}  }, \nonumber\\
	&&\pm \frac{\Phi}{C(r)}\sqrt{\mathcal{K}-L^2 \cot^2\theta },\; \frac{L\Phi }{C(r)\sin^2\theta}\Big\},\label{r}
\end{eqnarray}
where $\dot{x^{\mu}}=d x^{\mu}/d\tau$, and $\tau$ is the affine parameter along the geodesics.  $E$ and $L$ represent the energy and angular momentum of a photon, respectively, while $\mathcal{K}$ stands for the Carter constant. For photons confined to geodesics along the equatorial plane $\theta=\pi/2$, $\mathcal{K}=0$. 
With this, the radial motion equation takes the following simpler form at $\theta=\pi/2$
\begin{eqnarray}
	-\tilde{g}_{tt}\tilde{g}_{rr}\dot{r}^2+V&=&E^2,\label{req}	
\end{eqnarray}	
such that the radial potential $V$ for the null geodesics of the effective metric
\begin{equation}
	V=-\frac{\tilde{g}_{tt}}{\tilde{g}_{\phi\phi}}L^2=\frac{\Phi A(r)}{\mathcal{L'} C(r)} L^2,\label{effectivepot1}
\end{equation}
and for the null geodesics of the background metric, it simplifies as
\begin{equation}
	V=\frac{ A(r)}{C(r)} L^2.\label{effectivepot}
\end{equation}
The radial potential vanishes at the black hole horizon $A(r)=0$ and at $\Phi(r)=0$. Radial motion is possible only when $\dot{r}^2\geq 0$ and any turning point $r=r_{\text{tp}}$ in trajectory can be obtained by solving $\dot{r}=0$ as a function of the impact parameter $b$ that is defined as $L/E$ [see Eqs.~(\ref{r}) and (\ref{effectivepot})]
\begin{equation}
	b=\frac{L}{E}=\left.\sqrt{ \frac{\tilde{g}_{\phi\phi}}{-\tilde{g}_{tt}}}\right|_{r=r_{\text{tp}}}= \left.\sqrt{\frac{\mathcal{L'} C(r)}{\Phi A(r)}}\right|_{r=r_{\text{tp}}}.\label{impact0}
\end{equation}	
We will focus on the unstable circular photon orbits, which correspond to the local extrema of the radial potential. The radius $r_{\rm p}$ of a circular orbit is determined by simultaneously solving the following equations for $r$
\begin{equation}
	p^r=0,\;\; \frac{d p^ r}{dr}=0,
\end{equation}
which can be cast in a single equation as follows
\begin{equation}
	\frac{d}{dr}\Big(\frac{\tilde{g}_{tt}}{\tilde{g}_{\phi\phi}}\Big)=0,\\
\end{equation}
or equivalently 
\begin{equation}	
	H(r) := \left(\frac{A'(r)}{A(r)}-\frac{C'(r)}{C(r)}\right)- \Big(\frac{\mathcal{L}'' }{\mathcal{L}'} -\frac{\Phi '}{\Phi}\Big)=0,\label{PhotonOrbit}
\end{equation}
such that $H(r=r_{\rm p})=0$ gives circular photon orbit radius $r_{\rm p}.$
Equation~(\ref{PhotonOrbit}) for Maxwell electrodynamics reduces to the simpler form
\begin{equation}
	A'(r)C(r)-A(r)C'(r)=0.\label{NeutriniOrbit}
\end{equation}
Real positive roots of Eq.~(\ref{PhotonOrbit}) and (\ref{NeutriniOrbit}) give radii of null circular orbits from the effective metric and the background metric, respectively. The critical value of the impact parameter, $b_{\text{cr}}$, is related with $r_{\rm p}$ as following
\begin{equation}
	b_{\text{cr}}=\left.\sqrt{ \frac{\tilde{g}_{\phi\phi}}{-\tilde{g}_{tt}}}\right|_{r=r_{\rm p}}= 
	\left. \sqrt{\frac{\mathcal{L'} C(r)}{\Phi A(r)}}\right|_{r=r_{\rm p}}.\label{impact}
\end{equation}
In other words, as the impact parameter reaches its critical value $b\to b_{\text{cr}}$, the turning radius approaches the null unstable circular orbit radius, $r_{\text{tp}}\to r_{\rm p}$ [cf. Eqs.~(\ref{impact0}) and (\ref{impact})]. Determining the shadow of a black hole requires careful use of both the $\tilde{g}_{\mu\nu}$ and $g_{\mu\nu}$ metrics: light rays forming the shadow boundary follow $\tilde{g}_{\mu\nu}$ metric, whereas the observer receives them in the metric $g_{\mu\nu}$. To illustrate this, let us consider a static observer located at $(r_o, \theta_o)$ shooting light rays into the past toward the black hole, as shown schematically in Fig.~\ref{fig:schematic}. Light rays emanating within a cone with semiopening angle $\Psi_{\text{sh}}$, the angle between a light ray tangent and the radial direction, at the observer's position falls into the black hole, forming the dark region on the image plane--``\textit{shadow}"~\cite{Perlick:2021aok}. Whereas, following the past-oriented light rays which asymptotically approach the unstable circular photon orbit with radius $r_{p}$, one can determine the shadow boundary curve, known as the ``critical curve" or ``light ring"  that is characterized by the angular radius $\Psi_{\text{sh}}$. 

To locate the shadow boundary, we project the four-momentum $p_{\mu}$ of these light rays onto the observer's tetrad frame defined by bases $\textbf{e}_{(\rho)}$ determined from $g_{\mu\nu}$.  These bases satisfy 
\begin{equation}
	g_{\mu\nu}\textbf{e}^{\mu}_{(\rho)}\textbf{e}^{\nu}_{(\sigma)}=\eta_{(\rho)(\sigma)}, 
\end{equation}
where $\eta_{(a)(b)}$ is the Minkowski metric in the observer's local frame. The basis vectors are defined such that $\textbf{e}_{(t)}$ is timelike, while $\textbf{e}_{(r)}$, $\textbf{e}_{(\theta)}$ and $\textbf{e}_{(\phi)}$ are spacelike and mutually orthonormal. Consider in the observer's frame, a light ray with locally measured four-momentum components $p_{(a)}=\textbf{e}^{\mu}_{(a)}p_{\mu}$ forms an angle $X$ from the $\textbf{e}_{(r)}-\textbf{e}_{(\phi)}$ plane, while its projection onto the $\textbf{e}_{(r)}-\textbf{e}_{(\phi)}$ plane forms an angle $\Psi$ from the $\textbf{e}_{(r)}$ direction, such that
\begin{eqnarray*}
	&&p^{(r)}=|P|\cos X \cos\Psi,\nonumber\\
	&&p^{(\theta)}=|P|\sin X,\nonumber\\
	&&p^{(\phi)}=|P|\cos X\sin\Psi,\label{Eq1}
\end{eqnarray*}
with  $|P|^2\equiv |p^{(r)}|^2+|p^{(\theta)}|^2+|p^{(\phi)}|^2$ and 
\begin{eqnarray}
	p^{(a)}p_{(a)}&=& -|p^{(t)}|^2+|P|^2,
\end{eqnarray}
where ($X, \Psi$) defines the celestial coordinates in the observer frame. 
From the normalization condition
\begin{eqnarray}
	p^{(a)}p_{(a)}&=&\eta^{(a)(b)}p_{(a)}p_{(b)},\nonumber\\
	&=& \eta^{(a)(b)}\textbf{e}^{\mu}_{(a)}\textbf{e}^{\nu}_{(b)}p_{\mu}p_{\nu},\nonumber\\
	&=& g^{\mu\nu}p_{\mu}p_{\nu}\neq 0,\label{normalization}
\end{eqnarray}
therefore, in a local intertial frame of the observer, the norm of light four-momentum does not vanish, implying that the light does not propagate along the null geodesics in the observer's frame. 

We establish an image plane centered at the black hole, and defined by Cartesian coordinate ($\alpha, \beta$)
\begin{equation}
	\alpha:=\left. -r \frac{p^{(\phi)}}{p^{(r)}}\right|_{(r_o,\theta_o)},\quad 
	\beta:=\left. r \frac{p^{(\theta)}}{p^{(r)}}\right|_{(r_o,\theta_o)}.\label{Eq2}
\end{equation}
This gives a relation between image plane coordinates ($\alpha, \beta$) and celestial coordinates ($X, \Psi$) as follows
\begin{equation}
	\alpha=-r_o\tan\Psi,\;\;\; \beta=r_o\tan X\sec\Psi,\label{Eq3}
\end{equation}
such that a light ray starting with local four-momentum $p^{(a)}$ or in the direction ($X, \Psi$) intersects the image plane on a point with coordinate $(\alpha,\beta)$. Using Eqs.~(\ref{Eq1})-(\ref{Eq3}),
\begin{eqnarray}
	\tan\Psi&=&\frac{p^{(\phi)}}{p^{(r)}}=\frac{\textbf{e}^{(\phi)}_{\phi}p^{\phi}}{\textbf{e}^{(r)}_{r} p^r}=\frac{\sqrt{g_{\phi\phi}}}{\sqrt{g_{rr}}}\frac{d\phi}{dr},	\label{Eq4}
\end{eqnarray}
where $d\phi/dr$ is determined from the effective metric. Because the shadow boundary is circular for a static black hole,  its radius on the image plane is defined as 
\begin{equation}
	R_{\text{sh}}=\sqrt{\alpha^2+\beta^2};
\end{equation}
and this is the only quantity characterizing the shadow. On the other hand,  additional features appear in the shadow in the case of stationary black holes and these could be generically captured via a Fourier expansion of the polar curve describing the shadow (see, \cite{Abdujabbarov:2015xqa}). Furthermore, the light trajectories are planar, i.e. $p^{(\theta)}=0$, this simplifies the shadow radius expression as
\begin{equation}
	R_{\text{sh}}=r_o\tan\Psi_{\text{sh}}.
\end{equation}
From the null geodesic equations, we get
\begin{equation}
	\left(\frac{dr}{d\phi}\right)^2=\frac{\tilde{g}_{\phi\phi}}{\tilde{g}_{rr}}\left(\frac{\tilde{g}_{\phi\phi}}{\tilde{g}_{tt}}\frac{1}{b^2}-1\right),
\end{equation}
and using this in the Eq.~(\ref{Eq4}), yields the shadow angular radius $\Psi_{\text{sh}}$ from
\begin{equation}
	\cot\Psi_{\text{sh}}=\frac{\sqrt{g_{rr}}}{\sqrt{g_{\phi\phi}}}\sqrt{\frac{\tilde{g}_{\phi\phi}}{\tilde{g}_{rr}}\left(\frac{\tilde{g}_{\phi\phi}}{\tilde{g}_{tt}}\frac{1}{b_{\text{cr}}^2}-1\right)},
\end{equation}
and the shadow radius as follows
\begin{eqnarray}
	R_{\text{sh}}&=&r_o\tan\Psi_{\text{sh}},\nonumber\\
	&=& r_o \frac{\sqrt{g_{\phi\phi}(r_o)}}{\sqrt{g_{rr}(r_o)}}\Bigg[\sqrt{\frac{\tilde{g}_{\phi\phi}(r_o)}{\tilde{g}_{rr}(r_o)}\left(\frac{\tilde{g}_{\phi\phi}(r_o)}{\tilde{g}_{tt}(r_o)}\frac{\tilde{g}_{tt}(r_{\rm p})}{\tilde{g}_{\phi\phi}(r_{\rm p})}-1\right)}\Bigg]^{-1/2},\label{shadowsize}
\end{eqnarray}
where we have used Eq.~(\ref{impact}) to substitute $b_{\text{cr}}$. Equation~(\ref{shadowsize}) gives shadow radius for arbitrary static spherically symmetric NED black hole, where photons follow effective metric. Note, shadow radius is not the same as the impact parameter. In the Maxwell electrodynamics, $\tilde{g}_{\mu\nu} \to g_{\mu\nu}$, and consequently the shadow radius simplifies to 
\begin{equation}
	R_{\text{sh}}= r_o \Bigg[\frac{g_{\phi\phi}(r_o)}{g_{tt}(r_o)}\frac{g_{tt}(r_{\rm p})}{g_{\phi\phi}(r_{\rm p})}-1\Bigg]^{-1/2},
\end{equation}
which, for an asymptotically flat spacetime and a far away observer $r_o\ggg M$, further simplifies to
\begin{equation}
	R_{\text{sh}}= \sqrt{\frac{g_{\phi\phi}(r_{\rm p})}{-g_{tt}(r_{\rm p})}}=b_{\text{cr}}.
\end{equation}
Only if the effective metric is asymptotically flat, i.e.,  $\displaystyle{\lim_{r_o\to \infty}} \{\tilde{g}_{tt}(r_o), \tilde{g}_{rr}(r_o), \tilde{g}_{\phi\phi}(r_o) \}\to \{-1, 1, r_o^2\sin^2\theta \}$, and the observer is in the asymptotically flat region, then shadow radius is
\begin{equation}
	R_{\text{sh}}=\sqrt{\frac{\tilde{g}_{\phi\phi}(r_{\rm p})}{-\tilde{g}_{tt}(r_{\rm p})}}=\left. \sqrt{\frac{\mathcal{L'} C(r)}{\Phi A(r)}}\right|_{r=r_{\rm p}}=b_{\text{cr}}.\label{shadowsize1}
\end{equation}
However, the effective metric is not asymptotically flat in most of the non-Maxwellian  magnetically charged NED spacetimes, and thus the shadow radius cannot be calculated from Eq.~(\ref{shadowsize1}); even though it is sometime used for NED black holes~\cite{Stuchlik:2019uvf}. Hereafter, we will use expression given in Eq.~(\ref{shadowsize}) to calculate the shadow size of black holes and HUCOs.

Two features of the spacetime are of particular importance: photon orbit radius $r_{\rm p}$ and shadow radius $R_{\rm sh}$. While the shadow radius solely depends on the black hole parameters and observer's location, the overall intensity profile additionally depends on the accretion details. In this paper, we consider two very simple accretion models: a spherically symmetric radially infalling Bondi-Michel accretion and a Novikov-Thorne type optically and geometrically thin accretion disk on the equatorial plane. Although these accretion models are simplistic representations and do not fully capture the complexity of supermassive black hole accretion \cite{EventHorizonTelescope:2022urf,EventHorizonTelescope:2019pgp}, they suffice to address the objectives outlined in the introduction. In particular, since we are interested in quantifying the effects of a NED effective metric on black hole shadows and comparing the images of regular black holes with those of the regular HUCOs, these two accretion  models serve the purpose well. Additionally, these models are simple enough to be analytically tractable, allowing us to compute the intensity profile semi-analytically. Moreover, we argue that the major findings are independent of the choice of accretion physics. 

\subsection{Radially Infalling Spherical Accretion Model}\label{sec-4a}
For our first accretion model, we consider that the regular black holes or the regular HUCOs are surrounded by a spherically symmetric, optically thin, and isotropically radiating matter \cite{Bambi:2013nla,Narayan:2019imo}. This matter, lacking azimuthal angular momentum, is radially infalling and following timelike geodesics of $g_{\mu\nu}$. For further insights into the interplay between the gravitational physics of black holes and the emission profile for spherically symmetric accretion in resulting shadows, refer to \cite{Kocherlakota:2022jnz,Kocherlakota:2023qgo,Kocherlakota:2024hyq}.  

The matter is emitting radiation, which follow null geodesics of the metric $\tilde{g}_{\mu\nu}$.  The shift in photon's frequency from the point of emission $r_{e}$ to the point of observation $r_{o}$ is defined by the redshift factor 
\begin{equation}
	z:=\frac{\nu_{o}}{\nu_{e}}=\frac{p_{\rho} u _o^{\rho}}{p_{\sigma } u _e^{\sigma }}, \label{EQ3.6}
\end{equation}
which includes both the gravitational and Doppler redshift from $\nu_e\to \nu_o$. Notice that the asymptotic observer is static and moves along a timelike word-line of the background metric, i.e., $g_{\rho\sigma}u _o^{\rho}u _{o}^{\sigma}=-1$, with four-velocity $u _o^{\rho}=\dot{x}^{\rho}=\frac{1}{\sqrt{-g_{tt}(r=r_o)}}\delta_t^{\rho}$. Similarly, the timelike unit normalization fixes the radial four-velocity of the emitter 
\begin{equation}
	u_e^{\sigma}=\left.\left(-\frac{1}{g_{tt}},-\sqrt{\frac{-1-g_{tt}}{g_{tt}\,g_{rr}}},\,0,\,0\right)\right|_{r=re}.
\end{equation}
Using the photon geodesics equations and the normalization condition with respect to the metric $\tilde{g}_{\mu\nu}p^{\mu}p^{\nu}=0$, the redshift factor becomes
\begin{eqnarray}
	z&=&\left.\sqrt{\frac{-1}{g_{tt}}}\right|_{r=r_o}\left.\left(\frac{-1}{g_{tt}} -\sqrt{\frac{-1-g_{tt}}{g_{tt}g_{rr}}} \frac{p_r}{p_t}\right)^{-1}\right|_{r=re},\nonumber\\
	z_{\epsilon}&=& \frac{A(r_e)}{\sqrt{A(r_o)}}\left.
	\Bigg[1-\epsilon \Big[(1-A(r)) (1-b^2\frac{\Phi}{\mathcal{L'}}\frac{A(r)}{C(r)})\Big]^{1/2}\Bigg]^{-1}
	\right|_{r=r_e}\label{redshift}
\end{eqnarray}
with
\begin{equation}
	p_r=\epsilon\, p_t\sqrt{\frac{-1}{\tilde{g}^{rr}}\left(\tilde{g}^{tt}+b^2 \tilde{g}^{\phi\phi}\right)},\label{pr}
\end{equation}
and we set $\epsilon$ to either $-1$ or $+1$, depending on whether the photon velocity is directed locally toward or away from the black hole, respectively. 

An infinitesimal path length for photons along emission point to the observer is given by
\begin{eqnarray}
	d\ell& =&-p_{\mu} u_e^{\mu}\, d\lambda= -\sqrt{\frac{-1}{g_{tt}(r_o)}} \frac{p_t}{z_{\epsilon}\, p^r} \, dr.
\end{eqnarray}
Additionally, we assume isotropic and monochromatic radiation emission in the rest frame of the accreting matter. While the matter may possess a finite absorption coefficient, for the purposes of EHT observing frequencies, we consider an optically-transparent accretion flow with zero absorption.
For emissivity $j(\nu)$,  we consider frequency-independent, corresponding to emission close to the peak of the synchrotron emissivity,  spherically symmetric profile which scales as  $j(\nu) \propto 1/r^2$ \citep{Falcke:1999pj,Bambi:2013nla}.
The observed specific intensity $\mathcal{I}_{\nu }^{\text{obs}}$ at the photon frequency $\nu_{o}$ can be obtained by integrating the emissivity along the photon path $\Gamma$ as follows \citep{Jaroszynski:1997bw,Bambi:2013nla}
\begin{equation}
	\mathcal{I}_{\nu }^{\text{obs}}=\int_{\Gamma}^{}z^3 j(\nu_e)\,d\ell.\label{emission}
\end{equation}

Integrating Eq.~(\ref{emission}) over all the observed frequencies, we obtain the observed photon intensity ~\cite{Bambi:2013nla}
\begin{equation}
	\mathcal{I}^{\text{obs}}\propto \int_{\Gamma}^{}\frac{z_{\epsilon}^4}{r^2}\,d\ell	\propto - \int_{\Gamma}^{}\frac{z_{\epsilon}^3}{r^2}\sqrt{\frac{-1}{g_{tt}(r_o)}}\frac{p_t}{p^r}\,dr,
\end{equation}
where for a far away observer $-g_{tt}(r_o)=1$,  and we transform an integral over the affine parameter, $\ell$, to an integral over the radial coordinate $r$.

For $b < b_{\rm c}$, light rays are traced backward in time from the observer to the central object. Photons along these rays experience a monotonic redshift forward in time. Conversely, for $b > b_{\rm c}$, light rays are first traced backward from $r_o$ to a turning point $r_{\text{tp}}$ along their trajectory, where photons are redshifted, and then from $r_{\text{tp}}$ to $r_e$, where photons experience a blueshift, so that
\begin{equation}
	\mathcal{I}_{\text{obs}}=
	\begin{cases}
		\int_{r_+}^{r_o}\frac{z_+^3}{r^2}\frac{p_t}{p^r}\,dr & \text{if } b< b_{\rm c}\\
		-\int_{r_e}^{r_{\text{tp}}}\frac{z_-^3}{r^2}\frac{p_t}{p^r}\,dr + \int_{r_{\text{tp}}}^{r_o}\frac{z_+^3}{r^2}\frac{p_t}{p^r}\,dr & \text{if } b\geq  b_{\rm c}
	\end{cases}\label{InfallInten}	
\end{equation}
Overall, the effects of the NED field are most evident in the redshift factor given by  Eq.~(\ref{redshift}) and in Eq.~(\ref{pr}). Thereafter, we will use Eq.~(\ref{InfallInten}) to calculate the images of regular black holes and their corresponding HUCOs within the effective metric. For the background metric, we simply use $\tilde{g}_{\mu\nu} \to g_{\mu\nu}$.

\subsection{Thin Disk Model}\label{sec-4b}
In our second model of accretion, we adopt a Novikov-Thorne like geometry \cite{Novikov:1973kta} for an optically thin accretion disk, neglecting absorptivity. 
The disk is made up of massive particles rotating along nearly Keplerian timelike circular orbits on the equatorial plane defined by the background metric. The emission region is assumed to be optically thin, allowing light to traverse it multiple times; otherwise, only a direct image of the disk is observable. We set the outer edge of the disk at a fixed radius of $r_{\text{out}}=20M$. While horizons exist only for $k\leq k_{\rm E}$, timelike circular orbits persist even for larger values $k\leq k_{\rm c}^{\text{TL}}$, where $k_{\rm c}^{\text{TL}}>k_{\rm E}$. For $k\leq k_{\rm c}^{\text{TL}}$, disk inner edge is at the $r_{\text{in}}=r_{\text{ISCO}}$, where ISCO is the innermost stable circular orbit, whereas for $k>k_{\rm c}^{\text{TL}}$, it is determined by solving $\left.\frac{d\Omega_o }{dr}\right|_{r=r_d}=0$, where $\Omega_o$ is disk angular velocity defined as follows \cite{Page:1974he}
\begin{equation}
	\Omega_o=\sqrt{\frac{-g_{tt,r}}{g_{\phi\phi,r}}}.
\end{equation}
For efficient accretion to happen, $\Omega_o$ should increase with decreasing $r$, therefore the inner edge is defined where $\Omega_o$ is globally maximum. Thus disk inner edge is fixed as \cite{Page:1974he,Harko:2009xf}
\begin{equation}
	r_{\text{in}}=
	\begin{cases}
		r_{\text{ISCO}}& \text{if } k\leq k_{\rm c}^{\text{TL}}\\
		r_d;\,\, \left.\Omega_o'\right|_{(r=r_d)} =0 & \text{if } k> k_{\rm c}^{\text{TL}}.
	\end{cases}\label{DiskInnerEdge}	
\end{equation}
Keplarian velocity $\Omega_o$ monotonically increases as we radially move inward along the accretion disk. Consequently, matter located closer to the disk's center orbits faster than matter farther out. This differential rotation induces tension within the disk, compelling matter in the inner regions to decelerate, thus leading to a loss of angular momentum in the outward direction and subsequent inward migration of matter toward lower orbits. However, if $\Omega_o$ were to decrease with $r$, it would be impossible for angular momentum to be transported outward, making the accretion process impossible. 

No emission is coming from $r>20M$ and within $r<r_{\text{in}}$. The flux of electromagnetic radiation emitted from a radial position $r_e$ within the accretion disk follows the standard formula \cite{Novikov:1973kta,Page:1974he}:
\begin{equation}
	\mathscr{F}(r) = -\frac{\dot{M}}{4\pi \sqrt{-g^{(3)}}}\frac{\Omega_o'}{(E_o-\Omega_o L_o)^2}\int_{r_{\text{in}}}^{r_e}(E_o-\Omega_o L_o) L_o' dr,\label{IntenAccDisk}
\end{equation}
where $\dot{M}$ is the mass accretion rate which is assumed to be time independent and fixed. $E_o$, $L_o$ and $\Omega_o$ are the energy, specific angular momentum and the orbital angular velocity of massive particles moving along the circular orbits in the accretion disk
\begin{equation}
	E_o= \frac{-g_{tt}}{\sqrt{-g_{tt}-\Omega_o^2 \,g_{\phi\phi}}},\quad
	L_o= \frac{\Omega_o\, g_{\phi\phi}}{\sqrt{-g_{tt}-\Omega_o ^2\, g_{\phi\phi}}}.
\end{equation}
$\sqrt{-g^{(3)}}$ is the determinant of the induced $3\times 3$ metric on the accretion disk plane.

The redshift factor is given in Eq.~(\ref{EQ3.6}), but now the four-velocity of the emitter is 
\begin{equation}
	u^{\sigma}_e=\left. u^t_e\left(1,0,0,\Omega_o\,\right)\right|_{r_e},
\end{equation}
and $g_{\mu\nu}u^{\mu}_eu^{\nu}_e=-1$ fixes the normalization coefficient as
$$u^{t}_e=\sqrt{\frac{-1}{g_{tt}+\Omega_o^2\, g_{\phi\phi}}}.$$
Redshift factor reads \cite{Schee:2019,Schee:2019gki}
\begin{equation}
	z=\left.\frac{1}{\tilde{g}_{tt}}\right|_{r_o}\left.\left(\frac{\sqrt{-g_{tt}-\Omega_o^2\, g_{\phi\phi}}}{\frac{-g_{tt}}{\tilde{g}g_{tt}}+b\Omega_o\, \frac{g_{\phi\phi}}{\tilde{g}_{\phi\phi}}}\right)\right|_{r_e},\label{redshiftAcc}
\end{equation}
which in the background metric simplifies to \cite{Schee:2019,Bambi:2012tg}
\begin{equation}
	z=\left.\frac{\sqrt{-g_{tt}-\Omega_o^2\, g_{\phi\phi}}}{(1-b\Omega_o)}\right|_{r_e}.
\end{equation}
$z$ determines how much the frequency of photons changes when they are emitted from a point $r_e$ on the disk and then observed far away at a distance $r_o$. The photon flux detected by the distant observer is quantified by
\begin{equation}
	\mathcal{I}_{obs}(r)=z^4 \mathscr{F}(r).\label{AccDiskfinal}
\end{equation}

To determine the observed flux at specific image screen coordinates ($\alpha_i, \beta_i$), we start by solving the null geodesic equations derived from the metric $\tilde{g}_{\mu\nu}$, with initial conditions fixed in terms of the coordinates ($\alpha_i, \beta_i$). Backtracking this light ray helps us determine the point of intersection or the emission point on the accretion disk. Then, we calculate the corresponding redshift to be used in Eq.~(\ref{AccDiskfinal}). On the other hand, if for some other specific values of ($\alpha_i, \beta_i$), null geodesics do not intersects the accretion disk,  we assign a flux value of zero to those points on the image plane. We systematically vary the ($\alpha_i, \beta_i$) values to cover the entire image plane, thus generating an intensity map of the accretion disk's images on the observer's image plane. For this, we consider that the observer is placed at the radial coordinate $r_o=10^4M$, which corresponds effectively to the asymptotic infinity. In the optically thin limit, each pixel represents the cumulative emission along an entire null geodesic

\section{Regular Black Hole Spacetimes}\label{sec-5}

\begin{table*}[ht!]
	\begin{tabular}{|c||c|c|c|c|}
		\hline
		\multirow{2}{*}{Spacetime}     & \multirow{2}{*}{Charge} 	&\multirow{2}{*}{$\displaystyle{\lim_{r \to 0}} \; g_{\mu\nu}$}  &\multirow{2}{*}{$\displaystyle{\lim_{r\gg M}}\; \mathcal{L(F)}$}  
		& \multirow{2}{*}{$\displaystyle{\lim_{k \to 0}} \;\tilde{g}_{\mu\nu}$}         \\	
		&        &             &             &                 \\ \hline
		
		Bardeen~\cite{Bardeen-RegularBH}             & M 		& de-Sitter   & $\times$    & $\times$	  	\\  
		Bardeen~\cite{Rodrigues:2018bdc}    & E 	    & de-Sitter	  & $\times$	& $\times$	  \\
		Hayward~\cite{Hayward:2005gi}       & M 	    & de-Sitter   & $\times$    & $\times$    \\
		Ghosh-Culetu~\cite{Culetu:2014lca,Ghosh:2014pba}& M 		  &Minkowski    & $\checkmark$ &$\checkmark$  \\
		Simpson-Visser~\cite{Simpson:2018tsi,Bronnikov:2021uta} 	  & M 		    & bouncing    & $\times$ 	  & $\times$	\\
		Polymerized LQG~\cite{KumarWalia:2022ddq}       & M 		  & bouncing    & $\times$ 	  & $\times$	\\ 
		Dymnikova~\cite{dePaula:2023ozi} 	& E   		& de-Sitter   &$\checkmark$ & $\checkmark$ \\
		ABG~\cite{Novello:2000km,Ayon-Beato:1998hmi} 	& E   		& de-Sitter   &$\times$ & $\times$ \\ \hline
	\end{tabular}
	\caption{A nonexhaustive list of NED-charged regular black holes. Table columns include: (i) NED spacetimes, (ii) magnetic (M) or electric (E) monopole charges of NED, (iii) black hole's core geometry, (iv) $\checkmark$ for Maxwell weak-field limit of NED model or $\times$ otherwise, (v) $\checkmark$ for Schwarzschild limit as $k\to 0$ for the effective metric or otherwise $\times$.} \label{BH Table}
\end{table*}

It is important to declare here that we refer ``regular black hole" as one that is free from the curvature singularity, i.e., black holes with globally finite spacetime curvature, while being aware that curvature singularities might not be synonymous with geodesic incompleteness in general; see, for instance, \cite{Carballo-Rubio:2019fnb,Bejarano:2017fgz}. However, for spherically symmetric black holes with one shape function in four-dimensional GR, the finiteness of curvature invariants and the geodesics completeness are equivalent ~\cite{Hu:2023iuw}. Nevertheless, the possibility of  black holes with finite curvature scalars and extendable geodesics, or those with curvature singularity and inextendable geodesics, remains open. 

While regular black holes mimic the Schwarzschild metric in the exterior weak-gravitational field limit and are asymptotically flat, their interiors can manifest a variety of geometries near the core, for a in-depth exploration, refer to the reviews by Sebastiani and Zerbini  ~\cite{Sebastiani:2022wbz} and Carballo-Rubio \textit{et al.}~\cite{Carballo-Rubio:2023mvr}. Noteworthy examples include: (i) a de-Sitter interior geometry with an even number of horizons, (ii) a Minkowski core with an even number of horizons, (iii) a spacelike wormhole throat hidden inside a trapping horizon or a black-bounce core, and (iv) discrete spacetime structure or the presence of a fundamental length. 
Table~\ref{BH Table} presents some static and spherically symmetric NED charged regular black holes, along with their key geometric features including interior geometry nature. Our selection of models is by no means complete and represent only a subset of NED regular spacetimes.

For the purposes of this study, we focus only on two particular models: Bardeen ~\cite{Bardeen-RegularBH} and Ghosh-Culetu~\cite{Culetu:2014lca, Ghosh:2014pba}. The spacetime curvature remains globally finite for $r\geq 0$, rendering both black holes geodesically complete and asymptotically flat. Both Bardeen and GC  metrics are characterized by a charge  $k$ in addition to mass $M$, such that in the limit $k\to 0$, both metrics converge to the Schwarzschild metric. Furthermore, within a limiting parameter space $k< k_{\rm E}$, these metrics depict a regular black hole with two distinct horizons $r_{\pm},$ where $r_+>r_-$; when $k=k_{\rm E}$, the horizons coincide $r_+=r_-\equiv r_{\rm E}$, representing an extremal regular black hole with degenerate horizons at $r_{\rm E}$. In the parameter space $k>k_{\rm E}$, the horizon disappears, and the metrics describe the exterior of regular HUCOs, which are similar to the Gravastar model of Mazur-Mottola~\cite{Mazur:2001fv}. Thus, Bardeen and GC metrics describe both black holes for $k\leq k_{\rm E}$ and the regular HUCOs spacetimes for $k>k_{\rm E}$.  These black hole models are further examined in detail in the subsequent sections, exploring both the background metric and the effective metric descriptions to address the questions raised in the introduction.

\subsection{Bardeen Spacetime}\label{sec-5a}
Solving field equations (\ref{EfE}) and (\ref{EfE2}) for the following NED Lagrangian density \cite{Ayon-Beato:2000mjt}
\begin{equation}
	\mathcal{L(F)}=\frac{6}{s k^2}\Big(\frac{\sqrt{k^2 \mathcal{F}/2}}{1+\sqrt{k^2 \mathcal{F}/2}} \Big)^{5/2}, \label{BarLag}
\end{equation}
with $\mathcal{F}=\frac{2 k^2}{r^4}$, we obtained the metric functions 
\begin{equation}
	A(r)=\frac{1}{B(r)}=1-\frac{2 M}{r}\Big(\frac{r^2}{r^2+k^2}\Big)^{3/2},\quad	C(r)=r^2,\label{BarMetric}
\end{equation}
which described both Bardeen-BH and Bardeen-HUCO for ~\cite{Bardeen-RegularBH}. Here, $s$ is a constant determined by the field equations to be $s=\frac{k}{2M}$. Utilizing  Eqs.~(\ref{effmetric}) and (\ref{phi}), and substituting $\mathcal{L(F)}$ from Eq.~(\ref{BarLag}) and the metric functions from Eq.~ (\ref{BarMetric}), we obtained the effective metric $\tilde{g}_{\mu\nu}$ for the Bardeen model. Some key properties of interest are listed as follow
\begin{itemize}
	\item The Bardeen spacetime center exhibits a de-Sitter geometry with an effective cosmological constant $\displaystyle{\lim_{r \to 0}} T^{\mu}_{\nu}= \frac{6M}{k^3}\delta^{\mu}_{\nu}$. 
	\item The strong and dominant energy conditions are satisfied only within, respectively, $r\geq \sqrt{\frac{2}{3}}k$ and $r\leq 2k$, whereas the weak and null energy conditions are satisfied globally $r\geq 0$.
	\item In the limit $k\to 0$, while $g_{\mu\nu}$ recovers the Schwarzschild metric,  $\tilde{g}_{\mu\nu}$ does not match.
	\item For $r\gg M$, $g_{\mu\nu}$ does not reduce to the RN metric.
	\item For $r\gg M$, $\tilde{g}_{\mu\nu}$ does not reduce to $g_{\mu\nu}$.
\end{itemize}
\begin{figure}
	\includegraphics[scale=0.09]{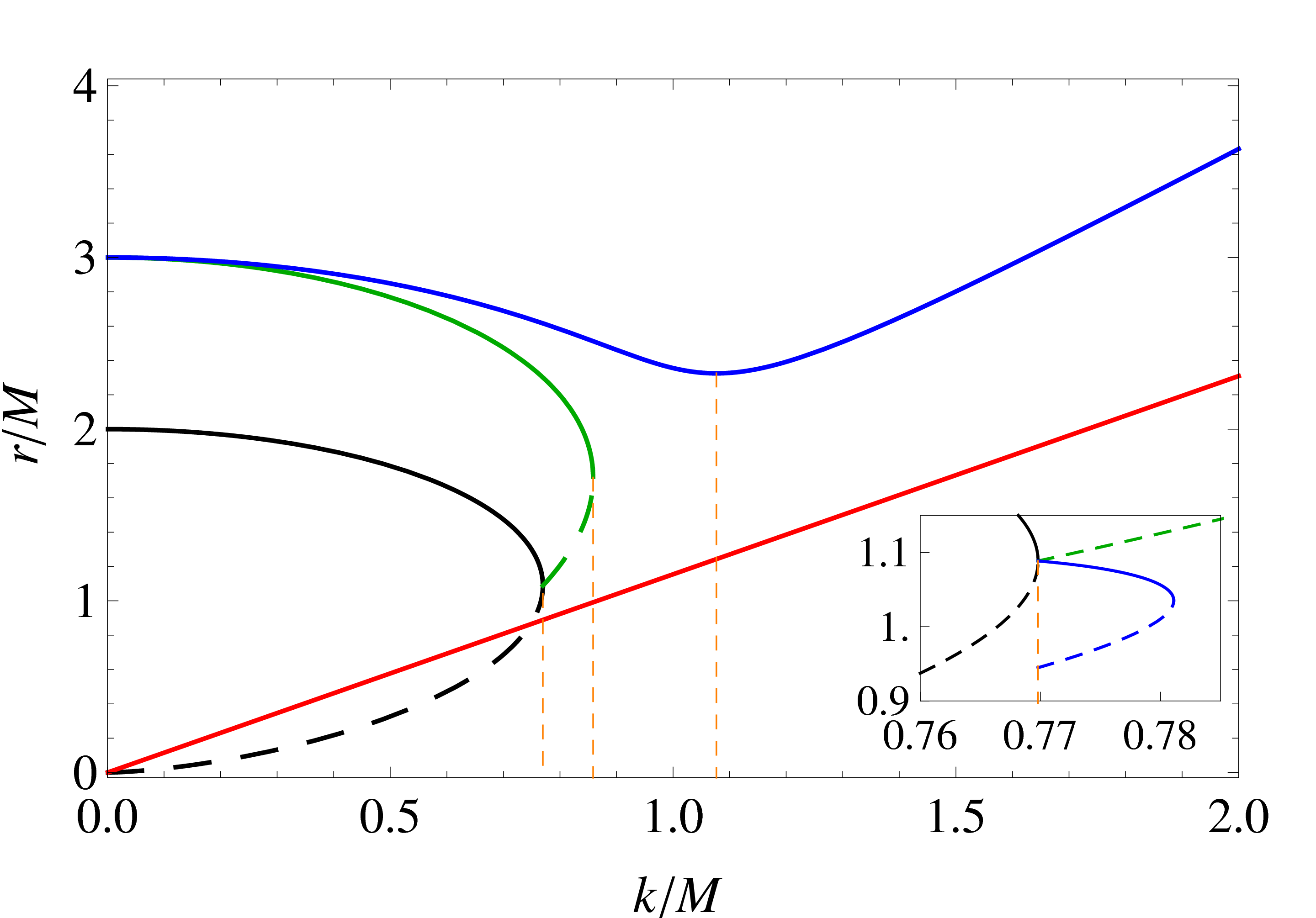}
	\caption{The radial coordinates of key features in the Bardeen spacetime are depicted in this figure: the event (Cauchy) horizon  by the black solid (dashed) curve, curvature singularity $r_{\text{sing}}$ by red solid curve, unstable (stable) null circular orbits from the background metric by the green solid (dashed) curve, and null (photon) circular orbits from the effective metric by the blue solid curve.  Three vertical orange dashed lines, from left to right, represent the extremal black hole case $k=k_{\rm E}$, extremal null circular orbits from the background metric case $k=k_{\rm c}$, and the minimum photon orbit radius case $k=k_{\rm p}$. While the effective metric allows for at least one photon circular orbit for all values of $k$, null circular orbits are only possible for $k\leq 0.85865M=k_{\rm c}$ in the background metric. The background metric is free of singularities, but a curvature singularity emerges in the effective metric followed by photons.} \label{fig:BarRadii}
\end{figure}

Various limits of the NED source and their repercussions on the $g_{\mu\nu}$ and $\tilde{g}_{\mu\nu}$ are further discussed in detail in Sec.~\ref{sec-6}. Some interesting geometrical features of the Bardeen spacetime are shown in Fig.~\ref{fig:BarRadii}. Both metrics, $g_{\mu\nu}$ and $\tilde{g}_{\mu\nu}$, predict identical black hole horizons radii. Extremal black hole exists for $k_{\rm E}=0.7698M$ with horizons at $r_{E}=1.0405M$. The metric $\tilde{g}_{\mu\nu}$ is singular at $\mathcal{L'}=0$ and $\Phi=0$. Here, $\mathcal{L'}=0$ ($\Rightarrow r=0$) is merely a coordinate singularity, while $\Phi=0$ gives a true curvature singularity at
\begin{equation}
	r=\frac{2}{\sqrt{3}}k\equiv r_{\text{sing}}.\label{BarSing}
\end{equation}
It is important to recall here that the Bardeen metric $g_{\mu\nu}$ is regular with globally finite curvature scalars, while this singularity manifests solely within the metric $\tilde{g}_{\mu\nu}$ seen only by photons.  The singularity resides on the surface of a two-sphere with radius $r_{\text{sing}}$, which is always enclosed by the event horizon when $k\leq k_{\rm E}$, but becomes globally naked for $k>k_{\rm E}$. Novello \textit{et. al.}~\cite{Novello:2000km} have identified similar curvature singularity in the effective metric of a electrically charged NED regular black hole. Curvature singularity in $\tilde{g}_{\mu\nu}$ metric is still a subtle issue being further discussed in detail in Sec.~\ref{sec-6}. For the present discussion, we consider that the range accessible to photon motion is within $r\geq r_{\text{sing}}$.

Null circular orbits of the metric $g_{\mu\nu}$ are compared with those of the metric $\tilde{g}_{\mu\nu}$ followed by photons. Within the background metric, the Bardeen-BH spacetimes accommodate two distinct null circular orbits, with the outer one being unstable outside the event horizon and the inner one being stable between the event and Cauchy horizons. Even within Bardeen-HUCOs spacetimes ($k>k_{\rm E}$), null circular orbits continue to exist (shown as green curves in Fig.~\ref{fig:BarRadii}), converging at $r=  1.717966M\equiv r_{\rm c}$ for $k= 0.85865M\equiv k_{\rm c}$. For $k>k_{\rm c}$, no null circular orbits exist. The existence of null circular orbits in regular HUCOs spacetimes leads to some interesting observational effects, as presented in Ref.~\cite{Kumar:2020ltt}. HUCOs have recently grabbed significant attention and are now being actively investigated in various settings~\cite{Eichhorn:2022fcl,Ayzenberg:2023hfw,Eichhorn:2022bbn,Carballo-Rubio:2022nuj, Arrechea:2023oax}. On the other hand, the effective metric $\tilde{g}_{\mu\nu}$ supports multiple null circular orbits for photons. For $k<0.781099M\equiv \tilde{k}_{\rm c}$, three null circular orbits exist, whereas for $k>\tilde{k}_{\rm c}$, only one null circular orbit exist that is unstable under radial perturbations. Interestingly, $\tilde{g}_{\mu\nu}$ ensures that the HUCOs always possess at least one null (photon) circular orbit. This is not surprising; indeed, a sufficiently strong NED field even in vacuum can compel photons to traverse circular orbits~\cite{Novello:2000xw}. The radial coordinate, $r_{\rm p}$, of the outermost photon circular orbit corresponding to the photon sphere initially decreases slowly with $k$, reaching a minimum value of $r_{\rm p}=2.32506M$ at $k=1.07663M\equiv k_{\rm p}$, and then monotonically increases with $k$. The minimum radius of photon orbits can be determined by simultaneously solving $H(r)=0$ and $\partial H/\partial k=0$ in Eq.~(\ref{PhotonOrbit}). It is important to note that, for HUCOs, the photon orbits always remain outside the curvature singularity, i.e., $r_{\rm p}>r_{\text{sing}}$ (cf. Fig.~\ref{fig:BarRadii}). In summary, even though Bardeen-BH horizons exist only for $k\leq 0.7698M=k_{\rm E}$ and null circular orbits for the background metric exist for $k\leq 0.85865M=k_{\rm c}$, timelike circular orbits for massive particles continue to exist in HUCO spacetime for even larger values of $k$, specifically, within $k\leq 0.95629M=k_{\rm c}^{\text{TL}}$ and atleast one null circular orbit for the effective metric exists for all values of $k$.

The shadows size depicted in Fig.~\ref{fig:Shadow-Eff} for Bardeen-BH and Bardeen-HUCO spacetimes do not align with the size constraints of M87* and Sgr A*. Consequently, we will not pursue further investigation into the ray-traced images of these spacetimes. Instead, our focus will be solely on discussing the GC-spacetime in detail.

\subsection{GC Spacetime}\label{sec-5b}
For GC model, we considered the following form of the NED field Lagrangian density:
\begin{equation}
	\mathcal{L(F)}=\mathcal{F}\,e^{-s\Big(\frac{k^2\mathcal{F}}{2}\Big)^{1/4}},\label{Lag2}
\end{equation}
which upon solving the field equations~(\ref{EfE}) and (\ref{EfE2}), give the Ghosh-Culetu black hole metric functions \cite{Culetu:2014lca,Ghosh:2014pba}
\begin{equation}
	A(r)=\frac{1}{B(r)}=1-\frac{2 M}{r} e^{-\frac{k^2}{2 M r}},\, C(r)=r^2.
\end{equation}
Here, we defined $s=\frac{k}{2M}$ and $\mathcal{F}=\frac{2 k^2}{r^4}$. From Eqs.~(\ref{effmetric}) and (\ref{phi}), we obtained the effective metric $\tilde{g}_{\mu\nu}$ for photon propagation in GC spacetime.
Some notable features of interest for this model are listed as follow
\begin{itemize}
	\item The GC spacetime center exhibits a Minkowski geometry such that $\displaystyle{\lim_{r \to 0}} (G^{\mu}_{\nu}=T^{\mu}_{\nu})= 0$. 
	\item All classical energy conditions are violated in deep core within $0<r< k^2/8M$.
	\item In the limit $k\to 0$, both the metrics $g_{\mu\nu}$ and $\tilde{g}_{\mu\nu}$ recover the Schwarzschild metric.
	\item For $r\gg M$, $g_{\mu\nu}$  metric reduces to the RN metric.
	\item For $r\gg M$, $\tilde{g}_{\mu\nu}$ metric reduces to $g_{\mu\nu}$.
\end{itemize}

\begin{figure}
	\includegraphics[scale=0.057]{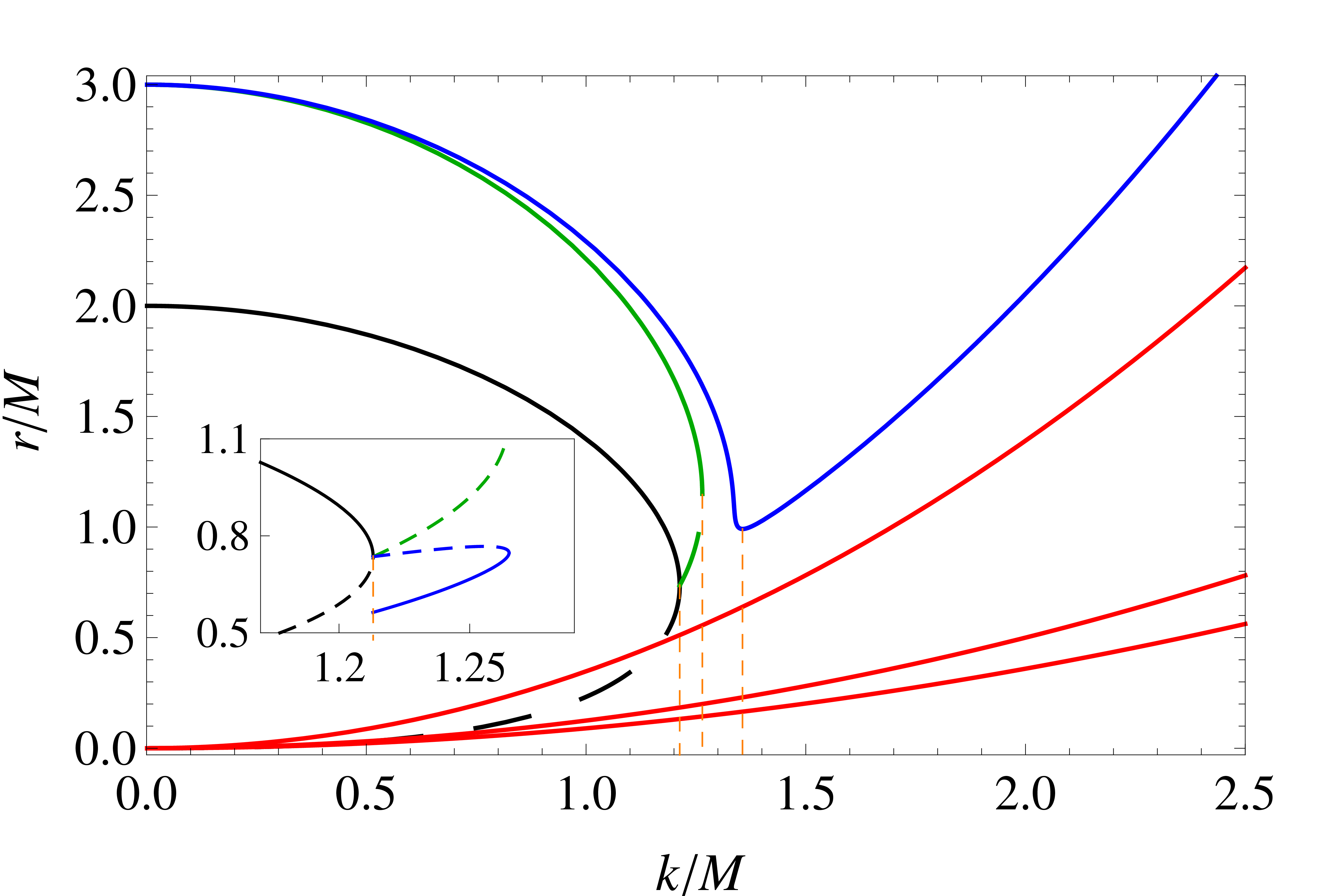}
	\caption{The radial coordinates of key features in the GC spacetime are depicted in this figure: the event (Cauchy) horizon by the black solid (dashed) curve, curvature singularities $r_{\text{sing}}$ by red solid curves, unstable (stable) null circular orbit from the background metric by the green solid (dashed) curve, and null (photon) circular orbits from the effective metric by the blue solid curve. Three vertical orange dashed lines, from left to right, represent the extremal black hole case $k=k_{\rm E}$, extremal null circular orbits of the background metric case $k=k_{\rm c}$, and the minimum photon orbit radius case $k=k_{\rm p}$. While the background metric is singularity-free, these curvature singularities appear only in the effective metric followed by photons.} \label{fig:NSRadii}
\end{figure}
\begin{figure*}
	\begin{tabular}{c c c}
		\includegraphics[scale=0.6]{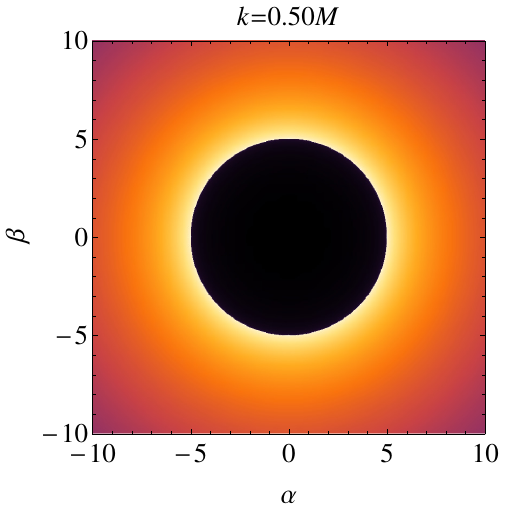}&
		\includegraphics[scale=0.6]{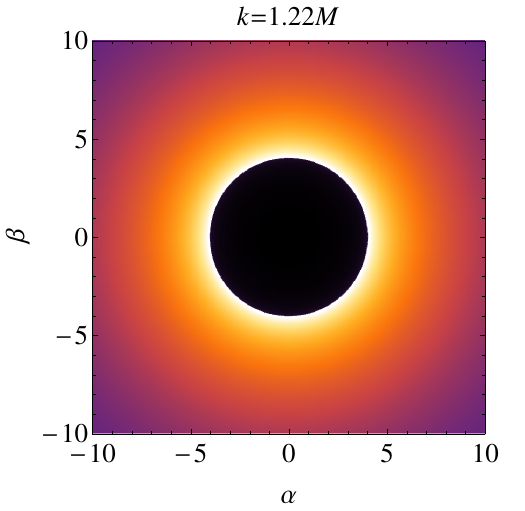}&
		\includegraphics[scale=0.6]{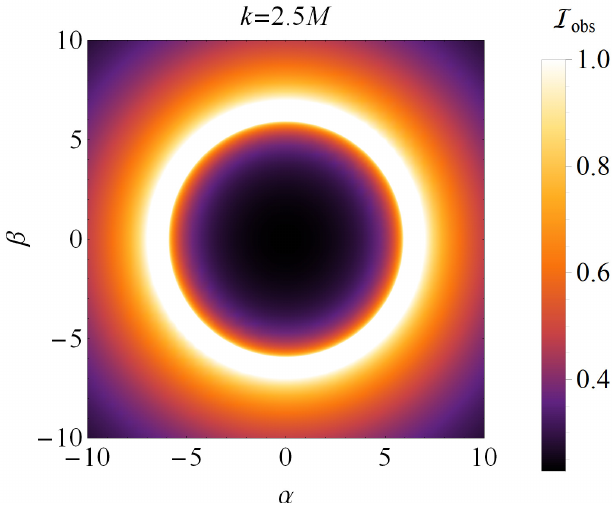}
	\end{tabular}	
	\caption{This figure illustrates shadows of GC-BH with $k=0.5M$ and GC-HUCOs with $k=0.80M$ and $k=2.0M$ determined from the effective metric under the radially infalling spherical accretion.}\label{fig:InfallNS-Eff}
\end{figure*}
\begin{figure*}
	\begin{tabular}{c c}
		\includegraphics[scale=0.68]{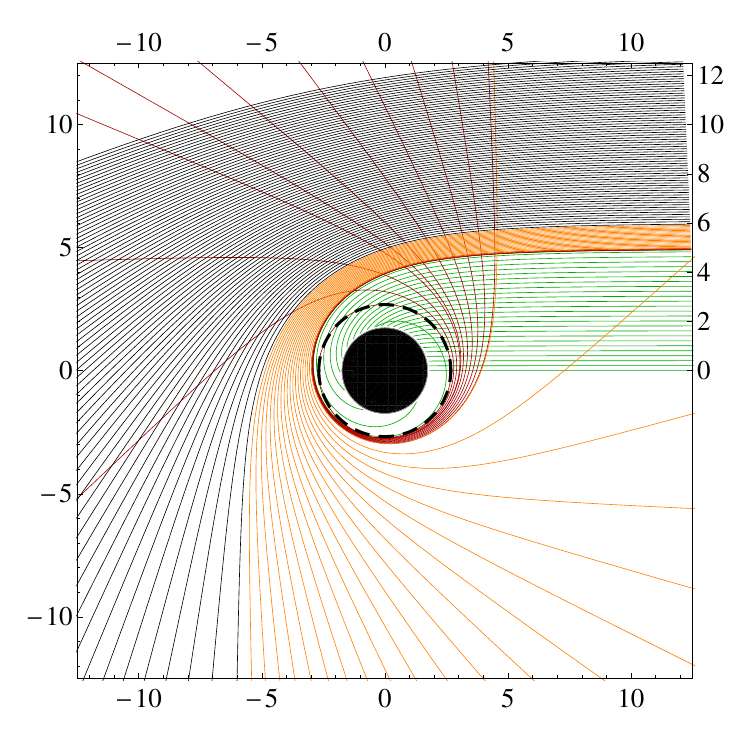} &
		\includegraphics[scale=0.68]{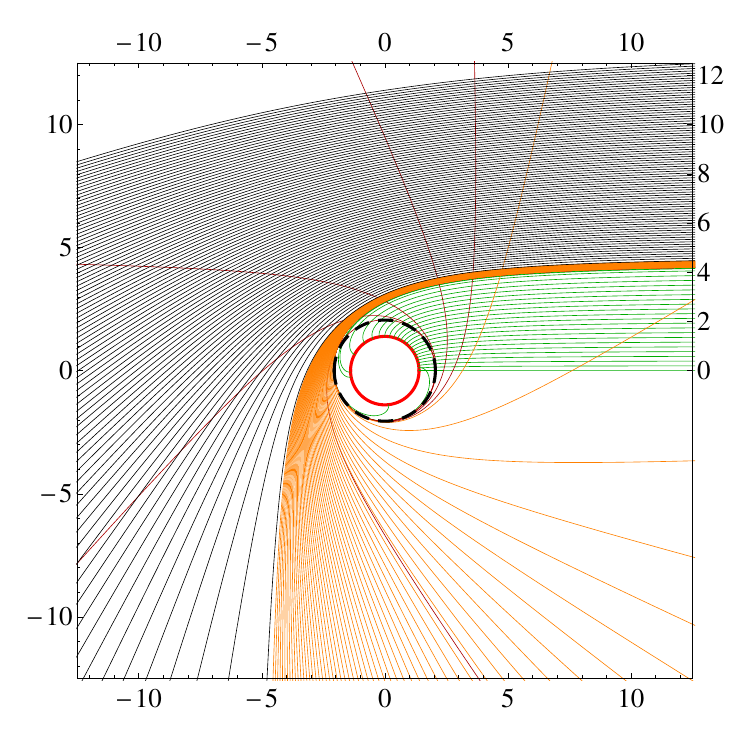}
	\end{tabular}	
	\caption{Photon orbits from the effective metric are shown for GC-BH with $k=0.7M$ (left) and GC-HUCO with $k=2.0M$ (right). The black dashed circle, red solid circle, and black disk represent the photon sphere, curvature singularity, and black hole event horizon, respectively.}\label{fig:OrbitEffNS}
\end{figure*}		
\begin{figure}
	\includegraphics[scale=0.41]{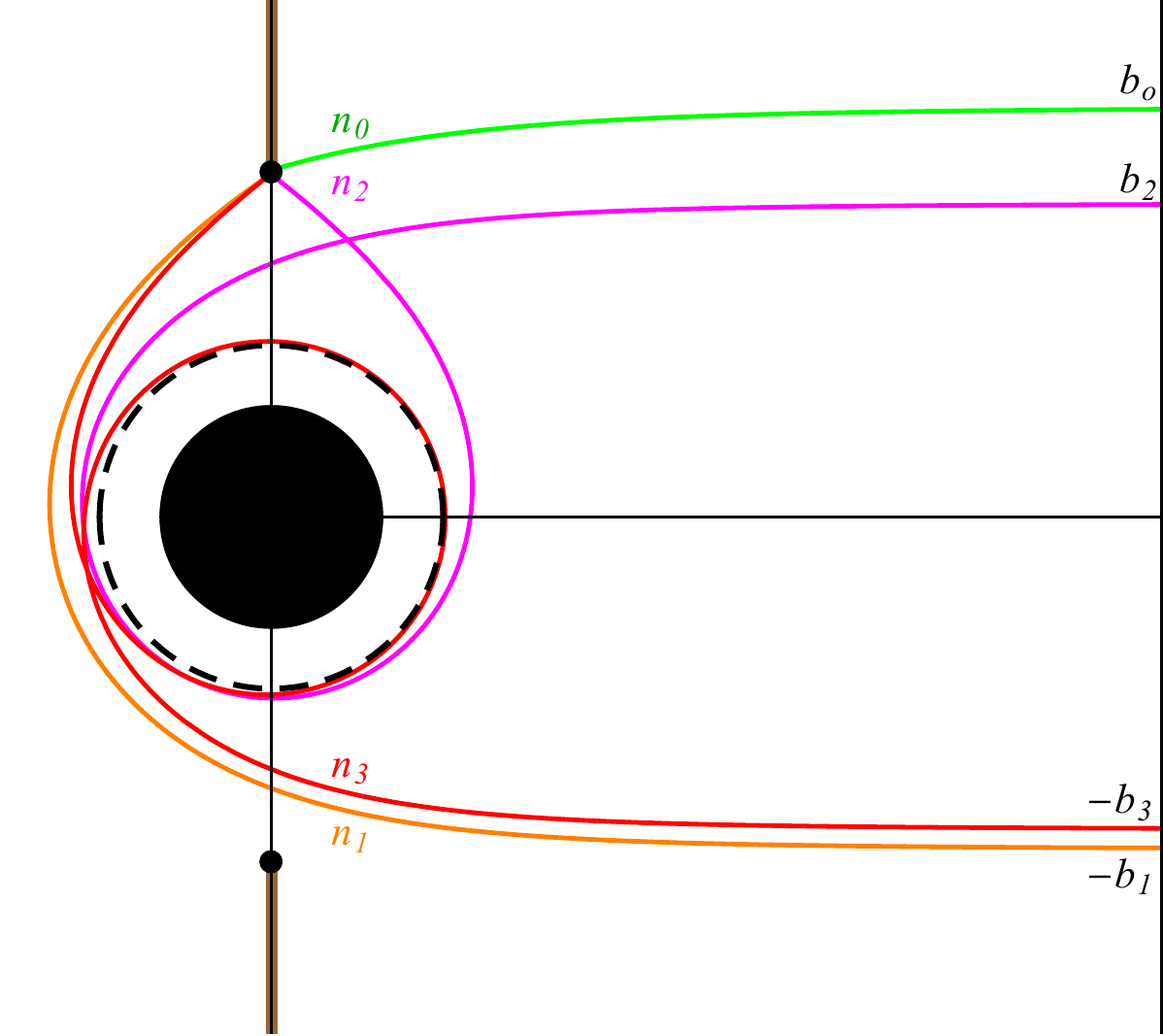}
	\caption{A schematic illustrating the identification of photon rings around a spherically symmetric static black hole. The black disk at the center represents the black hole, while the brown thick vertical line depicts the edge-on view of the accretion disk at $\theta=\pi/2$  viewed from an observer located to the right at $\theta_o=\pi/2$. Black dots mark the inner edge of the disk, defined as in Eq.~(\ref{DiskInnerEdge}). Photon rings are characterized by an order number$-n$, representing the number of half-orbits around the black hole. Orbits defining the inner diameter of the $n = 0$ image (green orbit), $n = 1$ ring (orange orbit), $n = 2$ ring (magenta orbit), and $n = 3$ ring (red orbit) are depicted. Black dashed circle represent the photon sphere around the black hole,}
	\label{Fig:PhotonRing}
\end{figure}
\begin{figure*}
	\begin{tabular}{c c c}
		\includegraphics[scale=0.6]{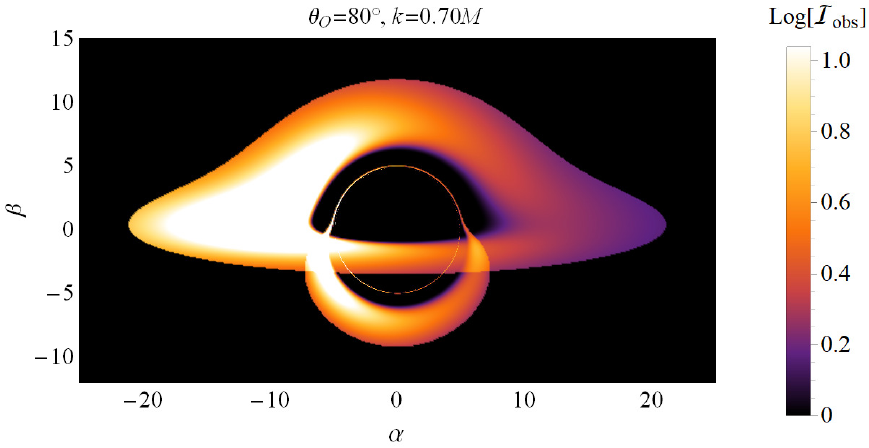}&
		\includegraphics[scale=0.6]{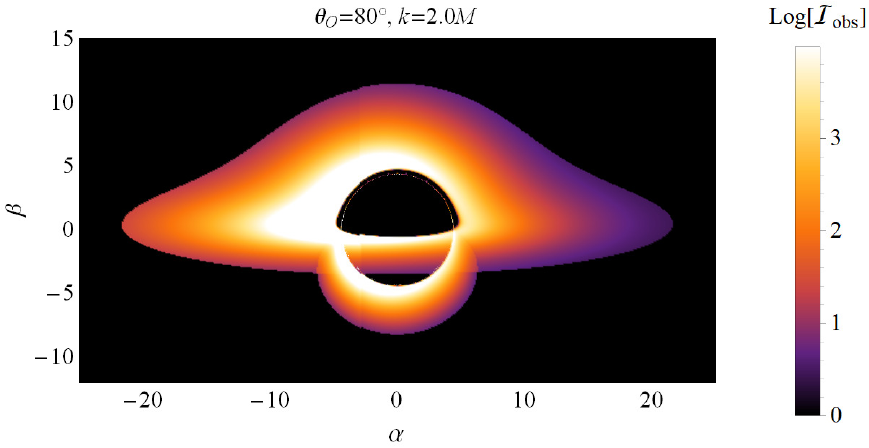}
	\end{tabular}	
	\caption{Shadows of GC-BH (left) and GC-HUCO (right) cast by photons tracing null geodesics of the effective metric $\tilde{g}_{\mu\nu}$ under an equatorial accretion disk model. Observed flux $\mathcal{I}_{\text{obs}}$ is represented on a logarithmic scale.}\label{fig:DiskEffectiveNS}
\end{figure*}

In the limit $r\gg M$, unlike the Bardeen background metric, the GC metric approaches the RN metric. This convergence was expected, as the NED source for the GC spacetime exactly satisfies a correspondence with the Maxwell linear electrodynamics in the weak-field regime.  Even within the effective metric $\tilde{g}_{\mu\nu}$, the radii of black hole horizons remain unchanged. The GC-BH exhibits two distinct horizons $r_{\pm}$ for $k<1.2130M\equiv k_{\rm E}$, with degenerate horizons at $r_- = r_+=0.73576M\equiv r_{\rm E}$ for the extremal case $k=k_{\rm E}$.  For $\mathcal{L}'=0$ and $\Phi=0$, the effective metric functions turn singular, respectively, at
\begin{equation}
	r=\frac{k^2}{8 M} \equiv r_2,\,\,\,\, r=\frac{(7 \pm \sqrt{17}) k^2}{32 M}\equiv r_{1,3} ,
\end{equation}
which give three curvature singularities $ r_1>r_2>r_3$ only for the photons. These singularities are on the surface of two-sphere with increasing radii $r_{\text{sing}}\in  \{r_1,r_2,r_3\}$, scaling proportionally to $k^2/M$, where curvature scalars diverge. While for $k<k_{\rm E}$, these singularities remain hidden behind the horizon, emerging as globally naked singularity for $k>k_{\rm E}$. The horizon radii, curvature singularities, and the null circular orbits of the background and the effective metric are shown in Fig.~\ref{fig:NSRadii}.  GC background spacetime exhibits two null circular orbits for $k<1.2645M\equiv k_{\rm c}$: an outer unstable orbit and an inner stable one, which merge together for $k=k_{\rm c} $ at $r=1.1467M\equiv r_{\rm c}$. These null circular orbits within GC-HUCO spacetimes were first reported in Ref.~\cite{Kumar:2020ltt}. Conversely, the metric  $\tilde{g}_{\mu\nu}$ facilitates three null (photon) circular orbits when $k< 1.2651M\equiv \tilde{k}_{\rm c}$, whereas for $k> \tilde{k}_{\rm c}$, only one such orbit exists. The outermost photon circular orbit that exists for all values of $k$ is radially unstable, and its radius initially decreases with $k$,  reaching a minimum value of $r_{\rm p}=0.99129M$ at $k=1.356402M\equiv k_{\rm p}$, and then monotonically increases with $k$. This outermost photon orbit is always outside the curvature singularity, i.e., $r_{\rm p}>r_{1,2,3}$. (cf. Fig.~\ref{fig:NSRadii}). In numerical ray-tracing, we consider the photon motion only outside the outermost singularity, i.e,  in the region limited to $r>r_1$. 

Here, we employed our accretion models to investigate the GC spacetime shadows predicted from the effective metric. 
We first considered the optically thin, radiating, radially infalling spherical accretion flow, as elaborated in the Sec.~\ref{sec-4a}. Shadows of GC-BHs and GC-HUCOs are shown in Fig.~\ref{fig:InfallNS-Eff}. Different colors correspond to different values of the observed intensity, and we used one color function for all shadow plots, where the greater (smaller) intensity means the brighter (darker) color. All observed intensities are normalized to one. With increasing $k$, photons redshift from the emission region to the observer decrease, and eventually, the central brightness depression in the image increases. This explains a brighter emission ring outside the shadow boundary for $k=2.5$ in Fig.~\ref{fig:InfallNS-Eff}. The observed intensity at the image plane exhibits circular symmetry due to the spherical symmetry of the emission flow in the bulk.

For a GC-BH with $k=0.50M$, the shadow exhibits an expected intensity peak at the lensed position of the photon orbit, resulting in the shadow critical curve at $\alpha^2+\beta^2=R_{\text{sh}}^2$ on the image plane. Within $\alpha^2+\beta^2\leq R_{\text{sh}}^2$, a prominent intensity depression represents the shadow feature. The observed intensity $\mathcal{I}_{\text{obs}}$ peaks immediately outside the shadow boundary, diminishing gradually with radial distance from it. However, the shadow inner region within $\alpha^2+\beta^2<R_{\text{sh}}^2$ exhibits a finite intensity and is not entirely dark, as would be observed if the accreting flow  were entirely behind the black hole. This arises because the accreting flow is also present along the lines of sight intersecting the surface of the black hole. A fraction of radiation within this region can escape to the distant observer, albeit highly redshifted, contributing to the non-zero intensity within shadow inner region. 
Interestingly, HUCOs, for e.g. GC-HUCO with $k=1.22M$ and $k=2.50M$, also display a similar intensity depression at the screen center, casting black hole-like shadows. This occurs because for $k>k_{\rm E}$, light rays with $b<b_{\text{cr}}$ intersect the singularity at $r=r_1 $ and do not experience any radial turning point outside of $r>r_1$. These light rays intersecting the singularity account for the dark patch at the image screen center. Similar features have been reported for the Janis-Newman-Winicour \cite{Janis:1968zz} naked singularity, where a naked curvature singularity appears at the 2-sphere boundary of  radial coordinate  $r_{\text{sing}}=\frac{2M}{\gamma}$, and a photon sphere form at $r_{\rm p}=\frac{M}{\gamma}(2\gamma+1)$; $\gamma$ is the scalar charge. For $0.5<\gamma\leq 1$, JNW naked singularity exhibit photon orbit outside the curvature singularity, thus casting a shadow reminiscent of a black hole  \cite{Gyulchev:2020cvo,Gyulchev:2019tvk,Deliyski:2024wmt}. In the framework of the electrically charged NED effective metric, photons approaching the singularity undergo infinite blueshift, potentially destabilizing the spacetime \cite{Novello:2000km,Bronnikov:2022ofk}.  Conversely, singularities in the magnetically charged NED effective metric have not been previously explored in detail. Interestingly, in this scenario, photons reaching the singularity at $r=r_1$ undergo only a finite blueshift in frequency.

For our second accretion model, we simulate images of thin equatorial disk accreting onto a GC spacetime as discussed in Sec.~\ref{sec-4b}. 
Photon orbits around the GC-BH and GC-HUCO are shown in Fig.~\ref{fig:OrbitEffNS}. Identifying photon orbits with multiple intersections of the accretion disk is further useful for determining the gravitationally lensed image of an accretion disk. 
Let the thin accretion disk is along the vertical line passing through the black hole center, with the observer positioned far away from it to the right. When tracing light rays backward in time, from the observer to the central object, our focus is on identifying the point of intersection on the accretion disk to project it onto the observer's screen. In summary, light rays intersecting the $\theta=\pi/2$ plane $n+1$ times complete $n$ half-orbits around the central object, generating the $n$th order image of the accretion disk on the observer's screen. $n=0$ and $n\to \infty$, respectively, correspond to the direct image of the disk and the black hole's critical curve. As shown in Fig.~\ref{fig:OrbitEffNS}, different colors distinguish orbits based on the number of  intersections of the  $\theta=\pi/2$ plane or the number of the half-orbits around the central object. Light rays of the black color intersect the $\theta=\pi/2$ plane only once without making any loop around the central object, directly imaging the front side of the accretion disk.   Orange-colored light rays intersect the plane twice, completing one half-orbit, thus forming the first-order or primary image of the backside of the accretion disk.  Red-colored light rays intersect the $\theta=\pi/2$ plane thrice, completing two half-orbits, resulting in the second-order image of the accretion disk, and so forth.  Light rays coming with the critical value of impact parameter $b=b_{\text{cr}}$ asymptotically approach the photon sphere, shown by the black dashed circle. The optically thin accretion flow enables such multiple crossing of disk, such that, the direct image is generically accompanied by a series of higher-order images. However, only when the emission is non-spherical do these successive higher-order images manifest as a discrete series of concentric exponentially demagnified ``\textit{photon rings}," labeled by their half-orbit number $n$, converging toward the ``critical curve"  \cite{Gralla:2019xty}. The higher the ring order, the less it depends on the accretion disk features and more on the black hole geometry.  Nevertheless, these higher-order orbits help identify the photon ring structures discussed subsequently.  Our convention for identifying the $n = 0, 1, 2, 3$ photon rings around a black hole is illustrated in Fig.~\ref{Fig:PhotonRing}. For simplicity, we have chosen the observer to be at $\theta_o = \pi/2$, but a similar convention holds for all inclination angles. For $k=2.0M$, we traced the photon geodesics \textit{only} outside the curvature singularity. 

To generate the images of the accretion disk, we defined an screen with coordinates $\alpha\in [-25M, 25M]$ and $\beta\in [-15M, 15M]$. This screen is perpendicular to the line of sight of the observer to the black hole in bulk, positioned at $r_o=5000M$, which, for practical purposes, can be treated as asymptotic infinity. The screen was further divided into  $500\times 300$ pixels ($\alpha_i,\beta_i$). While the outer radius of the accretion disk remained fixed at $r_{\text{out}}=20M$, the inner edge is at $r_{\text{in}}$ determined from Eq.~(\ref{DiskInnerEdge}). The observer inclination angle was set to $\theta_o=80\degree$ for all accretion disk images. We traced null geodesics backward in time from each pixel ($\alpha_i,\beta_i$) on the screen until they either:  (i) intersected the horizon in the case of a black hole, or reached the singularity of the effective metric, which is visible only to photons in HUCOs spacetimes;  or (ii) approached the asymptotic source. Along these trajectories, we tracked the point of intersection $r_e$ on the accretion disk, where $r_{\text{in}}\leq r_e\leq r_{\text{out}}$, to use in Eq.~(\ref{IntenAccDisk}) for calculating the observed intensity. The resulting intensity was computed for all pixels ($\alpha_i, \beta_i$) on the screen.

Figure~\ref{fig:DiskEffectiveNS} displays the resulting accretion disk images for GC-BH with $k=0.70M$  and GC-HUCO with $k=2.0M$. While for GC-BH, photon rings are clearly visible  and well within the emission ring, for GC-HUCO, photon rings are immensely thin and are close to the emission ring. The relative positions of the photon and emission rings depend on the inner and outer edges of the accretion disk. Nonetheless, the morphology of the direct and indirect images of the Keplerian disk closely resembles those associated with Schwarzschild black holes.  Diameters and widths of the first three photon rings of the Sgr A* black hole, as a GC-BH with $k=0.50M$, are calculated and summarized in Table~\ref{NSPhotonRing}. Precise measurements of radiation emanating from the innermost regions of Keplerian accretion disks near the black hole horizon could facilitate the potential validation or elimination of direct NED effects around supermassive black holes like Sgr A* and M87*. 
\begin{figure*}[ht]
	\begin{tabular}{c c c}
		\includegraphics[scale=0.5]{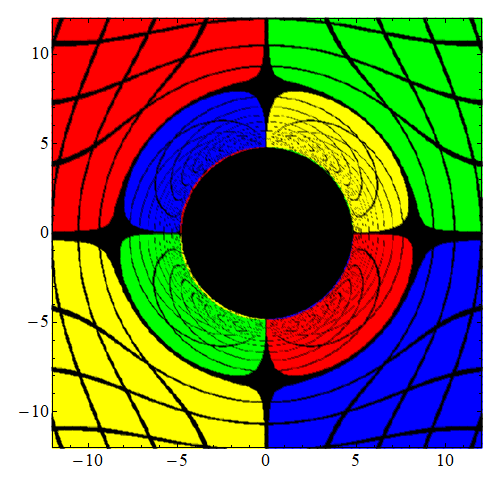}&
		\includegraphics[scale=0.5]{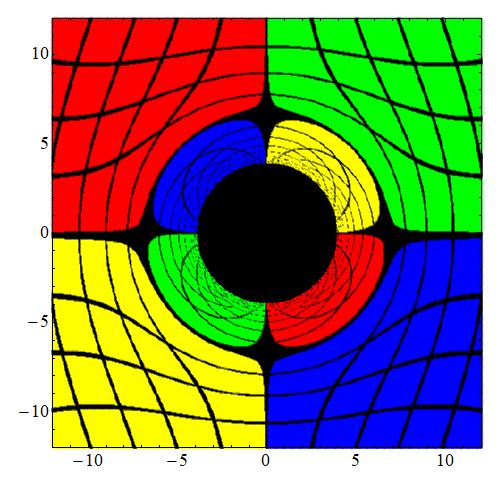}
	\end{tabular}	
	\caption{Gravitational lensing of light as predicted by the GC effective metric $\tilde{g}_{\mu\nu}$ for a black hole with $k=0.70M$ (left figure) and a HUCO  object with $k=2.0M$ (right figure) located at the center of the celestial sphere. The celestial sphere has a radius of $30M$, and the observer is situated at a distance of $15M$ from the center of the celestial sphere. The black lines denote lines of constant longitude and latitude, with the central black circular region representing the shadow.}\label{fig:CelesNS}
\end{figure*}

To further visualize the gravitational lensing effect, we established a celestial sphere of radius $r_s=30M$, centered on the GC spacetime, refer to Fig.\ref{fig:CelesNS}. The observer is situated within this sphere, positioned off-centered at a distance of $r_o=15M$ from the center.  This celestial sphere is partitioned into four quadrants, each painted with a distinct color; along the observer's line of sight, the top-left, top-right, bottom-left and bottom-right quadrants, are, respectively, seen as red, green, yellow and blue. We further subdivided the sphere boundary into evenly spaced grids using  latitude and longitude lines separated by $\pi/18$ degrees to aid in the interpretation of the distortion of the resulting images. In the absence of any object along the observer's line of sight, these latitude and longitude lines are nearly straight, resulting in colored grid of identical area. However, in the presence of a black hole or a HUCO, strong gravitational lensing causes multiple copies of a single grid to appear, each with decreasingly smaller area approaching the shadow boundary. To analyze these effects, we partitioned the observer screen into $800\times 800$ small pixels and traced the paths of light rays from each pixel, following the $\tilde{g}_{\mu\nu}$ metric, until they intersected either the celestial sphere or the horizon for $k\leq k_{\rm E}$ or the singularity for $k>k_{\rm E}$. Redshift effects were disregarded, focusing solely on the spatial distortion of the resulting images. Figure~\ref{fig:CelesNS} presents lensed images of GC-BH with $k=0.70M$ and GC-HUCO with $k=2.0M$. Dark regions indicate photons captured by the black hole or those plunging into the singularity of the HUCO, resulting in the observable shadow. The apparent angular size of the shadow appears larger in the observer's field of view, primarily due to geometric factors arising from the observer's proximity to the black hole. Strong gravitational lensing induced by the compact mass is evident in the warping of the grid lines, particularly pronounced around the shadow boundary. Additionally, the region of the celestial sphere located behind the observer undergoes lensing on the image plane. The lensing images of HUCOs closely resemble those of a black hole, differing primarily in the size of shadow and the width of higher-order photon rings.

\subsection{Observational Predictions of NED Spacetimes}\label{sec-4c}
In Sec. \ref{sec-3}, we presented the shadow radii of Bardeen and GC spacetimes derived from the effective metric and placed constraints on $k$ using EHT measurements of Sgr A* and M87*. In this section, we derive additional observables within two frameworks: the background metric, applicable to non-NED alternative theories of regular black holes, and the effective metric, relevant to the NED theory of the same black hole. This allows for a comparative analysis of how different theoretical models of a given regular black hole influence observable phenomena. 

\begin{figure}
	\includegraphics[scale=0.87]{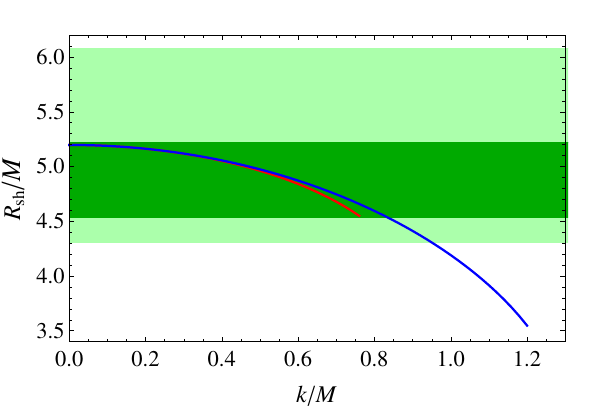}
	\caption{The shadow radii of Bardeen spacetime (red curve) and GC spacetime (blue curve) predicted by the background metric are shown. The darker green and lighter green regions indicate the EHT bounds for the Sgr A* and M87* black hole shadow radii, respectively.}
	\label{fig:Shadow-BG}
\end{figure}

The shadow sizes predicted by the null geodesics of the background metric for Bardeen and GC spacetimes are shown in Fig.~\ref{fig:Shadow-BG}. Firstly, the shadow sizes of both Bardeen-BHs and GC-BHs precisely match the Schwarzschild value  $R_{\text{sh}}=3\sqrt{3}M$ in the limit $k\to 0$, decreasing further with $k$. The background metrics predict no shadow like features for regular HUCOs. Notably, for a fixed value of $k$, the effective metrics predict larger radii of photon circular orbits and shadow size compared to that from the background metric (cf. Fig.~\ref{fig:Shadow-Eff} and Fig.~\ref{fig:Shadow-BG}). This discrepancy in the shadow size is also reported for electrically charged NED black holes \cite{dePaula:2023ozi}. The predicted $R_{\text{sh}}$ values for both Bardeen and GC spacetimes from the background metric are within the $1\sigma$ bounds  deduced by the EHT for Sgr A* and M87* black holes. Specifically, derived constraints from bounds (\ref{EHTbound}) on (Bardeen and GC) parameters are, respectively, ($k\leq 0.7634M$, $k\leq 0.8311M$) and ($k\leq k_{\rm E}=0.7698M$, $k\leq 0.948M$). Similar constraints on parameter $k$, deduced from the background metric, were reported by the EHT Collaboration \cite{EventHorizonTelescope:2021dqv,EventHorizonTelescope:2022xqj} and independently in Refs~\cite{KumarWalia:2022aop,Vagnozzi:2022moj,Kumar:2020yem}. Predictions from effective metric, especially for Bardeen spacetime, fail to reconcile with EHT measurements, unlike GC spacetime.

\begin{table}
	\begin{tabular}{|c||c|c|c|c|}  
		\hline	
		Photon Rings				            		& $n=1$  		 	 						   & $n=2$						    		& $n=3$\\
		\hline\hline
		\multirow{4}{*}{$g_{\mu\nu}$}	    & $d=10.988M$		      			    & $d=9.972M$			    	     & $d=9.923M$\\
		& $=52.994\mu$as					  & $=48.096\mu$as    	    		& $=47.86\mu$as\\
		& $w=0.4365M$ 							& $w=0.0175M$             		   & $w=0.00095M$ \\
		& $=2.10528\mu$as		    		  & $=0.0844\mu$as	 		        & $=0.00458\mu$as\\
		\hline											
		\multirow{4}{*}{$\tilde{g}_{\mu\nu}$}		 & $d=15.894M$		      & $d=12.829M$		         & $d=12.629M$\\
		& $=76.659\mu$as	    & $=61.876\mu$as	      & $=60.91\mu$as\\
		& $w=1.698M$ 			   & $w=0.0819M$              & $w=0.0067M$ \\
		& $=8.189\mu$as   		  & $=0.395\mu$as	  	    & $=0.0324\mu$as\\
		\hline
	\end{tabular}
	\caption{We model Sgr A* as Bardeen-BH with $k=0.50M$. Average diameter and width of first three photon rings are shown. From the background metric, the diameter of  $n=0$ inner ring is $d=13.014M$ ($62.767\mu$as) and critical curve is at $2R_{\text{sh}}=9.920M$ ($47.84\mu$as), whereas from the effective metric, diameter of $n=0$ inner ring is $d=14.48M$ ($69.842\mu$as) and critical curve is at $2R_{\text{sh}}=12.609M (60.819\mu$as).}\label{BarPhotonRing}
	\begin{tabular}{|c||c|c|c|c|}  
		\hline	
		Photon Rings				          					& $n=1$ 								 & $n=2$ 								   & $n=3$\\
		\hline\hline
		\multirow{4}{*}{$g_{\mu\nu}$}		 	    & $d=10.9304M$		   			  & $d=9.98748M$			 		& $d=9.9485M$\\
		& $=52.718\mu$as		   		 & $=48.1705\mu$as    			  & $=47.9826\mu$as\\
		& $w=0.4116M$ 					   & $w=0.01452M$           		 & $w=0.00067M$ \\
		& $=1.985\mu$as		    		  & $=0.07003\mu$as	      		  & $=0.00323\mu$as\\
		\hline											
		\multirow{4}{*}{$\tilde{g}_{\mu\nu}$}  		& $d=11.0517M$		   		& $d=10.1256M$		      & $d=10.0885M$\\
		& $=53.3033\mu$as	     & $=48.837\mu$as	      & $=48.6578\mu$as\\
		& $w=0.4079M$ 			   & $w=0.01396M$            & $w=0.00063M$ \\
		& $=1.967\mu$as   		   & $=0.0673\mu$as	  	    & $=0.0030\mu$as\\
		\hline
	\end{tabular}
	\caption{We model Sgr A* as GC-BH with $k=0.50M$. Average diameter $d$ and width $w$ of first three photon rings are shown. From the background metric, the diameter of $n=0$ inner ring is $d=13.084M$ ($63.1053\mu$as) and critical curve $2R_{\text{sh}}=9.9463M$ ($47.972\mu$as), whereas from the effective metric, the diameter of $n=0$ inner ring is $d=13.174M$ ($63.54\mu$as) and critical curve $2R_{\text{sh}}=10.087M (48.65\mu$as).}\label{NSPhotonRing}
\end{table}

\begin{figure*}
	\includegraphics[scale=0.38]{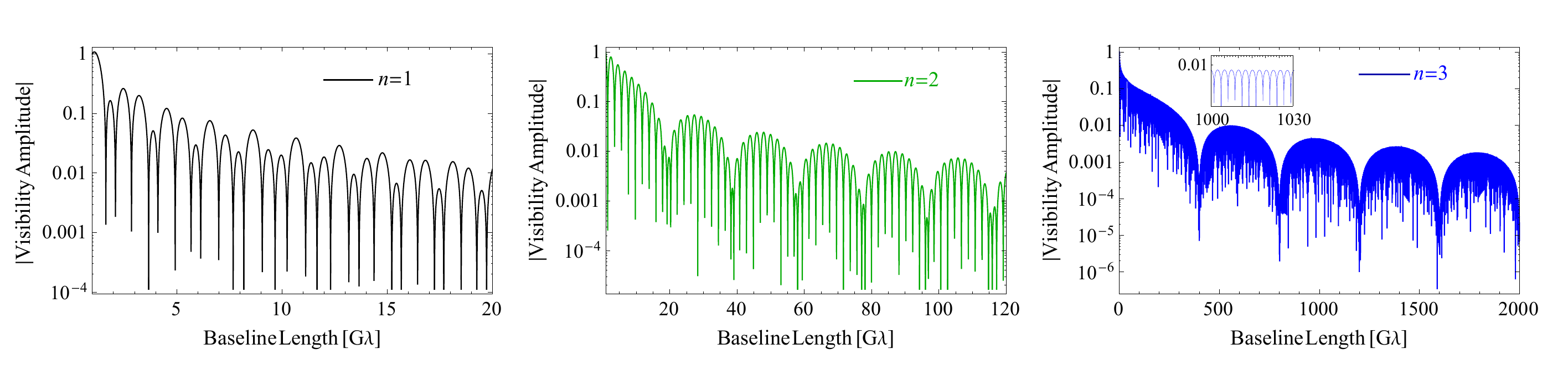}\\
	\includegraphics[scale=0.4]{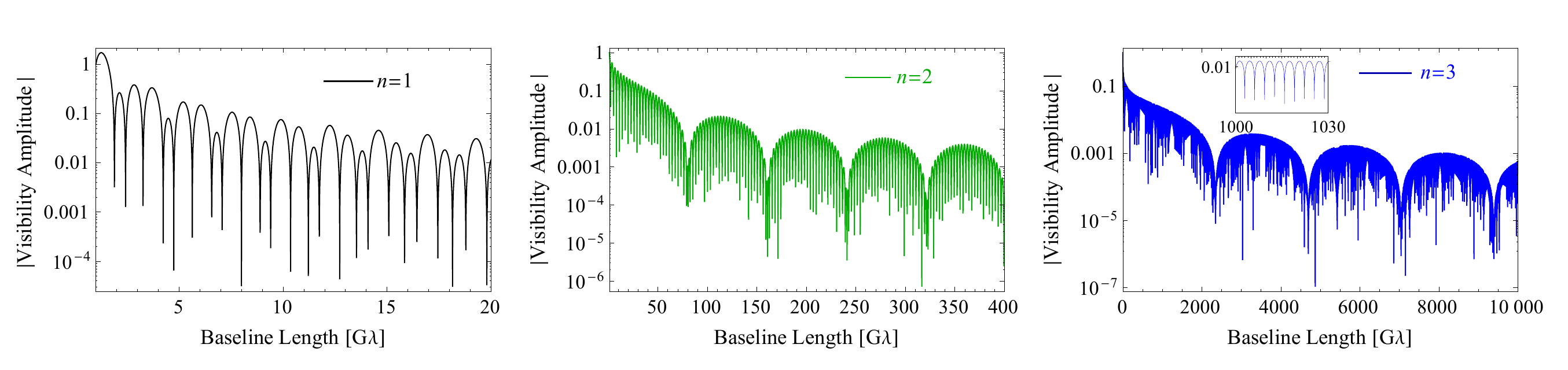}
	\caption{Normalized visibility amplitude as a function of EHT baseline length at 1.3mm observing wavelength for $n=1,2, 3$ photon rings from the effective metric of Bardeen-BH with $k=0.50M$ (top panel)  and GC-BH with $k=0.50M$ (bottom panel) for the Sgr A* black hole. The diameter and width of photon rings are presented in the Table~\ref{BarPhotonRing} and \ref{NSPhotonRing}. }
	\label{fig:PRfourier}
\end{figure*}
\begin{figure}
	\includegraphics[scale=0.87]{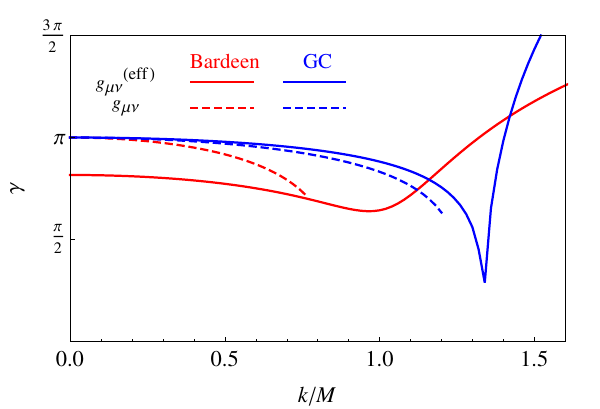}
	\caption{Lyapunov exponent for the null circular orbits of the background metric (dashed curve) and those from the effective metric (solid curve) for Bardeen and GC spacetimes.  }
	\label{fig:Gamma}
\end{figure}

Furthermore, our numerical ray-tracing code allows us to track the properties of each individual photon rings: radii in the image plane and in bulk, impact parameters, number of half-orbits, and their corresponding fluxes. We modeled the Sgr A* black hole as Bardeen-BH and GC-BH with $k=0.50M$, and calculated the diameters and widths of the first three photon rings, both from the background metric and the effective metric, using the accretion disk model with $\theta_o=80^o$, and summarized the results in Table~\ref{BarPhotonRing} and \ref{NSPhotonRing}. For the Bardeen-BH, the photon ring diameter from the effective metric is about 25$\%$ larger than those derived from the background metric. The width, and subsequently the flux, of these rings exponentially falls with order number $n$, making them challenging to observe. For instance, the $n=3$ ring itself is positioned already very close to the critical curve and has a width of the order of nano-arc-seconds.  While the critical curve is completely independent of the accretion and emission profiles and remains unobservable, our best bet are $n=1$ and $n=2$ rings for the precision test of gravity \cite{Gralla:2019xty,Gralla:2020srx}. Although these rings have not been observed directly in the shadows of M87* and Sgr A* due to the limited angular resolution of the EHT, preliminary studies suggested that they would produce a clean interferometric signature in the black hole images with a space-based interferometer with higher angular resolutions and/or flux-sensitivities, such as the Black Hole Explorer (BHEX) \cite{Johnson:2019ljv,Himwich:2020msm,Hadar:2020fda,Lupsasca:2024xhq}. The 230GHz  interferometric signatures of first three photon rings, derived from the effective metric, of diameters $d$ and width $w$, as presented in Table~\ref{BarPhotonRing} and \ref{NSPhotonRing}, are shown in Fig.~\ref{fig:PRfourier} as a function of baseline length. The visibility amplitude is normalized to have a value of unity at zero baseline length. Increasing the baseline improves angular resolution, enabling the resolution of diameter and width of higher-order photon rings. The important feature of our interest is two distinct periodicities in visibility amplitude for each ring: smaller and larger periodicities, respectively, encode the ring diameter and its width information. Specifically, to resolve image features of size $L$, baselines of length greater than $1/L$ are required. Therefore to resolve a diameter $d$, and both the diameter $d$ and width $w$, of a photon ring, we require baseline $u$ given as follows (i) and (ii), respectively \cite{Johnson:2019ljv}
\begin{equation}
	(i):\;\;\; \frac{1}{d}\ll u \ll  \frac{1}{w},\qquad
	(ii):\;\;\; \frac{1}{d}\ll  \frac{1}{w}\ll u.\\
\end{equation}
The ratio of the diameter to the width of a given photon ring is the same as the ratio of the two periodicities in the Fourier domain. While the diameter and width of the $n=1$ photon ring can be resolved with a small baseline such as with ngEHT, resolving the width of the $n=2$ ring necessitates an exceedingly large baseline as with BHEX. A larger baseline is required to resolve the photon ring of GC-BH compared to that of Bardeen-BH due to its narrower width.

The instability of null geodesics is better understood in terms of Lyapunov exponent $\gamma$ \cite{Cardoso:2008bp}, defined such that a nearly-bound photon starting at a radius $\delta r_0$ close to one of the photon orbit ends-up at $\delta r_n$ after making $n$-loops around the black hole 
\begin{equation}
	\delta r_n=e^{\gamma n}\delta r_0.
\end{equation}
For sufficiently higher-order photon rings,  $\gamma$ solely controls the relative width and essentially the flux between successive rings on the image plane and is an observable.  Similar to the shadow boundary curve, $\gamma$ is determined only by black hole parameters, completely independent of surrounding astrophysical processes. Therefore, a precise measurement of $\gamma$ gives shadow size \textit{independent} constrains on black holes' parameters. For instance, for the Schwarzschild black hole, $\gamma=\pi$, so each successive photon ring  is $e^{\pi}\sim 23.1$ times fainter than the previous one \cite{Bozza:2007gt, Johnson:2019ljv,Gralla:2019xty}.  Space-based interferometer is expected to measure the Lyapunov exponents of black holes \cite{Johnson:2019ljv}.   Figure~\ref{fig:Gamma} depicts the Lyapunov exponent of the Bardeen-BHs and GC-BHs determined from the background metric and the effective metric as a function of $k$. $\gamma$ is smaller than the Schwarzschild value, resulting in photon rings that are more wider and brighter than those for the Schwarzschild black hole. However, the effective metric predicts a non-monotonic behavior for $\gamma$, where HUCO with sufficiently large $k$ values have $\gamma > \pi$, resulting in successive photon rings becoming thinner and fainter compared to those in black holes. Note a discontinuous limit of $\gamma$ as $k\to 0$ for Bardeen-BH $\tilde{g}_{\mu\nu}$. 

\section{Discussion on NED induced effective spacetime}\label{sec-6}
We start by investigating the NED field within a Minkowski vacuum background, akin to the analysis conducted for Bardeen and GC NED  models  but without black holes in background. Our objective is to identify the unique characteristics of the resulting effective metric $\tilde{g}_{\mu\nu}$ for photon propagation in the absence of any external gravitational field. It turns out that the light propagation in Minkowski NED vacuum shows some profound and wide-ranging consequences due to the modification of the Minkowski geometry, effectively  simulating an external gravitational field:
\begin{equation}
	\tilde{g}^{\mu\nu}= \mathcal{L'} \eta^{\mu\nu} - 4 \mathcal{L''} F^{\mu\alpha} \tensor{F}{_\alpha^\nu},\label{EffMinkowski}
\end{equation}
where $\eta^{\mu\nu}$ is the Minkowski metric.

For Bardeen model, using $\mathcal{L(F)}$ from Eq.~(\ref{BarLag}) in (\ref{EffMinkowski}), the NED vacuum effective metric $\tilde{g}_{\mu\nu}$ reads
\begin{equation}
	\tilde{\rm ds}^2=\frac{4(k^2+r^2)^{7/2}s}{15 k r^6}\left(-dt^2 + dr^2+\frac{2(k^2+r^2)r^2}{3r^2-4k^2}\, d\Omega_2^2\right),\label{MinkowskiBH1}
\end{equation}
which mimics an ``effective" gravitational field with non-zero spacetime curvature guiding the photons trajectories.  While the metric (\ref{MinkowskiBH1}) does not replicate a black hole spacetime, i.e., lacking an event horizon, it does demonstrate a photon's unstable circular orbit at a radius of $r_{\rm p}=\sqrt{\frac{2}{3}(2+\sqrt{7})}k$, alongside a curvature singularity at $r_{\text{sing}}=\frac{2}{\sqrt{3}}k$. Both the curvature singularity and the photon orbit disappear as $k\to 0$.

Similarly, for GC NED model (\ref{Lag2}), the vacuum effective metric $\tilde{g}_{\mu\nu}$
\begin{equation}
	\tilde{\rm ds}^2=\frac{4r e^{k s/r}}{4r-ks}\left(-dt^2 + dr^2+\frac{2r^3(4r-ks)}{8r^2-7k r s  +k^2 s^2}\, d\Omega_2^2\right),\nonumber
\end{equation}
yields three curvature singularities located at $r_{\text{sing}}=\{\frac{k s }{4}, \frac{1}{16}(7\pm \sqrt{17})k s\}$, along with one photon circular orbit at $r_{\rm p}=1.02469 k s$. This illuminates intriguing aspects of solely NED fields and their impact on photon trajectories through $\tilde{g}_{\mu\nu}$ metric. 

\begin{figure}
	\begin{tabular}{c c}
		\includegraphics[scale=0.4]{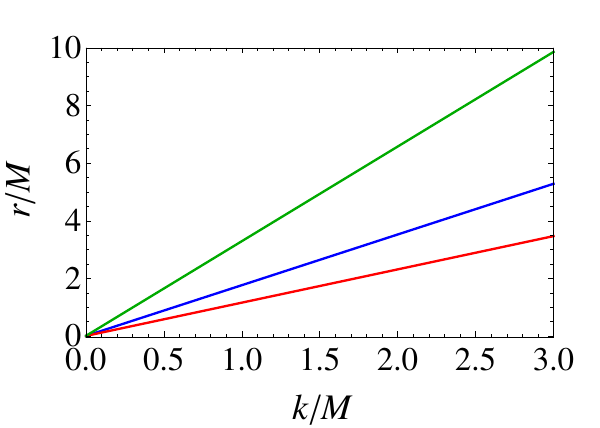}&
		\includegraphics[scale=0.4]{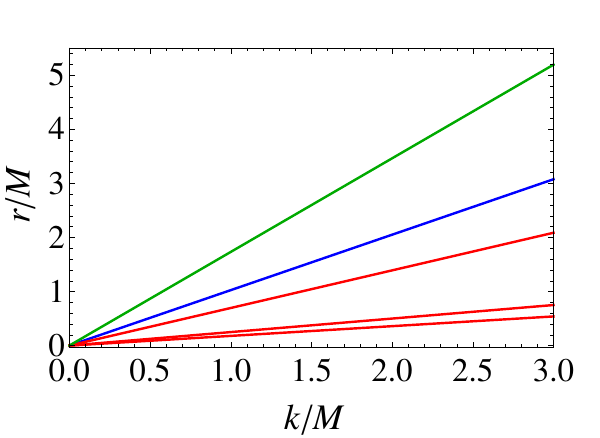}
	\end{tabular}	
	\caption{Effective metric $\tilde{g}_{\mu\nu}$ for NED fields even in purely Minkowski background, without black hole, lead to some interesting consequences. Both the Bardeen NED model (left figure) and the GC NED model (right figure) show that $\tilde{g}_{\mu\nu}$ exhibits a curvature singularity (red curve), a photon circular orbit (blue curve), and a shadow radius (green curve).} \label{fig:minkowski}
\end{figure}

If we compare the $\tilde{g}_{\mu\nu}$ metric predictions of a NED field in vacuum from Eq.~(\ref{EffMinkowski}) and the same NED field with a black hole background from Eq.~(\ref{EffMetric}), the findings are surprising.  In both scenarios, the metric $\tilde{g}_{\mu\nu}$ exhibits photon circular orbits, curvature singularities, and yields a finite photon-captured cross-section for all $k$ values as shown in Fig.~\ref{fig:minkowski}. Despite sharing the identical radial coordinates for the curvature singularities, the photon circular orbit and resulting shadow are notably smaller in NED vacuum compared to those in the presence of a black hole, particularly for $k\leq k_{\rm E}$. However, as $k$ increases significantly ($k\gg k_{\rm E}$), the photon circular orbit radius and shadow size in the presence of a black hole converge to their values in vacuum background. This convergence occurs because, at high charge values of $k$, the NED field dominates over gravitational effects in the metric $\tilde{g}_{\mu\nu}$, influencing photon trajectories and shadows primarily through the NED field rather than gravitational mass. In summary, the curvature singularities in the metric $\tilde{g}_{\mu\nu}$ for both Bardeen and GC models are attributed to their NED fields rather than black hole geometry. Furthermore, the circular photon orbits in Bardeen-HUCO and GC-HUCO spacetimes for $k>k_{\rm c}$ are not due to non-trivial features of HUCOs gravity, but rather a consequence of the NED field.  These effects highlight the important impact of the NED field on photon propagation.  Notably, in electrically charged NED vacuum spacetime, Novello \textit{et al.} have demonstrated the formation of trapped surfaces by NED fields, resulting in an electromagnetic analogue of gravitational black holes \cite{Novello:2002qg}, and the imposition of closed photon orbits \cite{Novello:2001fv}.

For Bardeen NED model (\ref{BarLag}), the weak-field expansion of Lagrangian density is as follows
\begin{equation}
	\mathcal{L(F)}\simeq \frac{3\times 2^{3/4} M}{\sqrt{k}}\mathcal{F}^{5/4} - \frac{15 \sqrt{k}M}{2^{3/4}}\mathcal{F}^{7/4} + \mathcal{O}(\mathcal{F}^{11/4}),\label{BH1Weak}
\end{equation}
with some interesting limits
\begin{eqnarray}
	&&\lim_{r\to 0}\{\mathcal{L(F)}, \mathcal{L'(F)}, \Phi(\mathcal{F})\}  \to \left\{\frac{12M}{k^3},0,0\right\},\nonumber\\
	&& \lim_{r\to \infty}\{\mathcal{L(F)}, \mathcal{L'(F)}, \Phi(\mathcal{F})\}  \to \{0,0,0\},\nonumber\\
	&& \lim_{k\to 0}\{\mathcal{L(F)}, \mathcal{L'(F)}, \Phi(\mathcal{F})\}  \to \left\{0,\frac{15M}{2 r},\frac{45M}{4r}\right\}.\label{BarWeak}
\end{eqnarray}
and the effective metric
\begin{eqnarray}
	\lim_{k \to 0}\tilde{\rm d s}^{2}&=&\frac{2r}{15}\Bigg(-\left(1-\frac{2M}{r}\right) dt^2,\nonumber\\
	&& +\left(1-\frac{2M}{r}\right)^{-1} dr^2  +\frac{2r^2}{3}d\Omega_2^2  \Bigg).\label{BH-Ilimit}
\end{eqnarray}
Because the leading-order term of $\mathcal{L(F)}$ in Eq.~(\ref{BH1Weak})  does not precisely match with the Maxwell's term, both the Bardeen metrics $g_{\mu\nu}$ and $\tilde{g}_{\mu\nu}$ fail to reproduce the RN metric in the weak electromagnetic field regime ($r\gg k$). In addition,  $\tilde{g}_{\mu\nu}$, neither in the vacuum nor in presence of a black hole, converge to the Minkowski metric or the Schwarzschild metric, respectively, as the limit $k\to 0$ approached. This deviation of $\tilde{g}_{\mu\nu}$ from GR reflects in the discontinuous limits of the black hole shadow size and the Lyapunov exponent as well, where $R_{\text{sh}}$ and $\gamma$ fail to converge to the Schwarzschild black hole values as $k\to0$ (cf. Fig.~\ref{fig:Shadow-Eff} and \ref{fig:Gamma}). 

For GC NED model, the weak-field expansion of Lagrangian density is
\begin{eqnarray}
	\mathcal{L(F)}&\simeq &\mathcal{F} - \frac{k^{3/2}}{2^{5/4}M}\mathcal{F}^{5/4} + \frac{k^3}{8\sqrt{2}M^2}\mathcal{F}^{3/2} +  \mathcal{O}(\mathcal{F}^{2})\label{MinkowskiBHII}
\end{eqnarray}	
with some interesting limits 
\begin{eqnarray}
	&&\lim_{r\to 0}\{\mathcal{L(F)}, \mathcal{L'(F)}, \Phi(\mathcal{F})\}  \to \{0,0,0\},\nonumber\\
	&& \lim_{r\to \infty}\{\mathcal{L(F)}, \mathcal{L'(F)}, \Phi(\mathcal{F})\}  \to \{0,1,1\},\nonumber\\
	&& \lim_{k\to 0}\{\mathcal{L(F)}, \mathcal{L'(F)}, \Phi(\mathcal{F})\}  \to \{0,1,1\},\label{NSWeak}
\end{eqnarray}
and the effective metric
\begin{eqnarray}
	\lim_{k \to 0}\tilde{\rm ds}^{2}&=&-\left(1-\frac{2M}{r}\right) dt^2 + \left(1-\frac{2M}{r}\right)^{-1} dr^2,\nonumber\\  &&+r^2d\Omega_2^2 .\label{BH-IIlimit}
\end{eqnarray}
The leading-order term of $\mathcal{L(F)}$ in Eq.~(\ref{MinkowskiBHII}) is exactly Maxwell, and thereby the GC metrics $g_{\mu\nu}$ and  $\tilde{g}_{\mu\nu}$ converge, to the RN metric in the limit $r\gg k$, to the Minkowski metric as  $r\to \infty$, and to the Schwarzschild metric as $k\to 0$. This limiting behavior of the $\tilde{g}_{\mu\nu}$ is apparent in the continuous limits of $R_{\text{sh}}$ and $\gamma$, where they converge to the Schwarzschild black hole values as $k\to 0$. 

For a general magnetically charged NED model,  its effects in the metric $\tilde{g}_{\mu\nu}$ directly appear through two terms: $\mathcal{L'}$ and $\Phi$. This means that we can predict certain features of the resulting metric $\tilde{g}_{\mu\nu}$ by analyzing $\mathcal{L}$, $\mathcal{L'}$, and $\Phi$, independent of the black hole metric $g_{\mu\nu}$. In particular, $\displaystyle{\lim_{r \to 0}} \;\mathcal{L}/2$ gives the energy density at the black hole center, revealing the black hole's interior  topology. As stated earlier, the de-Sitter and Minkowski natures of Bardeen and GC centers, respectively,  are attributed to the $\displaystyle{\lim_{r \to 0}}{T}{^\mu_\nu}$ values. Null and weak energy conditions hold if $\mathcal{L}\geq 0$ and $\mathcal{L'}\geq 0$ (cf. Eq.~(\ref{EMT})). We find that
\begin{itemize}
	\item For a black hole featuring a de-Sitter center and asymptotic flatness, $\mathcal{L}$ is a definite positive, monotonically decreasing function of $r$ without an extremum, ensuring $\mathcal{L'}>0$ for $r>0$ (e.g., Bardeen-BH).
	\item Conversely, for a black hole with a Minkowski core and asymptotic flatness, $\mathcal{L}$ vanishes as $r\to 0$ and $r\to \infty$, remaining nonzero in between, implying the existence of at least one extremum at $r>0$ where $\mathcal{L'} = 0$ (e.g., GC-BH).
	\item However, $\Phi$ always vanishes for magnetically charged NED black holes somewhere between the black hole center and the spatial infinity \cite{Bronnikov:2022ofk}. Notably, $\Phi=0$ results in a divergence in the angular part of the metric $\tilde{g}_{\mu\nu}$ (\ref{effmetric}).
	\item  $\mathcal{L'}= 0$ or $\Phi= 0$ generally occurs at distinct radii and introduce curvature  singularities to the effective metric $\tilde{g}_{\mu\nu}$.
\end{itemize}
Although at spacetime center,  $\mathcal{L'}, \Phi \to 0$, curvature scalars from the effective metric vanish. While, Bardeen $\tilde{g}_{\mu\nu}$ exhibits curvature singularity only from $\Phi=0$, GC exhibits from both $\mathcal{L}'=0$ and $\Phi=0$. Our analysis suggests that the effective metric of a \textit{magnetically} charged NED regular black hole exhibits at least one curvature singularity, which depending on the charge might appear inside or outside the horizon. Similar singularities in the NED effective metric for the electrically charged ABG black hole were reported in Ref.~\cite{Novello:2000km}, where both the curvature scalar and the photon radial potential diverge.  

At the singularities, photon geodesics equations take the following form
\begin{eqnarray}
	\lim_{\mathcal{L}'\to 0 }\{\dot{x}^{\mu} \}&=&\Big\{ 0,\; 0, \pm \frac{\Phi}{C(r)}\sqrt{\mathcal{K}-L^2 \cot^2\theta },\; \frac{L\Phi }{C(r)\sin^2\theta}\Big\},\nonumber\\
	\lim_{\Phi\to 0 }	\{\dot{x}^{\mu} \}&=&\Big\{ \frac{E \mathcal{L'}}{A(r)},\; \pm  \frac{E \mathcal{L'}}{\sqrt{A(r)B(r)}},\;0,\; 0 \Big\}.\label{GEq1}
\end{eqnarray}
The geodesic equations are integrable across the $\Phi=0$ singular surface. Particularly for Bardeen effective metric, ingoing null geodesics directed toward $r=0$ smoothly traverse this curvature singularity, crossing it within a finite affine parameter and finite radial velocity. Once past the singularity, these photons do not encounter any radial turning point and approach the center $r\to 0$ asymptotically in infinite affine time. However, across the $\mathcal{L'}=0$ singularity, geodesic equations lose integrability; $\dot{r}^2<0$ for $r<r_2$ where $\mathcal{L'}(r_2)=0$. Equations~(\ref{req}) and (\ref{effectivepot1}) illustrate that the  singularity at $r_2$  acts like a potential wall for the photons, where $\dot{r}\to 0$ as $r\to r_2$. Consequently, for BH-II, infalling photons from the exterior region $r> r_+$, except those with zero angular momentum, cannot approach the center \cite{Bronnikov:2000vy}. These photons, in principle, can either bounce back to the same spacetime or traverse to a distinct copy of the spacetime. Additionally, the $\tilde{g}_{\mu\nu}$ metric signature switches from $(-,+,+,+)$ to $(-,+,-,-)$ within the outermost singularity, where the angular coordinates ($\theta, \phi$) become timelike, facilitating compact and closed orbits. From the geodesic equations (\ref{GEq1}), the angular velocity vanish when $\Phi=0$ (i.e., at the outermost singularity) even for photons with non-zero angular momentum $L$. Therefore, in HUCO spacetimes, a given photon with $L>0$ have positive (negative) angular velocity outside (inside) the singularity, such that photons always cross the outermost singular surface orthogonally. The physical relevance of these curvature singularities in metric $\tilde{g}_{\mu\nu}$ and photon trajectories within it are not clear and require further investigation.  The causal structure of the metric $\tilde{g}_{\mu\nu}$ might help to better understand this. However, addressing this issue is  beyond the scope of this paper and will be explored in future work. Nevertheless, for black holes, these singularities are covered by the event horizon; thus, once photon trajectories intersect the event horizons, further tracing them to the singularity becomes irrelevant for black hole observations. However, for HUCOs, we exclusively consider photon motion only outside the outermost singularity.  Similar assumptions were made in the papers \cite{dePaula:2023ozi,Allahyari:2019jqz}. An interesting study on the photon geodesics in the metric $\tilde{g}_{\mu\nu}$ spacetime is presented in Ref.~\cite{Bronnikov:2022ofk}.

Now, let us examine a general spherically symmetric model of a single-invariant $\mathcal{F}$ dependent magnetically charged NED field $\mathcal{L(F)}$. In the weak-field ($k\ll M$) regime, the Lagrangian density and its derivatives approximately take the following forms
\begin{eqnarray}
	&&\mathcal{L(F)}\simeq w\mathcal{F}^{s+1}+....,\nonumber\\
	&&\mathcal{L'}\simeq w(s+1)\mathcal{F}^{s}+....,\nonumber\\
	&&\mathcal{L''}\simeq w s(s+1)\mathcal{F}^{s-1}+.....	\label{GenNED0}
\end{eqnarray}
Here, we consider that the NED field lacks a correct Maxwell weak-field limit, i.e., $s>0$, with $w$ being a positive constant.  Considering a two-parameter ($M,k$) background metric $g_{\mu\nu}$, such that $\displaystyle{\lim_{k \to 0}}g_{\mu\nu}\to g_{\mu\nu}^{\rm Schw}$, the effective metric in weak-field reads as:
\begin{eqnarray}
	\tilde{g}^{\mu\nu}&=&\mathcal{L'}\left(g^{\mu\nu} +2\frac{s}{r^2}\left( \delta^{\mu}_{\theta}\delta^{\nu}_{\theta} +\frac{1}{\sin^2\theta} \delta^{\mu}_{\phi}\delta^{\mu}_{\phi}\right)\right).\label{GenNED}
\end{eqnarray}
One immediate consequence is that, in general, the metric $\tilde{g}_{\mu\nu}$ is not conformal to the metric $g_{\mu\nu}$ unless $s=0$; otherwise, both the background and the effective metrics would give identical null geodesics, resulting in identical shadows. In contrary to GC ($s=0$), the Bardeen ($s\neq 0$) metric $\tilde{g}_{\mu\nu}$ is not conformal to the Schwarzschild metric, even in the limit $k\to 0$, explaining the discontinuous limits of the Bardeen-BH shadow size and the Lyapunov exponent. This establishes a general result: ``For spherically symmetric magnetically charged single Lorentz-invariant NED models $\mathcal{L(F)}$ with the correct Maxwell limit in the weak-field and a background metric $g_{\mu\nu}$, the effective metric $\tilde{g}_{\mu\nu}$ is conformal to the metric $\displaystyle{\lim_{k \to 0}}g_{\mu\nu}$ in the limit of vanishing magnetic charge." Additionally, for such NED black holes, the lensing observables smoothly approach the Schwarzschild value as the magnetic charge tends to vanish. This result is independent of the black hole interior geometry and solely depends on the weak-field limit in $r\ggg k$. 

Nevertheless, the $(t,r)$-part of a generic $\tilde{g}_{\mu\nu}$, for a magnetically charged NED field, is always conformal to the part of $g_{\mu\nu}$. Consequently, the \textit{radial} null geodesics of $\tilde{g}_{\mu\nu}$ and $g_{\mu\nu}$ are the same; radially moving photons follow the same null geodesics in both metrics. However, for $\dot{\theta}\neq 0$ and/or $\dot{\phi}\neq 0$, the null geodesics of the metric $\tilde{g}_{\mu\nu}$ are either timelike or spacelike trajectories of the metric $g_{\mu\nu}$.
From the null geodesics normalization Eq.~(\ref{normalization}), one  gets 
\begin{eqnarray}
	g^{\mu\nu}p_{\mu}p_{\nu}&=& -4\frac{\mathcal{L}''}{\mathcal{L}'}g^{\theta\theta}g^{\phi\phi}(F_{\theta\phi})^2(g^{\theta\theta} p_{\theta}^2 + g^{\phi\phi} p_{\phi}^2).	\label{norm}
\end{eqnarray}
This reaffirms that a radially directed light ray, with $p_{\theta}=p_{\phi}=0$, moves along the null directions of the background metric. Noticing that the background NED field is also radial, only photons propagating orthogonal to the NED field are subject to follow the effective metric. Utilizing Eq.~(\ref{norm}), we establish the causality conditions, ensuring that although light rays follow the null geodesics of the effective metric, they remain causal curves of the background metric.
\begin{equation}
	g^{\mu\nu}p_{\mu}p_{\nu}\leq 0\Rightarrow \frac{\mathcal{L''}}{\mathcal{L'}}\geq 0.
\end{equation}
In the Bardeen spacetime, light follows timelike geodesics of the metric $g_{\mu\nu}$ within $r\geq \sqrt{6}k$ and spacelike for $r<\sqrt{6}k$.  Conversely, in the GC spacetimes, light follows timelike geodesics of the metric $g_{\mu\nu}$ only within the range $k^2/10<r<k^2/8$. Bronnikov~\cite{Bronnikov:2022ofk} obtained similar results for a generically magnetically charged NED black hole, indicating a violation of causality near the black hole's center.  Additionally, Ref.~\cite{dePaula:2023ozi} demonstrated that photons follow spacelike geodesics of the background spacetime in the presence of an electrically charged NED field. The net effect of the NED field's force on photons resembles that of gravity  \cite{Novello:2000xw}. In this context, from the background metric perspective, photons move under a radially attractive force thus causing larger shadows than that for the null geodesics of the background metric \cite{dePaula:2023ozi,Novello:2000xw}.

Regarding stability, while the NED charged regular black holes demonstrate stability against gravitational and electromagnetic perturbations~\cite{Toshmatov:2019gxg}, regular HUCOs lack conclusive stability assessment due to limited dynamics understanding. In the context of the background metric, these regular HUCOs have pair of null circular orbits for $k<k_{\rm c}$, degenerate orbits for $k=k_{\rm c}$, and no orbits for $k>k_{\rm c}$, where $k_{\rm c}>k_{\rm E}$ (cf. Fig.~\ref{fig:BarRadii}). Indeed, it is now an established fact in GR that HUCOs spacetime, if sufficiently compact enough to develop null circular orbits, always accompanied orbits in pairs: a \textit{stable } circular orbit along with an unstable orbit~ \cite{Cardoso:2014sna,Cunha:2017qtt,Cardoso:2019rvt}. Notably, similar stable circular null orbit exists also for RN and regular black holes, however, being invariably located inside the event horizon, it does not raise concerns for black hole stability or observations. However, in HUCOs spacetimes, their presence has profound implications as slow decay of linearized fluctuations suggests potential nonlinear instability, potentially leading to collapse into a black hole~ \cite{Cunha:2017qtt,Cardoso:2019rvt}. This leads to an interesting result that ``HUCOs with null (photon) circular orbits are black hole" \cite{Cardoso:2014sna,Cunha:2017qtt}.  On the other hand, in the effective metric, Bardeen-HUCO and GC-HUCO exhibit an odd number of null (photon) circular orbits, specifically, three photon orbits  for $k_{\rm E}<k< \tilde{k}_{\rm c}$, comprising two unstable orbits and one stable orbit, and only one unstable photon orbit for $k> \tilde{k}_{\rm c}$. For $k_{\rm E}<k<\tilde{k}_{\rm c}$, while the presence of a stable photon orbit might lead to nonlinear instability due to photons accumulation and subsequent back-reaction on the spacetime,  a comprehensive stability understanding requires modeling the matter interaction with this photon orbit and calculating its reflectivity or absorption properties. This is certainly a feature worth investigating for NED charged spacetimes, but it  is beyond the scope of this paper, and we will not consider the effects of the stable photon orbits in the HUCOs spacetimes. Nevertheless, HUCOs with $k> \tilde{k}_{\rm c}$ possess only one photon circular orbit, and that too \emph{unstable} (cf. Fig.~\ref{fig:BarRadii}). This is in clear contrast with the GR predictions. Consequently, the HUCOs spacetime instabilities associated with stable photon orbits are absent in these effective metric spacetimes. As a result, unlike in GR, where stable photon orbits always accompanied unstable orbits and  yield spacetime instability on astrophysically short time scales and potentially  destroy HUCOs that could otherwise be plausible candidates for astrophysical black holes, the effective metric description suggests that regular NED charged HUCOs lack stable photon orbits (except for a small parameter space $k_{\rm E} < k < \tilde{k}_{\rm c}$ beyond the extremal limit) and are thereby free from associated instability.  Therefore, if we observe photon orbits around astrophysical HUCOs, then one possible alternative explanation to black holes is NED regular HUCOs.

In summary, these features of the NED fields and the distinctive behavior of photon geodesics  indicate that the NED description of regular black holes and the effective metric formalism for photon geodesics are not completely understood, and warrant further investigation into these topics.

\section{Summary}\label{sec-7}
NED fields significantly impact spacetime geometry, with important implications in astrophysics and cosmology. In black holes, NED fields, with a suitable Lagrangian-density $\mathcal{L(F)}$, can eliminate central curvature singularities, resulting in globally regular black hole geometries. Another relatively less explored consequence is that, in the presence of NED fields, photons do not follow the null geodesics of the background metric $g_{\mu\nu}$ but instead propagate along those determined by an ``effective metric" $\tilde{g}_{\mu\nu} (g_{\mu\nu}, \mathcal{L(F)})$. Notably, this effective metric is specific to only photons, while the trajectories of other particles remain governed by $g_{\mu\nu}$.

In this paper, we focused on these two primary aspects of NED fields: singularity-free regular black holes and the effective metric descriptions of photon propagation. We examined two well-studied NED charged regular black hole metrics, namely the Bardeen and Ghosh-Culetu (GC) models, both featuring two parameters--mass $M$ and magnetic charge $k$. Although both metrics interpolate between regular black holes ($k\leq k_{\rm E}$) and regular HUCOs ($k>k_{\rm E}$), they manifest considerable geometrical distinctions at their cores ($r\sim 0$) and in weak-field regimes ($r\gg M$). The effective metric $\tilde{g}_{\mu\nu}$ introduces intriguing effects in observational features, ranging from strong gravitational lensing to changes in the emitted photon frequency and variations in the size and shape of shadows.  We analyzed these effects using two accretion models: spherically symmetric and radially infalling flow, and Novikov-Thorne type optically and geometrically thin disk. Additionally, to assess NED effects, we compared null circular orbits and resulting shadows predicted by photons following the null geodesics of $\tilde{g}_{\mu\nu}$ with those predicted by the null geodesics of $g_{\mu\nu}$.  In both Bardeen and GC spacetimes, null circular orbits derived from the effective metric have larger radii than those from the background metric. While null circular orbits from the effective metric persist for arbitrarily large values of magnetic charge $k$, those from the background metric exist only within the parameter space $k\leq k_{\rm c}$.

We found that, from the background metric, HUCOs shadows are markedly different from black holes shadows  (see appendix~\ref{sec-8A}). Under the spherical accretion model, HUCOs shadows typically lack intensity depression at the image center, instead, the intensity grows toward the image center, resulting in ``full-moon" like shadows, unlike black hole shadows. However, HUCOs with null circular orbits exhibit an additional weak signature—a faint circular ring along with steady rise in central intensity. These differences in black holes and HUCOs shadows were anticipated because, in the absence of a horizon, radially ingoing null geodesics with small impact parameters always have a turning point at $r\sim 0$ that contribute to the intensity at the image center.  

Conversely, the effective metric predicted completely different shadows. The shadow sizes are larger than those predicted by the background metric, Additionally, the shadows of HUCOs displayed intensity depression at the image center resembling those of black holes. This difference arises from the fact that, now in HUCOs spacetimes, the ingoing null geodesics intersect a singularity at $r=r_{\text{sing}}>0$ with no turning point outside it, resulting in a dark patch at the image center. This outcome holds true for all magnetically charged NED spacetimes with a de-Sitter geometry at the center.

Accretion disk images have revealed additional effects of NED field on gravitational lensing. In the limit of large $r$ or small NED charge $k \ll M$, the  NED field for Bardeen spacetime is stronger than the Maxwell's field,  unlike for GC spacetime, where it correctly matches with the Maxwell's field. This residual NED field in Bardeen spacetime resulted in stronger gravitational lensing of photons in large-$r$ regime, in comparison to that for GC spacetime. This subsequently reflected as bulging in accretion disk images around Bardeen-BHs, as shown in \cite{Schee:2019}, but not around GC-BHs, as shown in \ref{fig:DiskEffectiveNS}. Additionally, images of HUCOs resemble those of black holes. Unlike the spherical accretion model, accretion disk images exhibit a sequence of discrete photon rings. For both Bardeen-BHs and GC-BHs, the diameter of photon rings from the effective metric is larger than those rings of the same order from the background metric.

We assessed the impact of NED field effects on photon propagation through the effective metric $\tilde{g}_{\mu\nu}$, determining shadow sizes and establishing new constraints on the NED charge parameter $k$ for Bardeen and GC spacetimes using EHT measurements of the Sgr A* and M87* black holes. Although shadow sizes predicted by the background metric and the effective metric both decrease with $k$ for $k<k_{\rm E}$,  the effective metric predicts larger shadows than the background metric for the same charge $k$. This confirms that the gravitational lensing of null geodesics in the effective metric is stronger than those in the background metric. Bardeen-BH shadow sizes from the background metric fall within EHT bounds, while those from the effective metric exceed $1\sigma$ bounds for all $k$. Therefore, with the assertion that the shadows are cast by photons following the $\tilde{g}_{\mu\nu}$, Bardeen NED model can be completely ruled out for Sgr A* and M87* black hole candidates. In contrary, shadows of GC-BHs from both metrics are consistent with the $1\sigma$ bound.  Interestingly, GC-HUCOs with $1.987M\leq k\leq 2.476M$ and $2.059M\leq k\leq 2.257M$ also cast shadows that are consistent with the EHT shadow size measurements for M87* and Sgr A*, respectively. Indeed, a shadow of a given angular size can correspond to either a GC-BH or a GC-HUCO. 

In GR, it is known that \textit{any} HUCOs, if compact enough to posses null circular orbits, exhibit stable photon rings along with the unstable rings.  These stable rings possess long-lived perturbations, leading to nonlinear instability of these objects. This leads to a beautiful and very strong result in favor of astrophysical black holes over  HUCOs ``objects with a light ring are black holes" \cite{Cardoso:2014sna,Keir:2014oka}.  However, we have shown that NED charged regular HUCOs, except for a small parameter space beyond the extremal limit $k_{\rm E}<k<\tilde{k}_{\rm c}$, do not posses any \textit{stable } null (photon) circular orbits, but only \textit{unstable} null (photon) circular orbits. Consequently, the absence of \textit{stable} photon rings renders NED charged regular black holes and HUCOs as viable alternatives to astrophysical black holes. In this case, ``objects with a light ring do not rule out the NED charged regular HUCOs". However, a  conclusive answer can be made after investigating the gravitational and  electromagnetic field stability on model-by-model basis of NED spacetimes. 

Based on these results, we aim to further explore the full general-relativistic magnetohydrodynamic (GRMHD) simulation of plasma flow and the subsequent general-relativistic radiative-transfer (GRRT) images around NED charged regular black holes and HUCOs. The NED field is expected to significantly enhance synchrotron light polarization. Estimating the effect of the NED field on GRMHD simulated images and light polarization is the focus of our future research. Furthermore, vacuum birefringence induced by the NED field leads to time delays in light signals polarized orthogonally, providing an avenue to independently constrain NED models in astrophysical contexts.

While this work was in preparation, some theoretical aspects of photon propagation in NED-charged regular black hole spacetimes were recently explored in Ref.~\cite{Murk:2024nod}.
  
\begin{acknowledgements} 
RKW expresses gratitude to Luciano Rezzolla for the invitation to visit Goethe University, where this project was conceived, and for continuous suggestions and feedback on the project.  RKW also acknowledges discussions with Rajibul Shaikh at various stages of the project, Prashant Kocherlakota for the invitation to visit BHI, Harvard University, and for insightful discussions on effective metrics for photon propagation and their causal structure, Ramesh Narayan for insights into the NED field and to Sam Gralla for discussions on ray tracing. RKW's research is supported by the Fulbright-Nehru Postdoctoral Research Fellowship (award 2847/FNPDR/2022)  from the United States-India Educational Foundation.
\end{acknowledgements} 

\begin{appendix}
\section{GC spacetime shadows from the background metric}\label{sec-8A}
As discussed in the Sec.~\ref{sec-3}, regular black holes can be motivated from various theories, such as from the quantum gravity with a minimum length parameter or a classical Einstein's gravity with some non-NED source of finite energy-momentum tensor. However, the Lagrangian formalism for both alternative theories is mostly missing. Within these alternate explanation of the regular black holes, photons continue to follow the null geodesics of the background metric $g_{\mu\nu}$. For a comparative study, we determine the shadows of GC spacetimes, originating from the non-NED theory, within the background spacetime. 

\begin{figure*}
	\begin{tabular}{c c c}
		\includegraphics[scale=0.6]{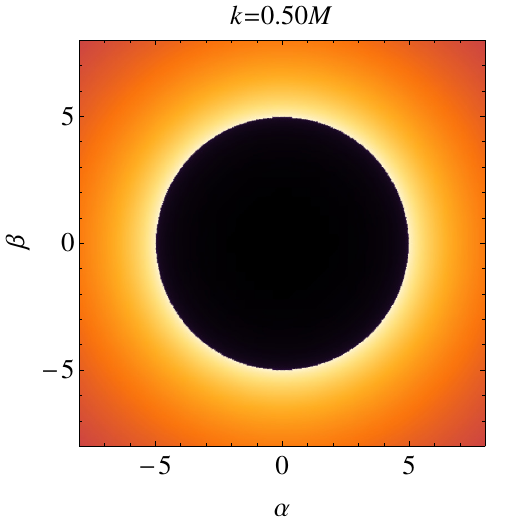}&
		\includegraphics[scale=0.6]{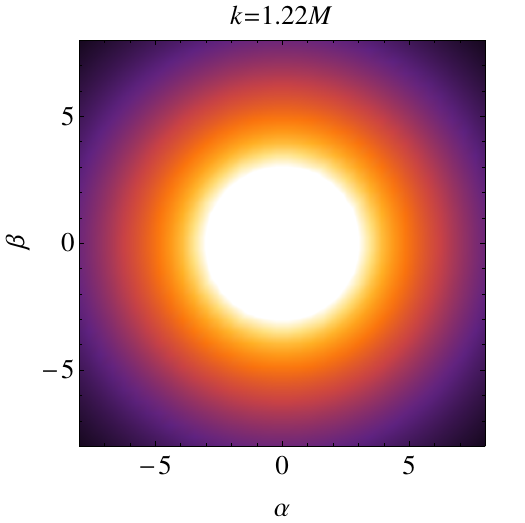}&
		\includegraphics[scale=0.6]{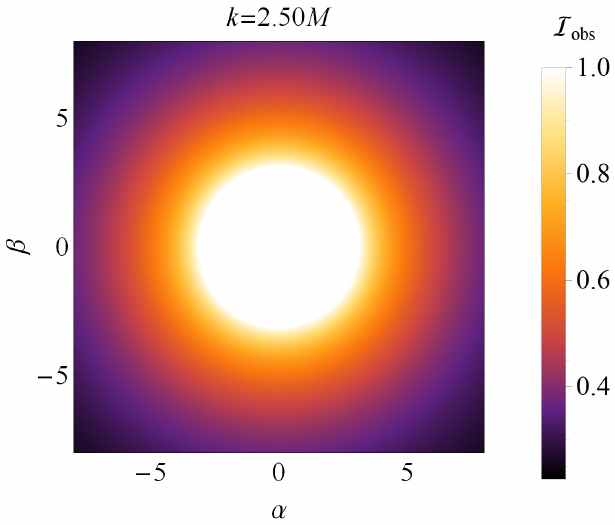}\\
	\end{tabular}	
	\caption{This figure illustrates shadows of GC-BH with $k=0.5M$ and GC-HUCOs with $k=0.80M$ and $k=2.0M$ from the background metric under the radially infalling spherical accretion.}\label{fig:InfallBackNS}
\end{figure*}
\begin{figure*}
	\includegraphics[scale=0.47]{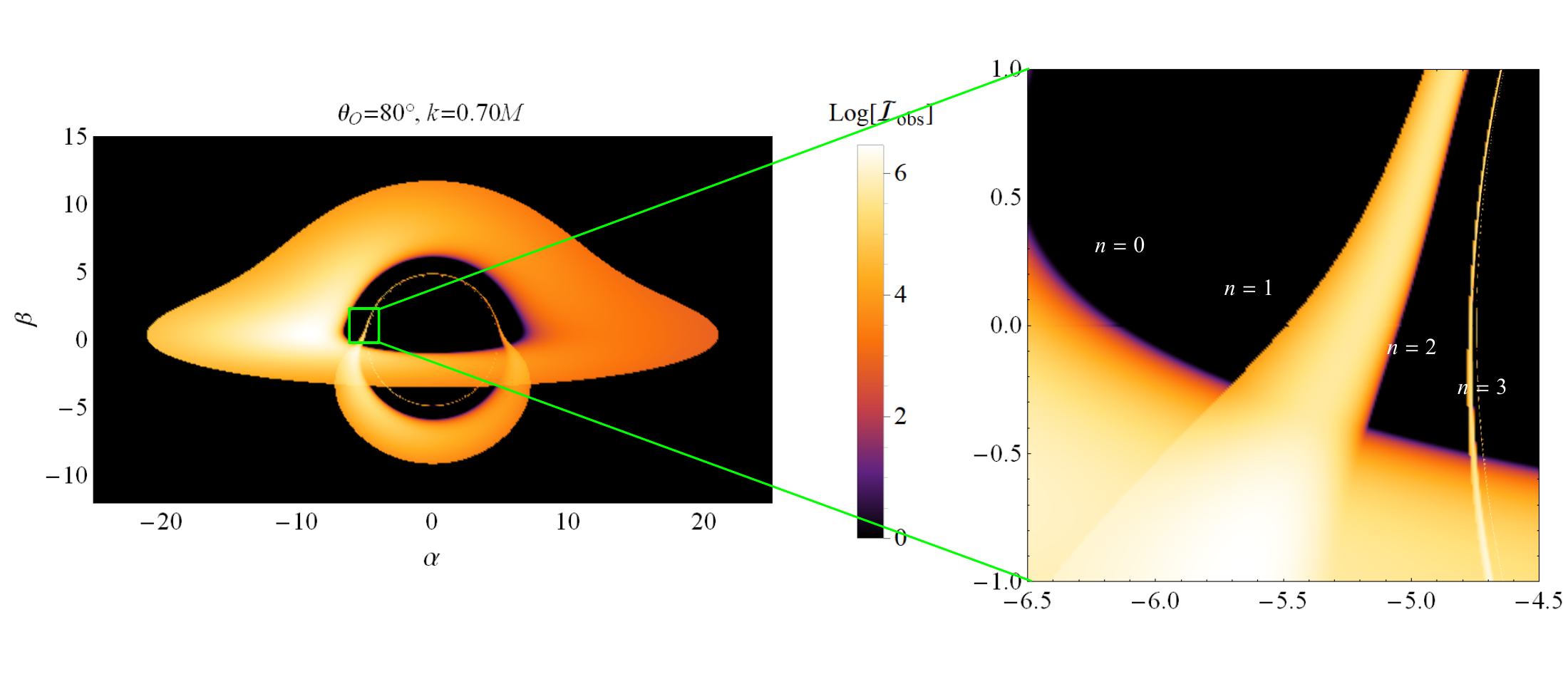}\\
	\includegraphics[scale=0.44]{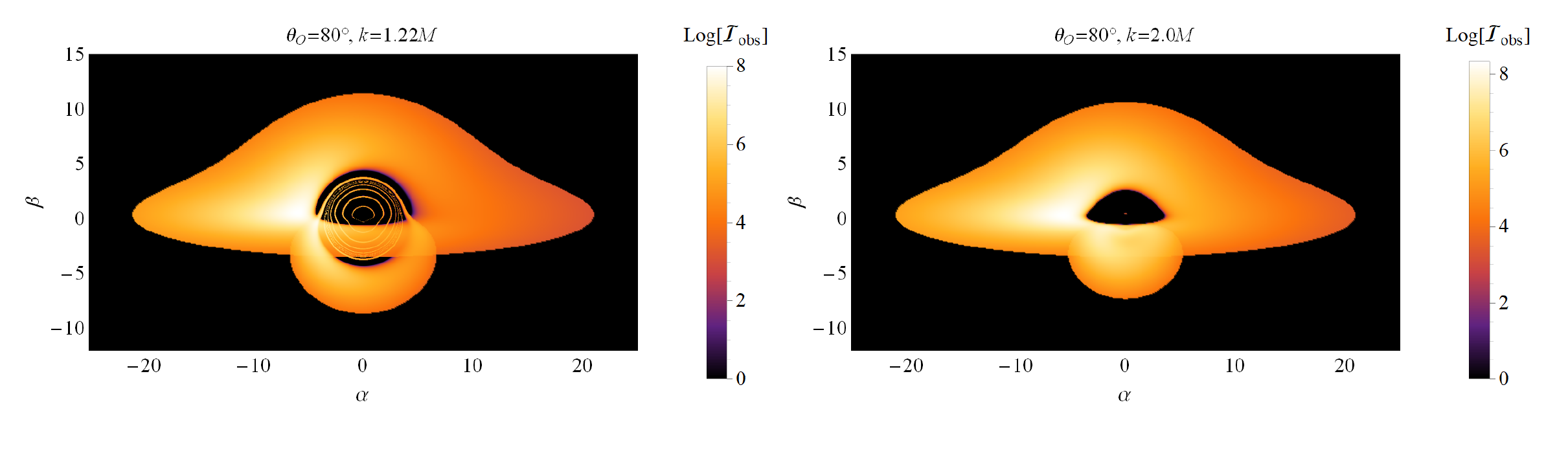}
	\caption{Shadows of GC-BH (top panel) and GC-HUCO (bottom panel) cast by null geodesics of the background metric $g_{\mu\nu}$ under an equatorial accretion disk model. $k=1.22M$ and $k=2.0M$, respectively, correspond to HUCOs with and without null circular orbits. Observed flux $\mathcal{I}_{\text{obs}}$ is represented on a logarithmic scale. }\label{fig:DiskBackgrNS}
\end{figure*}
Figure~\ref{fig:InfallBackNS} illustrates the shadows of GC-BHs and GC-HUCOs under spherical accretion. For GC-BH with $k<k_{\rm E}$, null geodesics lead to a strong intensity depression within $\alpha^2+\beta^2=b_{\text{cr}}^2$, which is a signature of black hole shadow. Note that the shadow radius is different from that determined for the effective metric, as explained previously. Conversely, for HUCOs with $k>k_{\rm E}$, the absence of a horizon allows accreting matter to exist up to $r=0$, resulting in a rising intensity toward the observer's screen center, akin to a ``full moon-like" image, attributed to the repulsive radial potential near $r\to 0$. Ingoing null geodesics experience a radial turning point $r_{\text{tp}}>0$, reflecting back toward the observer, contributing to image intensity. These new image features emerge because null geodesics can now pass through the spacetime region that was previously inside the horizon. Although radiation emitted from the close to  $r=0$ experiences high redshift,  this is compensated by the high emission rate of radiation ($j\propto 1/r^2$) from the accreting matter near $r=0$, thereby contributing to the intensity. For further analysis, we looked at two cases:  GC-HUCOs with ($k_{\rm E}<k<k_{\rm c}$) and without ($k>k_{\rm c}$) null circular orbits.  For $k_{\rm E}<k<k_{\rm c}$, ($k=1.22M$ in Fig.~\ref{fig:InfallBackNS}) we observed a secondary subdominant peak in the observed intensity, or a faint circular ring in the image, at the lensed position of null circular orbits, which disappeared for $k>k_{\rm c}$ ($k=2.50M$ in Fig.~\ref{fig:InfallBackNS}). Consequently, as predicted by the null geodesics of the background metric, shadows of GC-HUCOs  markedly differ from those of GC-BHs. This feature is in stark contrast with the behavior observed with null geodesics of the effective metric, where a ``full-moon" like shadow was absent. However, null geodesics in the effective metric undergo more redshift compared to those in the background metric, leading to a more pronounced intensity contrast in shadows from $\tilde{g}_{\mu\nu}$ than those from the $g_{\mu\nu}$.  Such ``full moon" images generically appear for RN naked singularity, and were recently reported for the JMN-1 and JMN-2 naked singularities by Shaikh \textit{et. al.}~\cite{Shaikh:2018lcc}.

In Fig.~\ref{fig:DiskBackgrNS}, the accretion disk images cast by null geodesics of the background metric of GC-BH and GC-HUCO are presented. In the case of a GC-BH, besides a central dark region, a bright and narrow ring appears that can be decomposed into a series of exponentially stacked $n = 1, 2, 3,...$ sub-rings, identified by the image order, converging to the black hole's critical curve. These subrings were absent for first accretion model discussed earlier. For $k=0.70M$, the inner edge of the disk is located at $r_{\text{in}}=5.227M$, and its direct image delineates the inner boundary of the emission ring on the image plane as shown in the top-left panel of Fig.~\ref{fig:DiskBackgrNS}. For example, along the $\alpha=0$ axis, the direct image of the disk's inner edge is mapped onto the screen coordinate at $\beta=6.084M$, and the null circular orbits at $r_{\rm p}=2.651M$ are mapped onto $\beta=\beta_{\rm c}=4.7438M$. Backtracing null geodesics from $0<\beta<\beta_{\rm c}$ end at the horizon without intersecting the accretion disk and therefore do not contribute to the intensity. However, for GC-HUCO, geodesics from $0<\beta<\beta_{\rm c}$ also play a role and yield distinctive additional signatures in the images. The shadows of GC-HUCO, both with ($k=1.22M<k_{\rm c}$) and without ($k=2.0M>k_{\rm c}$) null circular orbits, are shown in the bottom panel of Fig.~\ref{fig:DiskBackgrNS}.

For GC-HUCO with  $k=1.22M$, along the $\alpha=0$ axis on the screen, $\beta=4.385M$ and $\beta=3.45M$ map, respectively, the direct image of the inner edge of the disk at $r_{\text{in}}=3.124M$ and null circular orbits at $r_{\rm p}=1.577M$. Interestingly, unlike the black hole case, null geodesics starting from $\beta<3.45M$ still exhibit deflection angles exceeding $2\pi$ and intersect the accretion disk for discrete values of $\beta$, manifesting as an additional series of concentric circular rings at the center of the image plane. Consequently, two series of circular rings emerge for HUCO with $k<k_{\rm c}$: one for $b>b_{\text{cr}}$ resembling those found for black holes, and another for $b<b_{\text{cr}}$, representing a novel feature of HUCO. For $k=1.22M$ and for a very small values of $\beta$, the deflection angle is less than $\pi/2$, thus these geodesics neither make loops around the central object nor intersect the accretion disk, resulting in a dark region at the center of the image plane, as shown in the bottom-left panel of Fig.~\ref{fig:DiskBackgrNS}. However, these inner rings disappear in the images for GC-HUCO without null circular orbits, as shown for $k=2.0M$ in the bottom-right panel of Fig.~\ref{fig:DiskBackgrNS}.  This happens because light deflection angle decreases with increasing $k$ for a fixed value of $b$. For instance, for $k=2.0M$, the light deflection angle is less than even $\pi/2$, and for observer's inclination angle $\theta_o=80\degree$, light rays intersect the accretion disk at most twice, constructing only the direct image and the first-order image of the backside of the disk with no higher-order rings, as illustrated in the bottom-right panel of Fig.~\ref{fig:DiskBackgrNS}. The inner boundary of the emission ring on the screen corresponds to the inner edge of the disk in bulk. In summary, the intensity dip or the dark region at the image center results from the fact that the accretion disk could be extended only up to $r_{\text{in}}>0$. If $r_{\text{in}}\to 0$, then this dark region at the image center should disappear. Although  the angular frequency of Keplerian orbits becomes imaginary for $r<r_{\text{in}}$, accreting matter can radially fall into black holes within this region. GC-HUCO with and without null circular orbits manifest significantly distinct shadows compared to those of GC-BH. Additionally, the shape and size of these accretion disk shadows for GC-BH and GC-HUCO differ from those cast by null geodesics of the effective metric, as shown in Fig.~\ref{fig:DiskEffectiveNS}. For a detailed study on gravitational lensing in HUCOs spacetime, refer to Ref.~\cite{Schee:2015nua}.

\section{Schwarzschild black hole shadows}\label{sec-8B}
\begin{figure}
	\includegraphics[scale=0.67]{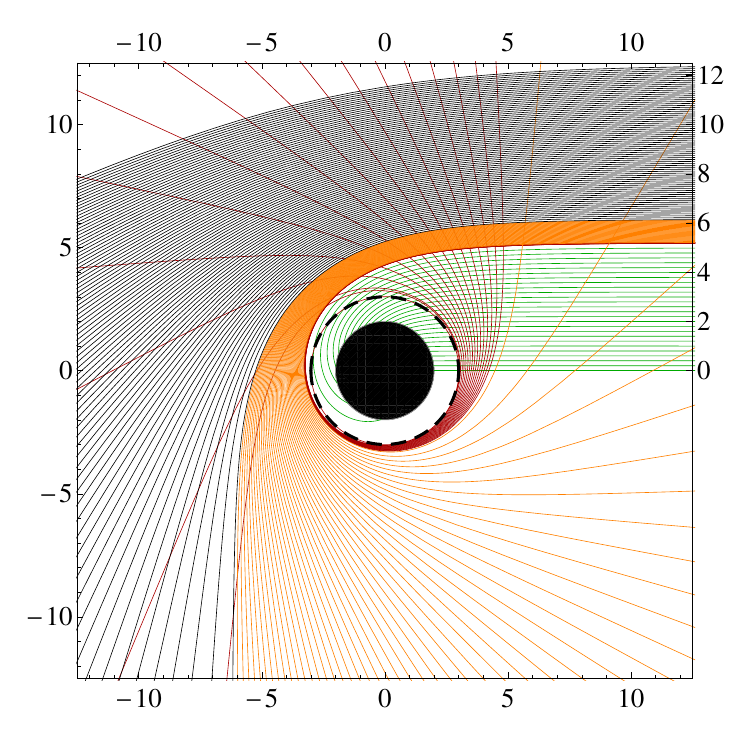}
	\caption{Gravitational lensing around Schwarzschild black hole. }
	\label{fig:orbitSchw}
\end{figure}

\begin{figure}	
	\includegraphics[scale=0.6]{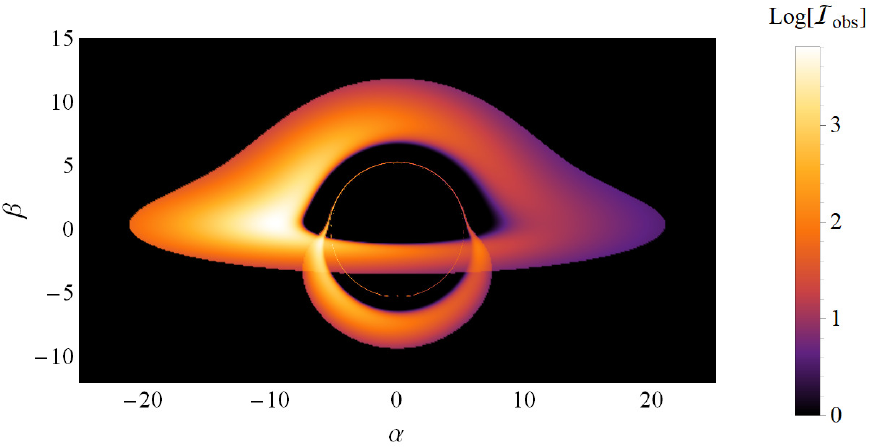}
	\caption{Shadow  of geometrically and optically thin accretion disk  around the Schwarzschild black hole.}\label{fig:AccSchw}
\end{figure}

\begin{figure}
	\includegraphics[scale=0.4]{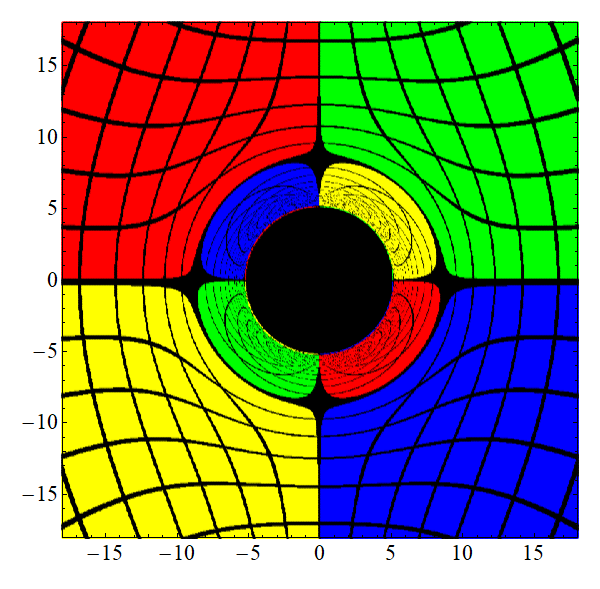}
	\caption{Gravitational lensing of light is shown as predicted by a Schwarzschild black hole at the center of a celestial sphere with a radius of $30M$. The observer is positioned at a distance of $15M$ from the center. Black lines represent lines of constant longitude and latitude, while the central black circular region indicates the shadow of the black hole.}
	\label{fig:CelesSchw}
\end{figure}

For a comparative study here we present the results for a Schwarzschild black hole. Figure~\ref{fig:orbitSchw} depict photon orbits around the Schwarzschild black hole, Fig.~\ref{fig:AccSchw} shows accretion disk image and Fig.~\ref{fig:CelesSchw} shows the celestial sphere image. Comparing the shadows of Schwarzschild black holes with those of Bardeen-BH  and GC-BH, as well as HUCOs, highlights the effects of NED fields in gravitational lensing.  This completes our comparative analysis between the shadows of regular black holes and Schwarzschild black holes, regular black holes and regular HUCOs, and regular black hole shadows predicted by the background metric and the effective metric.

\end{appendix}

\bibliography{NED_Charged_BHs}

\begin{thebibliography}{128}%
\makeatletter
\providecommand \@ifxundefined [1]{%
 \@ifx{#1\undefined}
}%
\providecommand \@ifnum [1]{%
 \ifnum #1\expandafter \@firstoftwo
 \else \expandafter \@secondoftwo
 \fi
}%
\providecommand \@ifx [1]{%
 \ifx #1\expandafter \@firstoftwo
 \else \expandafter \@secondoftwo
 \fi
}%
\providecommand \natexlab [1]{#1}%
\providecommand \enquote  [1]{``#1''}%
\providecommand \bibnamefont  [1]{#1}%
\providecommand \bibfnamefont [1]{#1}%
\providecommand \citenamefont [1]{#1}%
\providecommand \href@noop [0]{\@secondoftwo}%
\providecommand \href [0]{\begingroup \@sanitize@url \@href}%
\providecommand \@href[1]{\@@startlink{#1}\@@href}%
\providecommand \@@href[1]{\endgroup#1\@@endlink}%
\providecommand \@sanitize@url [0]{\catcode `\\12\catcode `\$12\catcode
  `\&12\catcode `\#12\catcode `\^12\catcode `\_12\catcode `\%12\relax}%
\providecommand \@@startlink[1]{}%
\providecommand \@@endlink[0]{}%
\providecommand \url  [0]{\begingroup\@sanitize@url \@url }%
\providecommand \@url [1]{\endgroup\@href {#1}{\urlprefix }}%
\providecommand \urlprefix  [0]{URL }%
\providecommand \Eprint [0]{\href }%
\providecommand \doibase [0]{https://doi.org/}%
\providecommand \selectlanguage [0]{\@gobble}%
\providecommand \bibinfo  [0]{\@secondoftwo}%
\providecommand \bibfield  [0]{\@secondoftwo}%
\providecommand \translation [1]{[#1]}%
\providecommand \BibitemOpen [0]{}%
\providecommand \bibitemStop [0]{}%
\providecommand \bibitemNoStop [0]{.\EOS\space}%
\providecommand \EOS [0]{\spacefactor3000\relax}%
\providecommand \BibitemShut  [1]{\csname bibitem#1\endcsname}%
\let\auto@bib@innerbib\@empty
\bibitem [{\citenamefont {Perez}(2017)}]{Perez:2017cmj}%
  \BibitemOpen
  \bibfield  {author} {\bibinfo {author} {\bibfnamefont {A.}~\bibnamefont
  {Perez}},\ }\bibfield  {title} {\bibinfo {title} {{Black Holes in Loop
  Quantum Gravity}},\ }\href {https://doi.org/10.1088/1361-6633/aa7e14}
  {\bibfield  {journal} {\bibinfo  {journal} {Rept. Prog. Phys.}\ }\textbf
  {\bibinfo {volume} {80}},\ \bibinfo {pages} {126901} (\bibinfo {year}
  {2017})}\BibitemShut {NoStop}%
\bibitem [{\citenamefont {Rovelli}(1998)}]{Rovelli:1997yv}%
  \BibitemOpen
  \bibfield  {author} {\bibinfo {author} {\bibfnamefont {C.}~\bibnamefont
  {Rovelli}},\ }\bibfield  {title} {\bibinfo {title} {{Loop quantum gravity}},\
  }\href {https://doi.org/10.12942/lrr-1998-1} {\bibfield  {journal} {\bibinfo
  {journal} {Living Rev. Rel.}\ }\textbf {\bibinfo {volume} {1}},\ \bibinfo
  {pages} {1} (\bibinfo {year} {1998})}\BibitemShut {NoStop}%
\bibitem [{\citenamefont {Bonanno}\ and\ \citenamefont
  {Reuter}(2000)}]{Bonanno:2000ep}%
  \BibitemOpen
  \bibfield  {author} {\bibinfo {author} {\bibfnamefont {A.}~\bibnamefont
  {Bonanno}}\ and\ \bibinfo {author} {\bibfnamefont {M.}~\bibnamefont
  {Reuter}},\ }\bibfield  {title} {\bibinfo {title} {{Renormalization group
  improved black hole space-times}},\ }\href
  {https://doi.org/10.1103/PhysRevD.62.043008} {\bibfield  {journal} {\bibinfo
  {journal} {Phys. Rev. D}\ }\textbf {\bibinfo {volume} {62}},\ \bibinfo
  {pages} {043008} (\bibinfo {year} {2000})}\BibitemShut {NoStop}%
\bibitem [{\citenamefont {Torres}(2014)}]{Torres:2014gta}%
  \BibitemOpen
  \bibfield  {author} {\bibinfo {author} {\bibfnamefont {R.}~\bibnamefont
  {Torres}},\ }\bibfield  {title} {\bibinfo {title} {{Singularity-free
  gravitational collapse and asymptotic safety}},\ }\href
  {https://doi.org/10.1016/j.physletb.2014.04.010} {\bibfield  {journal}
  {\bibinfo  {journal} {Phys. Lett. B}\ }\textbf {\bibinfo {volume} {733}},\
  \bibinfo {pages} {21} (\bibinfo {year} {2014})}\BibitemShut {NoStop}%
\bibitem [{\citenamefont {Nicolini}\ \emph {et~al.}(2019)\citenamefont
  {Nicolini}, \citenamefont {Spallucci},\ and\ \citenamefont
  {Wondrak}}]{Nicolini:2019irw}%
  \BibitemOpen
  \bibfield  {author} {\bibinfo {author} {\bibfnamefont {P.}~\bibnamefont
  {Nicolini}}, \bibinfo {author} {\bibfnamefont {E.}~\bibnamefont
  {Spallucci}},\ and\ \bibinfo {author} {\bibfnamefont {M.~F.}\ \bibnamefont
  {Wondrak}},\ }\bibfield  {title} {\bibinfo {title} {{Quantum Corrected Black
  Holes from String T-Duality}},\ }\href
  {https://doi.org/10.1016/j.physletb.2019.134888} {\bibfield  {journal}
  {\bibinfo  {journal} {Phys. Lett. B}\ }\textbf {\bibinfo {volume} {797}},\
  \bibinfo {pages} {134888} (\bibinfo {year} {2019})}\BibitemShut {NoStop}%
\bibitem [{\citenamefont {Cano}\ \emph {et~al.}(2019)\citenamefont {Cano},
  \citenamefont {Chimento}, \citenamefont {Ort\'\i{}n},\ and\ \citenamefont
  {Ruip\'erez}}]{Cano:2018aod}%
  \BibitemOpen
  \bibfield  {author} {\bibinfo {author} {\bibfnamefont {P.~A.}\ \bibnamefont
  {Cano}}, \bibinfo {author} {\bibfnamefont {S.}~\bibnamefont {Chimento}},
  \bibinfo {author} {\bibfnamefont {T.}~\bibnamefont {Ort\'\i{}n}},\ and\
  \bibinfo {author} {\bibfnamefont {A.}~\bibnamefont {Ruip\'erez}},\ }\bibfield
   {title} {\bibinfo {title} {{Regular Stringy Black Holes?}},\ }\href
  {https://doi.org/10.1103/PhysRevD.99.046014} {\bibfield  {journal} {\bibinfo
  {journal} {Phys. Rev. D}\ }\textbf {\bibinfo {volume} {99}},\ \bibinfo
  {pages} {046014} (\bibinfo {year} {2019})}\BibitemShut {NoStop}%
\bibitem [{\citenamefont {Nicolini}\ \emph {et~al.}(2006)\citenamefont
  {Nicolini}, \citenamefont {Smailagic},\ and\ \citenamefont
  {Spallucci}}]{Nicolini:2005vd}%
  \BibitemOpen
  \bibfield  {author} {\bibinfo {author} {\bibfnamefont {P.}~\bibnamefont
  {Nicolini}}, \bibinfo {author} {\bibfnamefont {A.}~\bibnamefont
  {Smailagic}},\ and\ \bibinfo {author} {\bibfnamefont {E.}~\bibnamefont
  {Spallucci}},\ }\bibfield  {title} {\bibinfo {title} {{Noncommutative
  geometry inspired Schwarzschild black hole}},\ }\href
  {https://doi.org/10.1016/j.physletb.2005.11.004} {\bibfield  {journal}
  {\bibinfo  {journal} {Phys. Lett. B}\ }\textbf {\bibinfo {volume} {632}},\
  \bibinfo {pages} {547} (\bibinfo {year} {2006})}\BibitemShut {NoStop}%
\bibitem [{\citenamefont {{Bardeen}}(1968)}]{Bardeen-RegularBH}%
  \BibitemOpen
  \bibfield  {author} {\bibinfo {author} {\bibfnamefont {J.}~\bibnamefont
  {{Bardeen}}},\ }\bibfield  {title} {\bibinfo {title} {{Non-singular general
  relativistic gravitational collapse}},\ }in\ \href
  {https://doi.org/https://ui.adsabs.harvard.edu/abs/1968qtr..conf...87B}
  {\emph {\bibinfo {booktitle} {Proceedings of the 5th International Conference
  on Gravitation and the Theory of Relativity}}}\ (\bibinfo {year} {1968})\
  p.~\bibinfo {pages} {87}\BibitemShut {NoStop}%
\bibitem [{\citenamefont {Pellicer}\ and\ \citenamefont
  {Torrence}(1969)}]{Pellicer:1969cf}%
  \BibitemOpen
  \bibfield  {author} {\bibinfo {author} {\bibfnamefont {R.}~\bibnamefont
  {Pellicer}}\ and\ \bibinfo {author} {\bibfnamefont {R.~J.}\ \bibnamefont
  {Torrence}},\ }\bibfield  {title} {\bibinfo {title} {{Nonlinear
  electrodynamics and general relativity}},\ }\href
  {https://doi.org/10.1063/1.1665019} {\bibfield  {journal} {\bibinfo
  {journal} {J. Math. Phys.}\ }\textbf {\bibinfo {volume} {10}},\ \bibinfo
  {pages} {1718} (\bibinfo {year} {1969})}\BibitemShut {NoStop}%
\bibitem [{\citenamefont {Bronnikov}(2001)}]{Bronnikov:2000vy}%
  \BibitemOpen
  \bibfield  {author} {\bibinfo {author} {\bibfnamefont {K.~A.}\ \bibnamefont
  {Bronnikov}},\ }\bibfield  {title} {\bibinfo {title} {{Regular magnetic black
  holes and monopoles from nonlinear electrodynamics}},\ }\href
  {https://doi.org/10.1103/PhysRevD.63.044005} {\bibfield  {journal} {\bibinfo
  {journal} {Phys. Rev. D}\ }\textbf {\bibinfo {volume} {63}},\ \bibinfo
  {pages} {044005} (\bibinfo {year} {2001})}\BibitemShut {NoStop}%
\bibitem [{\citenamefont {Bronnikov}(2000)}]{Bronnikov:2000yz}%
  \BibitemOpen
  \bibfield  {author} {\bibinfo {author} {\bibfnamefont {K.~A.}\ \bibnamefont
  {Bronnikov}},\ }\bibfield  {title} {\bibinfo {title} {{Comment on `Regular
  black hole in general relativity coupled to nonlinear electrodynamics'}},\
  }\href {https://doi.org/10.1103/PhysRevLett.85.4641} {\bibfield  {journal}
  {\bibinfo  {journal} {Phys. Rev. Lett.}\ }\textbf {\bibinfo {volume} {85}},\
  \bibinfo {pages} {4641} (\bibinfo {year} {2000})}\BibitemShut {NoStop}%
\bibitem [{\citenamefont {Ayon-Beato}\ and\ \citenamefont
  {Garcia}(1998)}]{Ayon-Beato:1998hmi}%
  \BibitemOpen
  \bibfield  {author} {\bibinfo {author} {\bibfnamefont {E.}~\bibnamefont
  {Ayon-Beato}}\ and\ \bibinfo {author} {\bibfnamefont {A.}~\bibnamefont
  {Garcia}},\ }\bibfield  {title} {\bibinfo {title} {{Regular black hole in
  general relativity coupled to nonlinear electrodynamics}},\ }\href
  {https://doi.org/10.1103/PhysRevLett.80.5056} {\bibfield  {journal} {\bibinfo
   {journal} {Phys. Rev. Lett.}\ }\textbf {\bibinfo {volume} {80}},\ \bibinfo
  {pages} {5056} (\bibinfo {year} {1998})}\BibitemShut {NoStop}%
\bibitem [{\citenamefont {Ayon-Beato}\ and\ \citenamefont
  {Garcia}(2000)}]{Ayon-Beato:2000mjt}%
  \BibitemOpen
  \bibfield  {author} {\bibinfo {author} {\bibfnamefont {E.}~\bibnamefont
  {Ayon-Beato}}\ and\ \bibinfo {author} {\bibfnamefont {A.}~\bibnamefont
  {Garcia}},\ }\bibfield  {title} {\bibinfo {title} {{The Bardeen model as a
  nonlinear magnetic monopole}},\ }\href
  {https://doi.org/10.1016/S0370-2693(00)01125-4} {\bibfield  {journal}
  {\bibinfo  {journal} {Phys. Lett. B}\ }\textbf {\bibinfo {volume} {493}},\
  \bibinfo {pages} {149} (\bibinfo {year} {2000})}\BibitemShut {NoStop}%
\bibitem [{\citenamefont {Born}\ and\ \citenamefont
  {Infeld}(1934)}]{Born:1934gh}%
  \BibitemOpen
  \bibfield  {author} {\bibinfo {author} {\bibfnamefont {M.}~\bibnamefont
  {Born}}\ and\ \bibinfo {author} {\bibfnamefont {L.}~\bibnamefont {Infeld}},\
  }\bibfield  {title} {\bibinfo {title} {{Foundations of the new field
  theory}},\ }\href {https://doi.org/10.1098/rspa.1934.0059} {\bibfield
  {journal} {\bibinfo  {journal} {Proc. Roy. Soc. Lond. A}\ }\textbf {\bibinfo
  {volume} {144}},\ \bibinfo {pages} {425} (\bibinfo {year}
  {1934})}\BibitemShut {NoStop}%
\bibitem [{\citenamefont {Heisenberg}\ and\ \citenamefont
  {Euler}(1936)}]{Heisenberg:1936nmg}%
  \BibitemOpen
  \bibfield  {author} {\bibinfo {author} {\bibfnamefont {W.}~\bibnamefont
  {Heisenberg}}\ and\ \bibinfo {author} {\bibfnamefont {H.}~\bibnamefont
  {Euler}},\ }\bibfield  {title} {\bibinfo {title} {{Consequences of Dirac's
  theory of positrons}},\ }\href {https://doi.org/10.1007/BF01343663}
  {\bibfield  {journal} {\bibinfo  {journal} {Z. Phys.}\ }\textbf {\bibinfo
  {volume} {98}},\ \bibinfo {pages} {714} (\bibinfo {year} {1936})}\BibitemShut
  {NoStop}%
\bibitem [{\citenamefont {Fan}\ and\ \citenamefont {Wang}(2016)}]{Fan:2016hvf}%
  \BibitemOpen
  \bibfield  {author} {\bibinfo {author} {\bibfnamefont {Z.-Y.}\ \bibnamefont
  {Fan}}\ and\ \bibinfo {author} {\bibfnamefont {X.}~\bibnamefont {Wang}},\
  }\bibfield  {title} {\bibinfo {title} {{Construction of Regular Black Holes
  in General Relativity}},\ }\href {https://doi.org/10.1103/PhysRevD.94.124027}
  {\bibfield  {journal} {\bibinfo  {journal} {Phys. Rev. D}\ }\textbf {\bibinfo
  {volume} {94}},\ \bibinfo {pages} {124027} (\bibinfo {year}
  {2016})}\BibitemShut {NoStop}%
\bibitem [{\citenamefont {Rodrigues}\ \emph {et~al.}(2016)\citenamefont
  {Rodrigues}, \citenamefont {Junior}, \citenamefont {Marques},\ and\
  \citenamefont {Zanchin}}]{Rodrigues:2015ayd}%
  \BibitemOpen
  \bibfield  {author} {\bibinfo {author} {\bibfnamefont {M.~E.}\ \bibnamefont
  {Rodrigues}}, \bibinfo {author} {\bibfnamefont {E.~L.~B.}\ \bibnamefont
  {Junior}}, \bibinfo {author} {\bibfnamefont {G.~T.}\ \bibnamefont
  {Marques}},\ and\ \bibinfo {author} {\bibfnamefont {V.~T.}\ \bibnamefont
  {Zanchin}},\ }\bibfield  {title} {\bibinfo {title} {{Regular black holes in
  $f(R)$ gravity coupled to nonlinear electrodynamics}},\ }\href
  {https://doi.org/10.1103/PhysRevD.94.024062} {\bibfield  {journal} {\bibinfo
  {journal} {Phys. Rev. D}\ }\textbf {\bibinfo {volume} {94}},\ \bibinfo
  {pages} {024062} (\bibinfo {year} {2016})},\ \bibinfo {note} {[Addendum:
  Phys.Rev.D 94, 049904 (2016)]}\BibitemShut {NoStop}%
\bibitem [{\citenamefont {Ayon-Beato}\ and\ \citenamefont
  {Garcia}(1999)}]{Ayon-Beato:1999kuh}%
  \BibitemOpen
  \bibfield  {author} {\bibinfo {author} {\bibfnamefont {E.}~\bibnamefont
  {Ayon-Beato}}\ and\ \bibinfo {author} {\bibfnamefont {A.}~\bibnamefont
  {Garcia}},\ }\bibfield  {title} {\bibinfo {title} {{New regular black hole
  solution from nonlinear electrodynamics}},\ }\href
  {https://doi.org/10.1016/S0370-2693(99)01038-2} {\bibfield  {journal}
  {\bibinfo  {journal} {Phys. Lett. B}\ }\textbf {\bibinfo {volume} {464}},\
  \bibinfo {pages} {25} (\bibinfo {year} {1999})}\BibitemShut {NoStop}%
\bibitem [{\citenamefont {Bronnikov}(2018)}]{Bronnikov:2017sgg}%
  \BibitemOpen
  \bibfield  {author} {\bibinfo {author} {\bibfnamefont {K.~A.}\ \bibnamefont
  {Bronnikov}},\ }\bibfield  {title} {\bibinfo {title} {{Nonlinear
  electrodynamics, regular black holes and wormholes}},\ }\href
  {https://doi.org/10.1142/S0218271818410055} {\bibfield  {journal} {\bibinfo
  {journal} {Int. J. Mod. Phys. D}\ }\textbf {\bibinfo {volume} {27}},\
  \bibinfo {pages} {1841005} (\bibinfo {year} {2018})}\BibitemShut {NoStop}%
\bibitem [{\citenamefont {Burinskii}\ and\ \citenamefont
  {Hildebrandt}(2002)}]{Burinskii:2002pz}%
  \BibitemOpen
  \bibfield  {author} {\bibinfo {author} {\bibfnamefont {A.}~\bibnamefont
  {Burinskii}}\ and\ \bibinfo {author} {\bibfnamefont {S.~R.}\ \bibnamefont
  {Hildebrandt}},\ }\bibfield  {title} {\bibinfo {title} {{New type of regular
  black holes and particle - like solutions from NED}},\ }\href
  {https://doi.org/10.1103/PhysRevD.65.104017} {\bibfield  {journal} {\bibinfo
  {journal} {Phys. Rev. D}\ }\textbf {\bibinfo {volume} {65}},\ \bibinfo
  {pages} {104017} (\bibinfo {year} {2002})}\BibitemShut {NoStop}%
\bibitem [{\citenamefont {Dymnikova}\ and\ \citenamefont
  {Galaktionov}(2015)}]{Dymnikova:2015hka}%
  \BibitemOpen
  \bibfield  {author} {\bibinfo {author} {\bibfnamefont {I.}~\bibnamefont
  {Dymnikova}}\ and\ \bibinfo {author} {\bibfnamefont {E.}~\bibnamefont
  {Galaktionov}},\ }\bibfield  {title} {\bibinfo {title} {{Regular rotating
  electrically charged black holes and solitons in non-linear electrodynamics
  minimally coupled to gravity}},\ }\href
  {https://doi.org/10.1088/0264-9381/32/16/165015} {\bibfield  {journal}
  {\bibinfo  {journal} {Class. Quant. Grav.}\ }\textbf {\bibinfo {volume}
  {32}},\ \bibinfo {pages} {165015} (\bibinfo {year} {2015})}\BibitemShut
  {NoStop}%
\bibitem [{\citenamefont {Bronnikov}(2017)}]{Bronnikov:2017tnz}%
  \BibitemOpen
  \bibfield  {author} {\bibinfo {author} {\bibfnamefont {K.~A.}\ \bibnamefont
  {Bronnikov}},\ }\bibfield  {title} {\bibinfo {title} {{Comment on
  \textquotedblleft{}Construction of regular black holes in general
  relativity\textquotedblright{}}},\ }\href
  {https://doi.org/10.1103/PhysRevD.96.128501} {\bibfield  {journal} {\bibinfo
  {journal} {Phys. Rev. D}\ }\textbf {\bibinfo {volume} {96}},\ \bibinfo
  {pages} {128501} (\bibinfo {year} {2017})}\BibitemShut {NoStop}%
\bibitem [{\citenamefont {Bokuli\'c}\ \emph
  {et~al.}(2022{\natexlab{a}})\citenamefont {Bokuli\'c}, \citenamefont
  {Smoli\'c},\ and\ \citenamefont {Juri\'c}}]{Bokulic:2022cyk}%
  \BibitemOpen
  \bibfield  {author} {\bibinfo {author} {\bibfnamefont {A.}~\bibnamefont
  {Bokuli\'c}}, \bibinfo {author} {\bibfnamefont {I.}~\bibnamefont
  {Smoli\'c}},\ and\ \bibinfo {author} {\bibfnamefont {T.}~\bibnamefont
  {Juri\'c}},\ }\bibfield  {title} {\bibinfo {title} {{Constraints on
  singularity resolution by nonlinear electrodynamics}},\ }\href
  {https://doi.org/10.1103/PhysRevD.106.064020} {\bibfield  {journal} {\bibinfo
   {journal} {Phys. Rev. D}\ }\textbf {\bibinfo {volume} {106}},\ \bibinfo
  {pages} {064020} (\bibinfo {year} {2022}{\natexlab{a}})}\BibitemShut
  {NoStop}%
\bibitem [{\citenamefont {Bokuli\'c}\ \emph
  {et~al.}(2022{\natexlab{b}})\citenamefont {Bokuli\'c}, \citenamefont
  {Juri\'c},\ and\ \citenamefont {Smoli\'c}}]{Bokulic:2021xom}%
  \BibitemOpen
  \bibfield  {author} {\bibinfo {author} {\bibfnamefont {A.}~\bibnamefont
  {Bokuli\'c}}, \bibinfo {author} {\bibfnamefont {T.}~\bibnamefont {Juri\'c}},\
  and\ \bibinfo {author} {\bibfnamefont {I.}~\bibnamefont {Smoli\'c}},\
  }\bibfield  {title} {\bibinfo {title} {{Nonlinear electromagnetic fields in
  strictly stationary spacetimes}},\ }\href
  {https://doi.org/10.1103/PhysRevD.105.024067} {\bibfield  {journal} {\bibinfo
   {journal} {Phys. Rev. D}\ }\textbf {\bibinfo {volume} {105}},\ \bibinfo
  {pages} {024067} (\bibinfo {year} {2022}{\natexlab{b}})}\BibitemShut
  {NoStop}%
\bibitem [{\citenamefont {Bronnikov}(2022)}]{Bronnikov:2022ofk}%
  \BibitemOpen
  \bibfield  {author} {\bibinfo {author} {\bibfnamefont {K.~A.}\ \bibnamefont
  {Bronnikov}},\ }\bibfield  {title} {\bibinfo {title} {{Regular black holes
  sourced by nonlinear electrodynamics}},\ }\href@noop {} {\  (\bibinfo {year}
  {2022})}\BibitemShut {NoStop}%
\bibitem [{\citenamefont {Sebastiani}\ and\ \citenamefont
  {Zerbini}(2022)}]{Sebastiani:2022wbz}%
  \BibitemOpen
  \bibfield  {author} {\bibinfo {author} {\bibfnamefont {L.}~\bibnamefont
  {Sebastiani}}\ and\ \bibinfo {author} {\bibfnamefont {S.}~\bibnamefont
  {Zerbini}},\ }\bibfield  {title} {\bibinfo {title} {{Some Remarks on
  Non-Singular Spherically Symmetric Space-Times}},\ }\href
  {https://doi.org/10.3390/astronomy1020010} {\bibfield  {journal} {\bibinfo
  {journal} {Astronomy}\ }\textbf {\bibinfo {volume} {1}},\ \bibinfo {pages}
  {99} (\bibinfo {year} {2022})}\BibitemShut {NoStop}%
\bibitem [{\citenamefont {Carballo-Rubio}\ \emph
  {et~al.}(2023{\natexlab{a}})\citenamefont {Carballo-Rubio}, \citenamefont
  {Di~Filippo}, \citenamefont {Liberati},\ and\ \citenamefont
  {Visser}}]{Carballo-Rubio:2023mvr}%
  \BibitemOpen
  \bibfield  {author} {\bibinfo {author} {\bibfnamefont {R.}~\bibnamefont
  {Carballo-Rubio}}, \bibinfo {author} {\bibfnamefont {F.}~\bibnamefont
  {Di~Filippo}}, \bibinfo {author} {\bibfnamefont {S.}~\bibnamefont
  {Liberati}},\ and\ \bibinfo {author} {\bibfnamefont {M.}~\bibnamefont
  {Visser}},\ }\bibfield  {title} {\bibinfo {title} {{Singularity-free
  gravitational collapse: From regular black holes to horizonless objects}},\
  }\href@noop {} {\  (\bibinfo {year} {2023}{\natexlab{a}})}\BibitemShut
  {NoStop}%
\bibitem [{\citenamefont {Sorokin}(2022)}]{Sorokin:2021tge}%
  \BibitemOpen
  \bibfield  {author} {\bibinfo {author} {\bibfnamefont {D.~P.}\ \bibnamefont
  {Sorokin}},\ }\bibfield  {title} {\bibinfo {title} {{Introductory Notes on
  Non-linear Electrodynamics and its Applications}},\ }\href
  {https://doi.org/10.1002/prop.202200092} {\bibfield  {journal} {\bibinfo
  {journal} {Fortsch. Phys.}\ }\textbf {\bibinfo {volume} {70}},\ \bibinfo
  {pages} {2200092} (\bibinfo {year} {2022})}\BibitemShut {NoStop}%
\bibitem [{\citenamefont {Bambi}(2023)}]{Bambi:2023try}%
  \BibitemOpen
  \bibfield  {author} {\bibinfo {author} {\bibfnamefont {C.}~\bibnamefont
  {Bambi}},\ }\href {https://doi.org/10.1007/978-981-99-1596-5} {\emph
  {\bibinfo {title} {{Regular Black Holes}}}},\ Springer Series in Astrophysics
  and Cosmology\ (\bibinfo  {publisher} {Springer Singapore},\ \bibinfo {year}
  {2023})\ \Eprint {https://arxiv.org/abs/2307.13249} {arXiv:2307.13249
  [gr-qc]} \BibitemShut {NoStop}%
\bibitem [{\citenamefont {Kim}\ and\ \citenamefont {Kim}(2023)}]{Kim:2022fkt}%
  \BibitemOpen
  \bibfield  {author} {\bibinfo {author} {\bibfnamefont {C.~M.}\ \bibnamefont
  {Kim}}\ and\ \bibinfo {author} {\bibfnamefont {S.~P.}\ \bibnamefont {Kim}},\
  }\bibfield  {title} {\bibinfo {title} {{Vacuum birefringence at one-loop in a
  supercritical magnetic field superposed with a weak electric field and
  application to pulsar magnetosphere}},\ }\href
  {https://doi.org/10.1140/epjc/s10052-023-11243-1} {\bibfield  {journal}
  {\bibinfo  {journal} {Eur. Phys. J. C}\ }\textbf {\bibinfo {volume} {83}},\
  \bibinfo {pages} {104} (\bibinfo {year} {2023})}\BibitemShut {NoStop}%
\bibitem [{\citenamefont {Nathanail}\ \emph {et~al.}(2017)\citenamefont
  {Nathanail}, \citenamefont {Most},\ and\ \citenamefont
  {Rezzolla}}]{Nathanail:2017wly}%
  \BibitemOpen
  \bibfield  {author} {\bibinfo {author} {\bibfnamefont {A.}~\bibnamefont
  {Nathanail}}, \bibinfo {author} {\bibfnamefont {E.~R.}\ \bibnamefont
  {Most}},\ and\ \bibinfo {author} {\bibfnamefont {L.}~\bibnamefont
  {Rezzolla}},\ }\bibfield  {title} {\bibinfo {title} {{Gravitational collapse
  to a Kerr\textendash{}Newman black hole}},\ }\href
  {https://doi.org/10.1093/mnrasl/slx035} {\bibfield  {journal} {\bibinfo
  {journal} {Mon. Not. Roy. Astron. Soc.}\ }\textbf {\bibinfo {volume} {469}},\
  \bibinfo {pages} {L31} (\bibinfo {year} {2017})}\BibitemShut {NoStop}%
\bibitem [{\citenamefont {Ghosh}(2015)}]{Ghosh:2014pba}%
  \BibitemOpen
  \bibfield  {author} {\bibinfo {author} {\bibfnamefont {S.~G.}\ \bibnamefont
  {Ghosh}},\ }\bibfield  {title} {\bibinfo {title} {{A nonsingular rotating
  black hole}},\ }\href {https://doi.org/10.1140/epjc/s10052-015-3740-y}
  {\bibfield  {journal} {\bibinfo  {journal} {Eur. Phys. J. C}\ }\textbf
  {\bibinfo {volume} {75}},\ \bibinfo {pages} {532} (\bibinfo {year}
  {2015})}\BibitemShut {NoStop}%
\bibitem [{\citenamefont {Culetu}(2015)}]{Culetu:2014lca}%
  \BibitemOpen
  \bibfield  {author} {\bibinfo {author} {\bibfnamefont {H.}~\bibnamefont
  {Culetu}},\ }\bibfield  {title} {\bibinfo {title} {{On a regular charged
  black hole with a nonlinear electric source}},\ }\href
  {https://doi.org/10.1007/s10773-015-2521-6} {\bibfield  {journal} {\bibinfo
  {journal} {Int. J. Theor. Phys.}\ }\textbf {\bibinfo {volume} {54}},\
  \bibinfo {pages} {2855} (\bibinfo {year} {2015})}\BibitemShut {NoStop}%
\bibitem [{\citenamefont {Ruffini}\ \emph {et~al.}(2010)\citenamefont
  {Ruffini}, \citenamefont {Vereshchagin},\ and\ \citenamefont
  {Xue}}]{Ruffini:2009hg}%
  \BibitemOpen
  \bibfield  {author} {\bibinfo {author} {\bibfnamefont {R.}~\bibnamefont
  {Ruffini}}, \bibinfo {author} {\bibfnamefont {G.}~\bibnamefont
  {Vereshchagin}},\ and\ \bibinfo {author} {\bibfnamefont {S.-S.}\ \bibnamefont
  {Xue}},\ }\bibfield  {title} {\bibinfo {title} {{Electron-positron pairs in
  physics and astrophysics: from heavy nuclei to black holes}},\ }\href
  {https://doi.org/10.1016/j.physrep.2009.10.004} {\bibfield  {journal}
  {\bibinfo  {journal} {Phys. Rept.}\ }\textbf {\bibinfo {volume} {487}},\
  \bibinfo {pages} {1} (\bibinfo {year} {2010})}\BibitemShut {NoStop}%
\bibitem [{ELI()}]{ELI}%
  \BibitemOpen
  \href@noop {} {}\bibinfo {howpublished}
  {\url{https://eli-laser.eu/}}\BibitemShut {NoStop}%
\bibitem [{XFE()}]{XFEL}%
  \BibitemOpen
  \href@noop {} {}\bibinfo {howpublished}
  {\url{http://xfel.desy.de/}}\BibitemShut {NoStop}%
\bibitem [{LUX()}]{LUXE}%
  \BibitemOpen
  \href@noop {} {}\bibinfo {howpublished}
  {\url{https://luxe.desy.de/}}\BibitemShut {NoStop}%
\bibitem [{\citenamefont {Aaboud}\ \emph {et~al.}(2017)\citenamefont {Aaboud}
  \emph {et~al.}}]{ATLAS:2017fur}%
  \BibitemOpen
  \bibfield  {author} {\bibinfo {author} {\bibfnamefont {M.}~\bibnamefont
  {Aaboud}} \emph {et~al.} (\bibinfo {collaboration} {ATLAS}),\ }\bibfield
  {title} {\bibinfo {title} {{Evidence for light-by-light scattering in
  heavy-ion collisions with the ATLAS detector at the LHC}},\ }\href
  {https://doi.org/10.1038/nphys4208} {\bibfield  {journal} {\bibinfo
  {journal} {Nature Phys.}\ }\textbf {\bibinfo {volume} {13}},\ \bibinfo
  {pages} {852} (\bibinfo {year} {2017})}\BibitemShut {NoStop}%
\bibitem [{\citenamefont {Aad}\ \emph {et~al.}(2019)\citenamefont {Aad} \emph
  {et~al.}}]{ATLAS:2019azn}%
  \BibitemOpen
  \bibfield  {author} {\bibinfo {author} {\bibfnamefont {G.}~\bibnamefont
  {Aad}} \emph {et~al.} (\bibinfo {collaboration} {ATLAS}),\ }\bibfield
  {title} {\bibinfo {title} {{Observation of light-by-light scattering in
  ultraperipheral Pb+Pb collisions with the ATLAS detector}},\ }\href
  {https://doi.org/10.1103/PhysRevLett.123.052001} {\bibfield  {journal}
  {\bibinfo  {journal} {Phys. Rev. Lett.}\ }\textbf {\bibinfo {volume} {123}},\
  \bibinfo {pages} {052001} (\bibinfo {year} {2019})}\BibitemShut {NoStop}%
\bibitem [{\citenamefont {Ejlli}\ \emph {et~al.}(2020)\citenamefont {Ejlli},
  \citenamefont {Della~Valle}, \citenamefont {Gastaldi}, \citenamefont
  {Messineo}, \citenamefont {Pengo}, \citenamefont {Ruoso},\ and\ \citenamefont
  {Zavattini}}]{Ejlli:2020yhk}%
  \BibitemOpen
  \bibfield  {author} {\bibinfo {author} {\bibfnamefont {A.}~\bibnamefont
  {Ejlli}}, \bibinfo {author} {\bibfnamefont {F.}~\bibnamefont {Della~Valle}},
  \bibinfo {author} {\bibfnamefont {U.}~\bibnamefont {Gastaldi}}, \bibinfo
  {author} {\bibfnamefont {G.}~\bibnamefont {Messineo}}, \bibinfo {author}
  {\bibfnamefont {R.}~\bibnamefont {Pengo}}, \bibinfo {author} {\bibfnamefont
  {G.}~\bibnamefont {Ruoso}},\ and\ \bibinfo {author} {\bibfnamefont
  {G.}~\bibnamefont {Zavattini}},\ }\bibfield  {title} {\bibinfo {title} {{The
  PVLAS experiment: A 25 year effort to measure vacuum magnetic
  birefringence}},\ }\href {https://doi.org/10.1016/j.physrep.2020.06.001}
  {\bibfield  {journal} {\bibinfo  {journal} {Phys. Rept.}\ }\textbf {\bibinfo
  {volume} {871}},\ \bibinfo {pages} {1} (\bibinfo {year} {2020})}\BibitemShut
  {NoStop}%
\bibitem [{IXP()}]{IXPE}%
  \BibitemOpen
  \href@noop {} {}\bibinfo {howpublished}
  {\url{https://ixpe.msfc.nasa.gov/}}\BibitemShut {NoStop}%
\bibitem [{\citenamefont {Taverna}\ \emph {et~al.}(2022)\citenamefont {Taverna}
  \emph {et~al.}}]{Taverna:2022jgl}%
  \BibitemOpen
  \bibfield  {author} {\bibinfo {author} {\bibfnamefont {R.}~\bibnamefont
  {Taverna}} \emph {et~al.},\ }\bibfield  {title} {\bibinfo {title} {{Polarized
  x-rays from a magnetar}}\ }\href {https://doi.org/10.1126/science.add0080}
  {10.1126/science.add0080} (\bibinfo {year} {2022})\BibitemShut {NoStop}%
\bibitem [{\citenamefont {Mignani}\ \emph {et~al.}(2017)\citenamefont
  {Mignani}, \citenamefont {Testa}, \citenamefont {Caniulef}, \citenamefont
  {Taverna}, \citenamefont {Turolla}, \citenamefont {Zane},\ and\ \citenamefont
  {Wu}}]{Mignani:2016fwz}%
  \BibitemOpen
  \bibfield  {author} {\bibinfo {author} {\bibfnamefont {R.~P.}\ \bibnamefont
  {Mignani}}, \bibinfo {author} {\bibfnamefont {V.}~\bibnamefont {Testa}},
  \bibinfo {author} {\bibfnamefont {D.~G.}\ \bibnamefont {Caniulef}}, \bibinfo
  {author} {\bibfnamefont {R.}~\bibnamefont {Taverna}}, \bibinfo {author}
  {\bibfnamefont {R.}~\bibnamefont {Turolla}}, \bibinfo {author} {\bibfnamefont
  {S.}~\bibnamefont {Zane}},\ and\ \bibinfo {author} {\bibfnamefont
  {K.}~\bibnamefont {Wu}},\ }\bibfield  {title} {\bibinfo {title} {{Evidence
  for vacuum birefringence from the first optical-polarimetry measurement of
  the isolated neutron star RX J1856.5\ensuremath{-}3754}},\ }\href
  {https://doi.org/10.1093/mnras/stw2798} {\bibfield  {journal} {\bibinfo
  {journal} {Mon. Not. Roy. Astron. Soc.}\ }\textbf {\bibinfo {volume} {465}},\
  \bibinfo {pages} {492} (\bibinfo {year} {2017})}\BibitemShut {NoStop}%
\bibitem [{\citenamefont {Kumar~Walia}\ \emph {et~al.}(2022)\citenamefont
  {Kumar~Walia}, \citenamefont {Ghosh},\ and\ \citenamefont
  {Maharaj}}]{KumarWalia:2022aop}%
  \BibitemOpen
  \bibfield  {author} {\bibinfo {author} {\bibfnamefont {R.}~\bibnamefont
  {Kumar~Walia}}, \bibinfo {author} {\bibfnamefont {S.~G.}\ \bibnamefont
  {Ghosh}},\ and\ \bibinfo {author} {\bibfnamefont {S.~D.}\ \bibnamefont
  {Maharaj}},\ }\bibfield  {title} {\bibinfo {title} {{Testing Rotating Regular
  Metrics with EHT Results of Sgr A*}},\ }\href
  {https://doi.org/10.3847/1538-4357/ac9623} {\bibfield  {journal} {\bibinfo
  {journal} {Astrophys. J.}\ }\textbf {\bibinfo {volume} {939}},\ \bibinfo
  {pages} {77} (\bibinfo {year} {2022})}\BibitemShut {NoStop}%
\bibitem [{\citenamefont {Kumar}\ \emph {et~al.}(2020)\citenamefont {Kumar},
  \citenamefont {Kumar},\ and\ \citenamefont {Ghosh}}]{Kumar:2020yem}%
  \BibitemOpen
  \bibfield  {author} {\bibinfo {author} {\bibfnamefont {R.}~\bibnamefont
  {Kumar}}, \bibinfo {author} {\bibfnamefont {A.}~\bibnamefont {Kumar}},\ and\
  \bibinfo {author} {\bibfnamefont {S.~G.}\ \bibnamefont {Ghosh}},\ }\bibfield
  {title} {\bibinfo {title} {{Testing Rotating Regular Metrics as Candidates
  for Astrophysical Black Holes}},\ }\href
  {https://doi.org/10.3847/1538-4357/ab8c4a} {\bibfield  {journal} {\bibinfo
  {journal} {Astrophys. J.}\ }\textbf {\bibinfo {volume} {896}},\ \bibinfo
  {pages} {89} (\bibinfo {year} {2020})}\BibitemShut {NoStop}%
\bibitem [{\citenamefont {Vagnozzi}\ \emph {et~al.}(2023)\citenamefont
  {Vagnozzi} \emph {et~al.}}]{Vagnozzi:2022moj}%
  \BibitemOpen
  \bibfield  {author} {\bibinfo {author} {\bibfnamefont {S.}~\bibnamefont
  {Vagnozzi}} \emph {et~al.},\ }\bibfield  {title} {\bibinfo {title}
  {{Horizon-scale tests of gravity theories and fundamental physics from the
  Event Horizon Telescope image of Sagittarius A}},\ }\href
  {https://doi.org/10.1088/1361-6382/acd97b} {\bibfield  {journal} {\bibinfo
  {journal} {Class. Quant. Grav.}\ }\textbf {\bibinfo {volume} {40}},\ \bibinfo
  {pages} {165007} (\bibinfo {year} {2023})}\BibitemShut {NoStop}%
\bibitem [{\citenamefont {Ghosh}\ \emph {et~al.}(2021)\citenamefont {Ghosh},
  \citenamefont {Kumar},\ and\ \citenamefont {Islam}}]{Ghosh:2020spb}%
  \BibitemOpen
  \bibfield  {author} {\bibinfo {author} {\bibfnamefont {S.~G.}\ \bibnamefont
  {Ghosh}}, \bibinfo {author} {\bibfnamefont {R.}~\bibnamefont {Kumar}},\ and\
  \bibinfo {author} {\bibfnamefont {S.~U.}\ \bibnamefont {Islam}},\ }\bibfield
  {title} {\bibinfo {title} {{Parameters estimation and strong gravitational
  lensing of nonsingular Kerr-Sen black holes}},\ }\href
  {https://doi.org/10.1088/1475-7516/2021/03/056} {\bibfield  {journal}
  {\bibinfo  {journal} {JCAP}\ }\textbf {\bibinfo {volume} {03}},\ \bibinfo
  {pages} {056}}\BibitemShut {NoStop}%
\bibitem [{\citenamefont {Kumar}\ and\ \citenamefont
  {Ghosh}(2020)}]{Kumar:2018ple}%
  \BibitemOpen
  \bibfield  {author} {\bibinfo {author} {\bibfnamefont {R.}~\bibnamefont
  {Kumar}}\ and\ \bibinfo {author} {\bibfnamefont {S.~G.}\ \bibnamefont
  {Ghosh}},\ }\bibfield  {title} {\bibinfo {title} {{Black Hole Parameter
  Estimation from Its Shadow}},\ }\href
  {https://doi.org/10.3847/1538-4357/ab77b0} {\bibfield  {journal} {\bibinfo
  {journal} {Astrophys. J.}\ }\textbf {\bibinfo {volume} {892}},\ \bibinfo
  {pages} {78} (\bibinfo {year} {2020})}\BibitemShut {NoStop}%
\bibitem [{\citenamefont {Kar}\ and\ \citenamefont {Kar}(2024)}]{Kar:2023dko}%
  \BibitemOpen
  \bibfield  {author} {\bibinfo {author} {\bibfnamefont {A.}~\bibnamefont
  {Kar}}\ and\ \bibinfo {author} {\bibfnamefont {S.}~\bibnamefont {Kar}},\
  }\bibfield  {title} {\bibinfo {title} {{Novel regular black holes: geometry,
  source and shadow}},\ }\href {https://doi.org/10.1007/s10714-024-03238-4}
  {\bibfield  {journal} {\bibinfo  {journal} {Gen. Rel. Grav.}\ }\textbf
  {\bibinfo {volume} {56}},\ \bibinfo {pages} {52} (\bibinfo {year}
  {2024})}\BibitemShut {NoStop}%
\bibitem [{\citenamefont {Uniyal}\ \emph {et~al.}(2023)\citenamefont {Uniyal},
  \citenamefont {Pantig},\ and\ \citenamefont {\"Ovg\"un}}]{Uniyal:2022vdu}%
  \BibitemOpen
  \bibfield  {author} {\bibinfo {author} {\bibfnamefont {A.}~\bibnamefont
  {Uniyal}}, \bibinfo {author} {\bibfnamefont {R.~C.}\ \bibnamefont {Pantig}},\
  and\ \bibinfo {author} {\bibfnamefont {A.}~\bibnamefont {\"Ovg\"un}},\
  }\bibfield  {title} {\bibinfo {title} {{Probing a non-linear electrodynamics
  black hole with thin accretion disk, shadow, and deflection angle with M87*
  and Sgr A* from EHT}},\ }\href {https://doi.org/10.1016/j.dark.2023.101178}
  {\bibfield  {journal} {\bibinfo  {journal} {Phys. Dark Univ.}\ }\textbf
  {\bibinfo {volume} {40}},\ \bibinfo {pages} {101178} (\bibinfo {year}
  {2023})}\BibitemShut {NoStop}%
\bibitem [{\citenamefont {Bambi}(2014)}]{Bambi:2014nta}%
  \BibitemOpen
  \bibfield  {author} {\bibinfo {author} {\bibfnamefont {C.}~\bibnamefont
  {Bambi}},\ }\bibfield  {title} {\bibinfo {title} {{Testing the Bardeen metric
  with the black hole candidate in Cygnus X-1}},\ }\href
  {https://doi.org/10.1016/j.physletb.2014.01.037} {\bibfield  {journal}
  {\bibinfo  {journal} {Phys. Lett. B}\ }\textbf {\bibinfo {volume} {730}},\
  \bibinfo {pages} {59} (\bibinfo {year} {2014})}\BibitemShut {NoStop}%
\bibitem [{\citenamefont {Schee}\ and\ \citenamefont
  {Stuchl\'ik}(2019)}]{Schee:2019}%
  \BibitemOpen
  \bibfield  {author} {\bibinfo {author} {\bibfnamefont {J.}~\bibnamefont
  {Schee}}\ and\ \bibinfo {author} {\bibfnamefont {Z.}~\bibnamefont
  {Stuchl\'ik}},\ }\bibfield  {title} {\bibinfo {title} {{Effective Geometry of
  the Bardeen Spacetimes: Gravitational Lensing and Frequency Mapping of
  Keplerian Disks}},\ }\href {https://doi.org/10.3847/1538-4357/ab04f3}
  {\bibfield  {journal} {\bibinfo  {journal} {Astrophys. J.}\ }\textbf
  {\bibinfo {volume} {874}},\ \bibinfo {pages} {12} (\bibinfo {year}
  {2019})}\BibitemShut {NoStop}%
\bibitem [{\citenamefont {Schee}\ and\ \citenamefont
  {Stuchlik}(2019)}]{Schee:2019gki}%
  \BibitemOpen
  \bibfield  {author} {\bibinfo {author} {\bibfnamefont {J.}~\bibnamefont
  {Schee}}\ and\ \bibinfo {author} {\bibfnamefont {Z.}~\bibnamefont
  {Stuchlik}},\ }\bibfield  {title} {\bibinfo {title} {{Profiled spectral lines
  of Keplerian rings orbiting in the regular Bardeen black hole spacetimes}},\
  }\href {https://doi.org/10.1140/epjc/s10052-019-7420-1} {\bibfield  {journal}
  {\bibinfo  {journal} {Eur. Phys. J. C}\ }\textbf {\bibinfo {volume} {79}},\
  \bibinfo {pages} {988} (\bibinfo {year} {2019})}\BibitemShut {NoStop}%
\bibitem [{\citenamefont {Rayimbaev}\ \emph {et~al.}(2020)\citenamefont
  {Rayimbaev}, \citenamefont {Figueroa}, \citenamefont {Stuchl\'\i{}k},\ and\
  \citenamefont {Juraev}}]{Rayimbaev:2020hjs}%
  \BibitemOpen
  \bibfield  {author} {\bibinfo {author} {\bibfnamefont {J.}~\bibnamefont
  {Rayimbaev}}, \bibinfo {author} {\bibfnamefont {M.}~\bibnamefont {Figueroa}},
  \bibinfo {author} {\bibfnamefont {Z.}~\bibnamefont {Stuchl\'\i{}k}},\ and\
  \bibinfo {author} {\bibfnamefont {B.}~\bibnamefont {Juraev}},\ }\bibfield
  {title} {\bibinfo {title} {{Test particle orbits around regular black holes
  in general relativity combined with nonlinear electrodynamics}},\ }\href
  {https://doi.org/10.1103/PhysRevD.101.104045} {\bibfield  {journal} {\bibinfo
   {journal} {Phys. Rev. D}\ }\textbf {\bibinfo {volume} {101}},\ \bibinfo
  {pages} {104045} (\bibinfo {year} {2020})}\BibitemShut {NoStop}%
\bibitem [{\citenamefont {Toshmatov}\ \emph {et~al.}(2021)\citenamefont
  {Toshmatov}, \citenamefont {Ahmedov},\ and\ \citenamefont
  {Malafarina}}]{Toshmatov:2021fgm}%
  \BibitemOpen
  \bibfield  {author} {\bibinfo {author} {\bibfnamefont {B.}~\bibnamefont
  {Toshmatov}}, \bibinfo {author} {\bibfnamefont {B.}~\bibnamefont {Ahmedov}},\
  and\ \bibinfo {author} {\bibfnamefont {D.}~\bibnamefont {Malafarina}},\
  }\bibfield  {title} {\bibinfo {title} {{Can a light ray distinguish charge of
  a black hole in nonlinear electrodynamics?}},\ }\href
  {https://doi.org/10.1103/PhysRevD.103.024026} {\bibfield  {journal} {\bibinfo
   {journal} {Phys. Rev. D}\ }\textbf {\bibinfo {volume} {103}},\ \bibinfo
  {pages} {024026} (\bibinfo {year} {2021})}\BibitemShut {NoStop}%
\bibitem [{\citenamefont {de~Paula}\ \emph {et~al.}(2023)\citenamefont
  {de~Paula}, \citenamefont {Lima~Junior}, \citenamefont {Cunha},\ and\
  \citenamefont {Crispino}}]{dePaula:2023ozi}%
  \BibitemOpen
  \bibfield  {author} {\bibinfo {author} {\bibfnamefont {M.~A.~A.}\
  \bibnamefont {de~Paula}}, \bibinfo {author} {\bibfnamefont {H.~C.~D.}\
  \bibnamefont {Lima~Junior}}, \bibinfo {author} {\bibfnamefont {P.~V.~P.}\
  \bibnamefont {Cunha}},\ and\ \bibinfo {author} {\bibfnamefont {L.~C.~B.}\
  \bibnamefont {Crispino}},\ }\bibfield  {title} {\bibinfo {title}
  {{Electrically charged regular black holes in nonlinear electrodynamics:
  Light rings, shadows, and gravitational lensing}},\ }\href
  {https://doi.org/10.1103/PhysRevD.108.084029} {\bibfield  {journal} {\bibinfo
   {journal} {Phys. Rev. D}\ }\textbf {\bibinfo {volume} {108}},\ \bibinfo
  {pages} {084029} (\bibinfo {year} {2023})}\BibitemShut {NoStop}%
\bibitem [{\citenamefont {Stuchl\'\i{}k}\ and\ \citenamefont
  {Schee}(2019)}]{Stuchlik:2019uvf}%
  \BibitemOpen
  \bibfield  {author} {\bibinfo {author} {\bibfnamefont {Z.}~\bibnamefont
  {Stuchl\'\i{}k}}\ and\ \bibinfo {author} {\bibfnamefont {J.}~\bibnamefont
  {Schee}},\ }\bibfield  {title} {\bibinfo {title} {{Shadow of the regular
  Bardeen black holes and comparison of the motion of photons and neutrinos}},\
  }\href {https://doi.org/10.1140/epjc/s10052-019-6543-8} {\bibfield  {journal}
  {\bibinfo  {journal} {Eur. Phys. J. C}\ }\textbf {\bibinfo {volume} {79}},\
  \bibinfo {pages} {44} (\bibinfo {year} {2019})}\BibitemShut {NoStop}%
\bibitem [{\citenamefont {Akiyama}\ \emph
  {et~al.}(2022{\natexlab{a}})\citenamefont {Akiyama} \emph
  {et~al.}}]{EventHorizonTelescope:2022wkp}%
  \BibitemOpen
  \bibfield  {author} {\bibinfo {author} {\bibfnamefont {K.}~\bibnamefont
  {Akiyama}} \emph {et~al.} (\bibinfo {collaboration} {Event Horizon
  Telescope}),\ }\bibfield  {title} {\bibinfo {title} {{First Sagittarius A*
  Event Horizon Telescope Results. I. The Shadow of the Supermassive Black Hole
  in the Center of the Milky Way}},\ }\href
  {https://doi.org/10.3847/2041-8213/ac6674} {\bibfield  {journal} {\bibinfo
  {journal} {Astrophys. J. Lett.}\ }\textbf {\bibinfo {volume} {930}},\
  \bibinfo {pages} {L12} (\bibinfo {year} {2022}{\natexlab{a}})}\BibitemShut
  {NoStop}%
\bibitem [{\citenamefont {Akiyama}\ \emph
  {et~al.}(2022{\natexlab{b}})\citenamefont {Akiyama} \emph
  {et~al.}}]{EventHorizonTelescope:2022urf}%
  \BibitemOpen
  \bibfield  {author} {\bibinfo {author} {\bibfnamefont {K.}~\bibnamefont
  {Akiyama}} \emph {et~al.} (\bibinfo {collaboration} {Event Horizon
  Telescope}),\ }\bibfield  {title} {\bibinfo {title} {{First Sagittarius A*
  Event Horizon Telescope Results. V. Testing Astrophysical Models of the
  Galactic Center Black Hole}},\ }\href
  {https://doi.org/10.3847/2041-8213/ac6672} {\bibfield  {journal} {\bibinfo
  {journal} {Astrophys. J. Lett.}\ }\textbf {\bibinfo {volume} {930}},\
  \bibinfo {pages} {L16} (\bibinfo {year} {2022}{\natexlab{b}})}\BibitemShut
  {NoStop}%
\bibitem [{\citenamefont {Akiyama}\ \emph
  {et~al.}(2022{\natexlab{c}})\citenamefont {Akiyama} \emph
  {et~al.}}]{EventHorizonTelescope:2022xqj}%
  \BibitemOpen
  \bibfield  {author} {\bibinfo {author} {\bibfnamefont {K.}~\bibnamefont
  {Akiyama}} \emph {et~al.} (\bibinfo {collaboration} {Event Horizon
  Telescope}),\ }\bibfield  {title} {\bibinfo {title} {{First Sagittarius A*
  Event Horizon Telescope Results. VI. Testing the Black Hole Metric}},\ }\href
  {https://doi.org/10.3847/2041-8213/ac6756} {\bibfield  {journal} {\bibinfo
  {journal} {Astrophys. J. Lett.}\ }\textbf {\bibinfo {volume} {930}},\
  \bibinfo {pages} {L17} (\bibinfo {year} {2022}{\natexlab{c}})}\BibitemShut
  {NoStop}%
\bibitem [{\citenamefont {Akiyama}\ \emph {et~al.}(2024)\citenamefont {Akiyama}
  \emph {et~al.}}]{Collaboration:2024unf}%
  \BibitemOpen
  \bibfield  {author} {\bibinfo {author} {\bibfnamefont {K.}~\bibnamefont
  {Akiyama}} \emph {et~al.} (\bibinfo {collaboration} {Event Horizon
  Telescope}),\ }\bibfield  {title} {\bibinfo {title} {{First Sagittarius A*
  Event Horizon Telescope Results. VII. Polarization of the Ring}},\ }\href
  {https://doi.org/10.3847/2041-8213/ad2df0} {\bibfield  {journal} {\bibinfo
  {journal} {Astrophys. J. Lett.}\ }\textbf {\bibinfo {volume} {964}},\
  \bibinfo {pages} {L25} (\bibinfo {year} {2024})}\BibitemShut {NoStop}%
\bibitem [{\citenamefont {Akiyama}\ \emph
  {et~al.}(2019{\natexlab{a}})\citenamefont {Akiyama} \emph
  {et~al.}}]{EventHorizonTelescope:2019dse}%
  \BibitemOpen
  \bibfield  {author} {\bibinfo {author} {\bibfnamefont {K.}~\bibnamefont
  {Akiyama}} \emph {et~al.} (\bibinfo {collaboration} {Event Horizon
  Telescope}),\ }\bibfield  {title} {\bibinfo {title} {{First M87 Event Horizon
  Telescope Results. I. The Shadow of the Supermassive Black Hole}},\ }\href
  {https://doi.org/10.3847/2041-8213/ab0ec7} {\bibfield  {journal} {\bibinfo
  {journal} {Astrophys. J. Lett.}\ }\textbf {\bibinfo {volume} {875}},\
  \bibinfo {pages} {L1} (\bibinfo {year} {2019}{\natexlab{a}})}\BibitemShut
  {NoStop}%
\bibitem [{\citenamefont {Akiyama}\ \emph
  {et~al.}(2019{\natexlab{b}})\citenamefont {Akiyama} \emph
  {et~al.}}]{EventHorizonTelescope:2019pgp}%
  \BibitemOpen
  \bibfield  {author} {\bibinfo {author} {\bibfnamefont {K.}~\bibnamefont
  {Akiyama}} \emph {et~al.} (\bibinfo {collaboration} {Event Horizon
  Telescope}),\ }\bibfield  {title} {\bibinfo {title} {{First M87 Event Horizon
  Telescope Results. V. Physical Origin of the Asymmetric Ring}},\ }\href
  {https://doi.org/10.3847/2041-8213/ab0f43} {\bibfield  {journal} {\bibinfo
  {journal} {Astrophys. J. Lett.}\ }\textbf {\bibinfo {volume} {875}},\
  \bibinfo {pages} {L5} (\bibinfo {year} {2019}{\natexlab{b}})}\BibitemShut
  {NoStop}%
\bibitem [{\citenamefont {Akiyama}\ \emph {et~al.}(2023)\citenamefont {Akiyama}
  \emph {et~al.}}]{EHT:2023ujh}%
  \BibitemOpen
  \bibfield  {author} {\bibinfo {author} {\bibfnamefont {K.}~\bibnamefont
  {Akiyama}} \emph {et~al.} (\bibinfo {collaboration} {EHT}),\ }\bibfield
  {title} {\bibinfo {title} {{First M87 Event Horizon Telescope Results. IX.
  Detection of Near-horizon Circular Polarization}},\ }\href
  {https://doi.org/10.3847/2041-8213/acff70} {\bibfield  {journal} {\bibinfo
  {journal} {Astrophys. J. Lett.}\ }\textbf {\bibinfo {volume} {957}},\
  \bibinfo {pages} {L20} (\bibinfo {year} {2023})}\BibitemShut {NoStop}%
\bibitem [{\citenamefont {Kocherlakota}\ \emph {et~al.}(2021)\citenamefont
  {Kocherlakota} \emph {et~al.}}]{EventHorizonTelescope:2021dqv}%
  \BibitemOpen
  \bibfield  {author} {\bibinfo {author} {\bibfnamefont {P.}~\bibnamefont
  {Kocherlakota}} \emph {et~al.} (\bibinfo {collaboration} {Event Horizon
  Telescope}),\ }\bibfield  {title} {\bibinfo {title} {{Constraints on
  black-hole charges with the 2017 EHT observations of M87*}},\ }\href
  {https://doi.org/10.1103/PhysRevD.103.104047} {\bibfield  {journal} {\bibinfo
   {journal} {Phys. Rev. D}\ }\textbf {\bibinfo {volume} {103}},\ \bibinfo
  {pages} {104047} (\bibinfo {year} {2021})}\BibitemShut {NoStop}%
\bibitem [{\citenamefont {Hadamard}(1903)}]{Hadamard}%
  \BibitemOpen
  \bibfield  {author} {\bibinfo {author} {\bibnamefont {Hadamard}},\ }\bibfield
   {title} {\bibinfo {title} {{Leçons sur la propagation des ondes et les
  équations de l'hydrodynamique}},\ }in\ \href@noop {} {\emph {\bibinfo
  {booktitle} {University of Michigan Historical Math Collection}}}\ (\bibinfo
  {year} {1903})\BibitemShut {NoStop}%
\bibitem [{\citenamefont {Novello}\ \emph
  {et~al.}(2000{\natexlab{a}})\citenamefont {Novello}, \citenamefont
  {De~Lorenci}, \citenamefont {Salim},\ and\ \citenamefont
  {Klippert}}]{Novello:1999pg}%
  \BibitemOpen
  \bibfield  {author} {\bibinfo {author} {\bibfnamefont {M.}~\bibnamefont
  {Novello}}, \bibinfo {author} {\bibfnamefont {V.~A.}\ \bibnamefont
  {De~Lorenci}}, \bibinfo {author} {\bibfnamefont {J.~M.}\ \bibnamefont
  {Salim}},\ and\ \bibinfo {author} {\bibfnamefont {R.}~\bibnamefont
  {Klippert}},\ }\bibfield  {title} {\bibinfo {title} {{Geometrical aspects of
  light propagation in nonlinear electrodynamics}},\ }\href
  {https://doi.org/10.1103/PhysRevD.61.045001} {\bibfield  {journal} {\bibinfo
  {journal} {Phys. Rev. D}\ }\textbf {\bibinfo {volume} {61}},\ \bibinfo
  {pages} {045001} (\bibinfo {year} {2000}{\natexlab{a}})}\BibitemShut
  {NoStop}%
\bibitem [{\citenamefont {Novello}\ \emph
  {et~al.}(2000{\natexlab{b}})\citenamefont {Novello}, \citenamefont
  {Perez~Bergliaffa},\ and\ \citenamefont {Salim}}]{Novello:2000km}%
  \BibitemOpen
  \bibfield  {author} {\bibinfo {author} {\bibfnamefont {M.}~\bibnamefont
  {Novello}}, \bibinfo {author} {\bibfnamefont {S.~E.}\ \bibnamefont
  {Perez~Bergliaffa}},\ and\ \bibinfo {author} {\bibfnamefont {J.~M.}\
  \bibnamefont {Salim}},\ }\bibfield  {title} {\bibinfo {title} {{Singularities
  in general relativity coupled to nonlinear electrodynamics}},\ }\href
  {https://doi.org/10.1088/0264-9381/17/18/316} {\bibfield  {journal} {\bibinfo
   {journal} {Class. Quant. Grav.}\ }\textbf {\bibinfo {volume} {17}},\
  \bibinfo {pages} {3821} (\bibinfo {year} {2000}{\natexlab{b}})}\BibitemShut
  {NoStop}%
\bibitem [{\citenamefont {De~Lorenci}\ and\ \citenamefont
  {Souza}(2001)}]{DeLorenci:2001gf}%
  \BibitemOpen
  \bibfield  {author} {\bibinfo {author} {\bibfnamefont {V.~A.}\ \bibnamefont
  {De~Lorenci}}\ and\ \bibinfo {author} {\bibfnamefont {M.~A.}\ \bibnamefont
  {Souza}},\ }\bibfield  {title} {\bibinfo {title} {{Electromagnetic wave
  propagation inside a material medium: An Effective geometry
  interpretation}},\ }\href {https://doi.org/10.1016/S0370-2693(01)00588-3}
  {\bibfield  {journal} {\bibinfo  {journal} {Phys. Lett. B}\ }\textbf
  {\bibinfo {volume} {512}},\ \bibinfo {pages} {417} (\bibinfo {year}
  {2001})}\BibitemShut {NoStop}%
\bibitem [{\citenamefont {Novello}\ and\ \citenamefont
  {Salim}(2001)}]{Novello:2001gk}%
  \BibitemOpen
  \bibfield  {author} {\bibinfo {author} {\bibfnamefont {M.}~\bibnamefont
  {Novello}}\ and\ \bibinfo {author} {\bibfnamefont {J.~M.}\ \bibnamefont
  {Salim}},\ }\bibfield  {title} {\bibinfo {title} {{Effective electromagnetic
  geometry}},\ }\href {https://doi.org/10.1103/PhysRevD.63.083511} {\bibfield
  {journal} {\bibinfo  {journal} {Phys. Rev. D}\ }\textbf {\bibinfo {volume}
  {63}},\ \bibinfo {pages} {083511} (\bibinfo {year} {2001})}\BibitemShut
  {NoStop}%
\bibitem [{\citenamefont {Novello}\ \emph
  {et~al.}(2000{\natexlab{c}})\citenamefont {Novello}, \citenamefont
  {De~Lorenci}, \citenamefont {Elbaz},\ and\ \citenamefont
  {Salim}}]{Novello:2000xw}%
  \BibitemOpen
  \bibfield  {author} {\bibinfo {author} {\bibfnamefont {M.}~\bibnamefont
  {Novello}}, \bibinfo {author} {\bibfnamefont {V.~A.}\ \bibnamefont
  {De~Lorenci}}, \bibinfo {author} {\bibfnamefont {E.}~\bibnamefont {Elbaz}},\
  and\ \bibinfo {author} {\bibfnamefont {J.~M.}\ \bibnamefont {Salim}},\
  }\bibfield  {title} {\bibinfo {title} {{Closed lightlike curves in nonlinear
  electrodynamics}},\ }\href@noop {} {\  (\bibinfo {year}
  {2000}{\natexlab{c}})}\BibitemShut {NoStop}%
\bibitem [{\citenamefont {Plebanski}(1996)}]{Plebanski}%
  \BibitemOpen
  \bibfield  {author} {\bibinfo {author} {\bibfnamefont {J.}~\bibnamefont
  {Plebanski}},\ }\href@noop {} {\emph {\bibinfo {title} {{Non-Linear
  Electrodynamics -- A Study }}}}\ (\bibinfo  {publisher} {C.I.E.A. del I.P.N.,
  Mexico City},\ \bibinfo {year} {1996})\BibitemShut {NoStop}%
\bibitem [{\citenamefont {Plebanski}(1970)}]{Plebanski:1970zz}%
  \BibitemOpen
  \bibfield  {author} {\bibinfo {author} {\bibfnamefont {J.}~\bibnamefont
  {Plebanski}},\ }\bibfield  {title} {\bibinfo {title} {{Lectures on non linear
  electrodynsmics}},\ }\href@noop {} {\  (\bibinfo {year} {1970})}\BibitemShut
  {NoStop}%
\bibitem [{\citenamefont {Novello}\ and\ \citenamefont
  {Perez~Bergliaffa}(2003)}]{Novello:2003je}%
  \BibitemOpen
  \bibfield  {author} {\bibinfo {author} {\bibfnamefont {M.}~\bibnamefont
  {Novello}}\ and\ \bibinfo {author} {\bibfnamefont {S.~E.}\ \bibnamefont
  {Perez~Bergliaffa}},\ }\bibfield  {title} {\bibinfo {title} {{Effective
  geometry}},\ }\href {https://doi.org/10.1063/1.1587103} {\bibfield  {journal}
  {\bibinfo  {journal} {AIP Conf. Proc.}\ }\textbf {\bibinfo {volume} {668}},\
  \bibinfo {pages} {288} (\bibinfo {year} {2003})}\BibitemShut {NoStop}%
\bibitem [{\citenamefont {Gibbons}(2001)}]{Gibbons:2001gy}%
  \BibitemOpen
  \bibfield  {author} {\bibinfo {author} {\bibfnamefont {G.~W.}\ \bibnamefont
  {Gibbons}},\ }\bibfield  {title} {\bibinfo {title} {{Aspects of Born-Infeld
  theory and string / M theory}},\ }\href {https://doi.org/10.1063/1.1419338}
  {\bibfield  {journal} {\bibinfo  {journal} {AIP Conf. Proc.}\ }\textbf
  {\bibinfo {volume} {589}},\ \bibinfo {pages} {324} (\bibinfo {year}
  {2001})}\BibitemShut {NoStop}%
\bibitem [{\citenamefont {Maluf}\ and\ \citenamefont
  {Neves}(2018)}]{Maluf:2018ksj}%
  \BibitemOpen
  \bibfield  {author} {\bibinfo {author} {\bibfnamefont {R.~V.}\ \bibnamefont
  {Maluf}}\ and\ \bibinfo {author} {\bibfnamefont {J.~C.~S.}\ \bibnamefont
  {Neves}},\ }\bibfield  {title} {\bibinfo {title} {{Bardeen regular black hole
  as a quantum-corrected Schwarzschild black hole}},\ }\href
  {https://doi.org/10.1142/S0218271819500482} {\bibfield  {journal} {\bibinfo
  {journal} {Int. J. Mod. Phys. D}\ }\textbf {\bibinfo {volume} {28}},\
  \bibinfo {pages} {1950048} (\bibinfo {year} {2018})}\BibitemShut {NoStop}%
\bibitem [{\citenamefont {Konoplya}\ \emph {et~al.}(2023)\citenamefont
  {Konoplya}, \citenamefont {Ovchinnikov},\ and\ \citenamefont
  {Ahmedov}}]{Konoplya:2023ahd}%
  \BibitemOpen
  \bibfield  {author} {\bibinfo {author} {\bibfnamefont {R.~A.}\ \bibnamefont
  {Konoplya}}, \bibinfo {author} {\bibfnamefont {D.}~\bibnamefont
  {Ovchinnikov}},\ and\ \bibinfo {author} {\bibfnamefont {B.}~\bibnamefont
  {Ahmedov}},\ }\bibfield  {title} {\bibinfo {title} {{Bardeen spacetime as a
  quantum corrected Schwarzschild black hole: Quasinormal modes and Hawking
  radiation}},\ }\href {https://doi.org/10.1103/PhysRevD.108.104054} {\bibfield
   {journal} {\bibinfo  {journal} {Phys. Rev. D}\ }\textbf {\bibinfo {volume}
  {108}},\ \bibinfo {pages} {104054} (\bibinfo {year} {2023})}\BibitemShut
  {NoStop}%
\bibitem [{\citenamefont {Perlick}\ and\ \citenamefont
  {Tsupko}(2022)}]{Perlick:2021aok}%
  \BibitemOpen
  \bibfield  {author} {\bibinfo {author} {\bibfnamefont {V.}~\bibnamefont
  {Perlick}}\ and\ \bibinfo {author} {\bibfnamefont {O.~Y.}\ \bibnamefont
  {Tsupko}},\ }\bibfield  {title} {\bibinfo {title} {{Calculating black hole
  shadows: Review of analytical studies}},\ }\href
  {https://doi.org/10.1016/j.physrep.2021.10.004} {\bibfield  {journal}
  {\bibinfo  {journal} {Phys. Rept.}\ }\textbf {\bibinfo {volume} {947}},\
  \bibinfo {pages} {1} (\bibinfo {year} {2022})}\BibitemShut {NoStop}%
\bibitem [{\citenamefont {Abdujabbarov}\ \emph {et~al.}(2015)\citenamefont
  {Abdujabbarov}, \citenamefont {Rezzolla},\ and\ \citenamefont
  {Ahmedov}}]{Abdujabbarov:2015xqa}%
  \BibitemOpen
  \bibfield  {author} {\bibinfo {author} {\bibfnamefont {A.~A.}\ \bibnamefont
  {Abdujabbarov}}, \bibinfo {author} {\bibfnamefont {L.}~\bibnamefont
  {Rezzolla}},\ and\ \bibinfo {author} {\bibfnamefont {B.~J.}\ \bibnamefont
  {Ahmedov}},\ }\bibfield  {title} {\bibinfo {title} {{A coordinate-independent
  characterization of a black hole shadow}},\ }\href
  {https://doi.org/10.1093/mnras/stv2079} {\bibfield  {journal} {\bibinfo
  {journal} {Mon. Not. Roy. Astron. Soc.}\ }\textbf {\bibinfo {volume} {454}},\
  \bibinfo {pages} {2423} (\bibinfo {year} {2015})}\BibitemShut {NoStop}%
\bibitem [{\citenamefont {Bambi}(2013)}]{Bambi:2013nla}%
  \BibitemOpen
  \bibfield  {author} {\bibinfo {author} {\bibfnamefont {C.}~\bibnamefont
  {Bambi}},\ }\bibfield  {title} {\bibinfo {title} {{Can the supermassive
  objects at the centers of galaxies be traversable wormholes? The first test
  of strong gravity for mm/sub-mm very long baseline interferometry
  facilities}},\ }\href {https://doi.org/10.1103/PhysRevD.87.107501} {\bibfield
   {journal} {\bibinfo  {journal} {Phys. Rev. D}\ }\textbf {\bibinfo {volume}
  {87}},\ \bibinfo {pages} {107501} (\bibinfo {year} {2013})}\BibitemShut
  {NoStop}%
\bibitem [{\citenamefont {Narayan}\ \emph {et~al.}(2019)\citenamefont
  {Narayan}, \citenamefont {Johnson},\ and\ \citenamefont
  {Gammie}}]{Narayan:2019imo}%
  \BibitemOpen
  \bibfield  {author} {\bibinfo {author} {\bibfnamefont {R.}~\bibnamefont
  {Narayan}}, \bibinfo {author} {\bibfnamefont {M.~D.}\ \bibnamefont
  {Johnson}},\ and\ \bibinfo {author} {\bibfnamefont {C.~F.}\ \bibnamefont
  {Gammie}},\ }\bibfield  {title} {\bibinfo {title} {{The Shadow of a
  Spherically Accreting Black Hole}},\ }\href
  {https://doi.org/10.3847/2041-8213/ab518c} {\bibfield  {journal} {\bibinfo
  {journal} {Astrophys. J. Lett.}\ }\textbf {\bibinfo {volume} {885}},\
  \bibinfo {pages} {L33} (\bibinfo {year} {2019})}\BibitemShut {NoStop}%
\bibitem [{\citenamefont {Kocherlakota}\ and\ \citenamefont
  {Rezzolla}(2022)}]{Kocherlakota:2022jnz}%
  \BibitemOpen
  \bibfield  {author} {\bibinfo {author} {\bibfnamefont {P.}~\bibnamefont
  {Kocherlakota}}\ and\ \bibinfo {author} {\bibfnamefont {L.}~\bibnamefont
  {Rezzolla}},\ }\bibfield  {title} {\bibinfo {title} {{Distinguishing
  gravitational and emission physics in black hole imaging: spherical
  symmetry}},\ }\href {https://doi.org/10.1093/mnras/stac891} {\bibfield
  {journal} {\bibinfo  {journal} {Mon. Not. Roy. Astron. Soc.}\ }\textbf
  {\bibinfo {volume} {513}},\ \bibinfo {pages} {1229} (\bibinfo {year}
  {2022})}\BibitemShut {NoStop}%
\bibitem [{\citenamefont {Kocherlakota}\ \emph
  {et~al.}(2024{\natexlab{a}})\citenamefont {Kocherlakota}, \citenamefont
  {Rezzolla}, \citenamefont {Roy},\ and\ \citenamefont
  {Wielgus}}]{Kocherlakota:2023qgo}%
  \BibitemOpen
  \bibfield  {author} {\bibinfo {author} {\bibfnamefont {P.}~\bibnamefont
  {Kocherlakota}}, \bibinfo {author} {\bibfnamefont {L.}~\bibnamefont
  {Rezzolla}}, \bibinfo {author} {\bibfnamefont {R.}~\bibnamefont {Roy}},\ and\
  \bibinfo {author} {\bibfnamefont {M.}~\bibnamefont {Wielgus}},\ }\bibfield
  {title} {\bibinfo {title} {{Prospects for future experimental tests of
  gravity with black hole imaging: Spherical symmetry}},\ }\href
  {https://doi.org/10.1103/PhysRevD.109.064064} {\bibfield  {journal} {\bibinfo
   {journal} {Phys. Rev. D}\ }\textbf {\bibinfo {volume} {109}},\ \bibinfo
  {pages} {064064} (\bibinfo {year} {2024}{\natexlab{a}})}\BibitemShut
  {NoStop}%
\bibitem [{\citenamefont {Kocherlakota}\ \emph
  {et~al.}(2024{\natexlab{b}})\citenamefont {Kocherlakota}, \citenamefont
  {Rezzolla}, \citenamefont {Roy},\ and\ \citenamefont
  {Wielgus}}]{Kocherlakota:2024hyq}%
  \BibitemOpen
  \bibfield  {author} {\bibinfo {author} {\bibfnamefont {P.}~\bibnamefont
  {Kocherlakota}}, \bibinfo {author} {\bibfnamefont {L.}~\bibnamefont
  {Rezzolla}}, \bibinfo {author} {\bibfnamefont {R.}~\bibnamefont {Roy}},\ and\
  \bibinfo {author} {\bibfnamefont {M.}~\bibnamefont {Wielgus}},\ }\bibfield
  {title} {\bibinfo {title} {{Hotspots and photon rings in spherically
  symmetric space\textendash{}times}},\ }\href
  {https://doi.org/10.1093/mnras/stae1321} {\bibfield  {journal} {\bibinfo
  {journal} {Mon. Not. Roy. Astron. Soc.}\ }\textbf {\bibinfo {volume} {531}},\
  \bibinfo {pages} {3606} (\bibinfo {year} {2024}{\natexlab{b}})}\BibitemShut
  {NoStop}%
\bibitem [{\citenamefont {Falcke}\ \emph {et~al.}(2000)\citenamefont {Falcke},
  \citenamefont {Melia},\ and\ \citenamefont {Agol}}]{Falcke:1999pj}%
  \BibitemOpen
  \bibfield  {author} {\bibinfo {author} {\bibfnamefont {H.}~\bibnamefont
  {Falcke}}, \bibinfo {author} {\bibfnamefont {F.}~\bibnamefont {Melia}},\ and\
  \bibinfo {author} {\bibfnamefont {E.}~\bibnamefont {Agol}},\ }\bibfield
  {title} {\bibinfo {title} {{Viewing the shadow of the black hole at the
  galactic center}},\ }\href {https://doi.org/10.1086/312423} {\bibfield
  {journal} {\bibinfo  {journal} {Astrophys. J. Lett.}\ }\textbf {\bibinfo
  {volume} {528}},\ \bibinfo {pages} {L13} (\bibinfo {year}
  {2000})}\BibitemShut {NoStop}%
\bibitem [{\citenamefont {Jaroszynski}\ and\ \citenamefont
  {Kurpiewski}()}]{Jaroszynski:1997bw}%
  \BibitemOpen
  \bibfield  {author} {\bibinfo {author} {\bibfnamefont {M.}~\bibnamefont
  {Jaroszynski}}\ and\ \bibinfo {author} {\bibfnamefont {A.}~\bibnamefont
  {Kurpiewski}},\ }\bibfield  {title} {\bibinfo {title} {{Optics near kerr
  black holes: spectra of advection dominated accretion flows}}\ }\href
  {https://doi.org/10.48550/arXiv.astro-ph/9705044}
  {10.48550/arXiv.astro-ph/9705044}\BibitemShut {NoStop}%
\bibitem [{\citenamefont {Novikov}\ and\ \citenamefont
  {Thorne}(1973)}]{Novikov:1973kta}%
  \BibitemOpen
  \bibfield  {author} {\bibinfo {author} {\bibfnamefont {I.~D.}\ \bibnamefont
  {Novikov}}\ and\ \bibinfo {author} {\bibfnamefont {K.~S.}\ \bibnamefont
  {Thorne}},\ }\bibfield  {title} {\bibinfo {title} {{Astrophysics and black
  holes}},\ }in\ \href {https://doi.org/6167207} {\emph {\bibinfo {booktitle}
  {{Les Houches Summer School of Theoretical Physics}: {Black Holes}}}}\
  (\bibinfo {year} {1973})\ pp.\ \bibinfo {pages} {343--550}\BibitemShut
  {NoStop}%
\bibitem [{\citenamefont {Page}\ and\ \citenamefont
  {Thorne}(1974)}]{Page:1974he}%
  \BibitemOpen
  \bibfield  {author} {\bibinfo {author} {\bibfnamefont {D.~N.}\ \bibnamefont
  {Page}}\ and\ \bibinfo {author} {\bibfnamefont {K.~S.}\ \bibnamefont
  {Thorne}},\ }\bibfield  {title} {\bibinfo {title} {{Disk-Accretion onto a
  Black Hole. Time-Averaged Structure of Accretion Disk}},\ }\href
  {https://doi.org/10.1086/152990} {\bibfield  {journal} {\bibinfo  {journal}
  {Astrophys. J.}\ }\textbf {\bibinfo {volume} {191}},\ \bibinfo {pages} {499}
  (\bibinfo {year} {1974})}\BibitemShut {NoStop}%
\bibitem [{\citenamefont {Harko}\ \emph {et~al.}(2009)\citenamefont {Harko},
  \citenamefont {Kovacs},\ and\ \citenamefont {Lobo}}]{Harko:2009xf}%
  \BibitemOpen
  \bibfield  {author} {\bibinfo {author} {\bibfnamefont {T.}~\bibnamefont
  {Harko}}, \bibinfo {author} {\bibfnamefont {Z.}~\bibnamefont {Kovacs}},\ and\
  \bibinfo {author} {\bibfnamefont {F.~S.~N.}\ \bibnamefont {Lobo}},\
  }\bibfield  {title} {\bibinfo {title} {{Thin accretion disks in stationary
  axisymmetric wormhole spacetimes}},\ }\href
  {https://doi.org/10.1103/PhysRevD.79.064001} {\bibfield  {journal} {\bibinfo
  {journal} {Phys. Rev. D}\ }\textbf {\bibinfo {volume} {79}},\ \bibinfo
  {pages} {064001} (\bibinfo {year} {2009})}\BibitemShut {NoStop}%
\bibitem [{\citenamefont {Bambi}(2012)}]{Bambi:2012tg}%
  \BibitemOpen
  \bibfield  {author} {\bibinfo {author} {\bibfnamefont {C.}~\bibnamefont
  {Bambi}},\ }\bibfield  {title} {\bibinfo {title} {{A code to compute the
  emission of thin accretion disks in non-Kerr space-times and test the nature
  of black hole candidates}},\ }\href
  {https://doi.org/10.1088/0004-637X/761/2/174} {\bibfield  {journal} {\bibinfo
   {journal} {Astrophys. J.}\ }\textbf {\bibinfo {volume} {761}},\ \bibinfo
  {pages} {174} (\bibinfo {year} {2012})}\BibitemShut {NoStop}%
\bibitem [{\citenamefont {Rodrigues}\ and\ \citenamefont
  {de~Sousa~Silva}(2018)}]{Rodrigues:2018bdc}%
  \BibitemOpen
  \bibfield  {author} {\bibinfo {author} {\bibfnamefont {M.~E.}\ \bibnamefont
  {Rodrigues}}\ and\ \bibinfo {author} {\bibfnamefont {M.~V.}\ \bibnamefont
  {de~Sousa~Silva}},\ }\bibfield  {title} {\bibinfo {title} {{Bardeen Regular
  Black Hole With an Electric Source}},\ }\href
  {https://doi.org/10.1088/1475-7516/2018/06/025} {\bibfield  {journal}
  {\bibinfo  {journal} {JCAP}\ }\textbf {\bibinfo {volume} {06}},\ \bibinfo
  {pages} {025}}\BibitemShut {NoStop}%
\bibitem [{\citenamefont {Hayward}(2006)}]{Hayward:2005gi}%
  \BibitemOpen
  \bibfield  {author} {\bibinfo {author} {\bibfnamefont {S.~A.}\ \bibnamefont
  {Hayward}},\ }\bibfield  {title} {\bibinfo {title} {{Formation and
  evaporation of regular black holes}},\ }\href
  {https://doi.org/10.1103/PhysRevLett.96.031103} {\bibfield  {journal}
  {\bibinfo  {journal} {Phys. Rev. Lett.}\ }\textbf {\bibinfo {volume} {96}},\
  \bibinfo {pages} {031103} (\bibinfo {year} {2006})}\BibitemShut {NoStop}%
\bibitem [{\citenamefont {Simpson}\ and\ \citenamefont
  {Visser}(2019)}]{Simpson:2018tsi}%
  \BibitemOpen
  \bibfield  {author} {\bibinfo {author} {\bibfnamefont {A.}~\bibnamefont
  {Simpson}}\ and\ \bibinfo {author} {\bibfnamefont {M.}~\bibnamefont
  {Visser}},\ }\bibfield  {title} {\bibinfo {title} {{Black-bounce to
  traversable wormhole}},\ }\href
  {https://doi.org/10.1088/1475-7516/2019/02/042} {\bibfield  {journal}
  {\bibinfo  {journal} {JCAP}\ }\textbf {\bibinfo {volume} {02}},\ \bibinfo
  {pages} {042}}\BibitemShut {NoStop}%
\bibitem [{\citenamefont {Bronnikov}\ and\ \citenamefont
  {Walia}(2022)}]{Bronnikov:2021uta}%
  \BibitemOpen
  \bibfield  {author} {\bibinfo {author} {\bibfnamefont {K.~A.}\ \bibnamefont
  {Bronnikov}}\ and\ \bibinfo {author} {\bibfnamefont {R.~K.}\ \bibnamefont
  {Walia}},\ }\bibfield  {title} {\bibinfo {title} {{Field sources for
  Simpson-Visser spacetimes}},\ }\href
  {https://doi.org/10.1103/PhysRevD.105.044039} {\bibfield  {journal} {\bibinfo
   {journal} {Phys. Rev. D}\ }\textbf {\bibinfo {volume} {105}},\ \bibinfo
  {pages} {044039} (\bibinfo {year} {2022})}\BibitemShut {NoStop}%
\bibitem [{\citenamefont {Kumar~Walia}(2023)}]{KumarWalia:2022ddq}%
  \BibitemOpen
  \bibfield  {author} {\bibinfo {author} {\bibfnamefont {R.}~\bibnamefont
  {Kumar~Walia}},\ }\bibfield  {title} {\bibinfo {title} {{Observational
  predictions of LQG motivated polymerized black holes and constraints from Sgr
  A* and M87*}},\ }\href {https://doi.org/10.1088/1475-7516/2023/03/029}
  {\bibfield  {journal} {\bibinfo  {journal} {JCAP}\ }\textbf {\bibinfo
  {volume} {03}},\ \bibinfo {pages} {029}}\BibitemShut {NoStop}%
\bibitem [{\citenamefont {Carballo-Rubio}\ \emph {et~al.}(2020)\citenamefont
  {Carballo-Rubio}, \citenamefont {Di~Filippo}, \citenamefont {Liberati},\ and\
  \citenamefont {Visser}}]{Carballo-Rubio:2019fnb}%
  \BibitemOpen
  \bibfield  {author} {\bibinfo {author} {\bibfnamefont {R.}~\bibnamefont
  {Carballo-Rubio}}, \bibinfo {author} {\bibfnamefont {F.}~\bibnamefont
  {Di~Filippo}}, \bibinfo {author} {\bibfnamefont {S.}~\bibnamefont
  {Liberati}},\ and\ \bibinfo {author} {\bibfnamefont {M.}~\bibnamefont
  {Visser}},\ }\bibfield  {title} {\bibinfo {title} {{Geodesically complete
  black holes}},\ }\href {https://doi.org/10.1103/PhysRevD.101.084047}
  {\bibfield  {journal} {\bibinfo  {journal} {Phys. Rev. D}\ }\textbf {\bibinfo
  {volume} {101}},\ \bibinfo {pages} {084047} (\bibinfo {year}
  {2020})}\BibitemShut {NoStop}%
\bibitem [{\citenamefont {Bejarano}\ \emph {et~al.}(2017)\citenamefont
  {Bejarano}, \citenamefont {Olmo},\ and\ \citenamefont
  {Rubiera-Garcia}}]{Bejarano:2017fgz}%
  \BibitemOpen
  \bibfield  {author} {\bibinfo {author} {\bibfnamefont {C.}~\bibnamefont
  {Bejarano}}, \bibinfo {author} {\bibfnamefont {G.~J.}\ \bibnamefont {Olmo}},\
  and\ \bibinfo {author} {\bibfnamefont {D.}~\bibnamefont {Rubiera-Garcia}},\
  }\bibfield  {title} {\bibinfo {title} {{What is a singular black hole beyond
  General Relativity?}},\ }\href {https://doi.org/10.1103/PhysRevD.95.064043}
  {\bibfield  {journal} {\bibinfo  {journal} {Phys. Rev. D}\ }\textbf {\bibinfo
  {volume} {95}},\ \bibinfo {pages} {064043} (\bibinfo {year}
  {2017})}\BibitemShut {NoStop}%
\bibitem [{\citenamefont {Hu}\ \emph {et~al.}(2023)\citenamefont {Hu},
  \citenamefont {Lan},\ and\ \citenamefont {Miao}}]{Hu:2023iuw}%
  \BibitemOpen
  \bibfield  {author} {\bibinfo {author} {\bibfnamefont {H.-W.}\ \bibnamefont
  {Hu}}, \bibinfo {author} {\bibfnamefont {C.}~\bibnamefont {Lan}},\ and\
  \bibinfo {author} {\bibfnamefont {Y.-G.}\ \bibnamefont {Miao}},\ }\bibfield
  {title} {\bibinfo {title} {{A regular black hole as the final state of
  evolution of a singular black hole}},\ }\href
  {https://doi.org/10.1140/epjc/s10052-023-12228-w} {\bibfield  {journal}
  {\bibinfo  {journal} {Eur. Phys. J. C}\ }\textbf {\bibinfo {volume} {83}},\
  \bibinfo {pages} {1047} (\bibinfo {year} {2023})}\BibitemShut {NoStop}%
\bibitem [{\citenamefont {Mazur}\ and\ \citenamefont
  {Mottola}(2023)}]{Mazur:2001fv}%
  \BibitemOpen
  \bibfield  {author} {\bibinfo {author} {\bibfnamefont {P.~O.}\ \bibnamefont
  {Mazur}}\ and\ \bibinfo {author} {\bibfnamefont {E.}~\bibnamefont
  {Mottola}},\ }\bibfield  {title} {\bibinfo {title} {{Gravitational Condensate
  Stars: An Alternative to Black Holes}},\ }\href
  {https://doi.org/10.3390/universe9020088} {\bibfield  {journal} {\bibinfo
  {journal} {Universe}\ }\textbf {\bibinfo {volume} {9}},\ \bibinfo {pages}
  {88} (\bibinfo {year} {2023})}\BibitemShut {NoStop}%
\bibitem [{\citenamefont {Kumar}\ and\ \citenamefont
  {Ghosh}(2021)}]{Kumar:2020ltt}%
  \BibitemOpen
  \bibfield  {author} {\bibinfo {author} {\bibfnamefont {R.}~\bibnamefont
  {Kumar}}\ and\ \bibinfo {author} {\bibfnamefont {S.~G.}\ \bibnamefont
  {Ghosh}},\ }\bibfield  {title} {\bibinfo {title} {{Photon ring structure of
  rotating regular black holes and no-horizon spacetimes}},\ }\href
  {https://doi.org/10.1088/1361-6382/abdd48} {\bibfield  {journal} {\bibinfo
  {journal} {Class. Quant. Grav.}\ }\textbf {\bibinfo {volume} {38}},\ \bibinfo
  {pages} {8} (\bibinfo {year} {2021})}\BibitemShut {NoStop}%
\bibitem [{\citenamefont {Eichhorn}\ \emph {et~al.}(2023)\citenamefont
  {Eichhorn}, \citenamefont {Gold},\ and\ \citenamefont
  {Held}}]{Eichhorn:2022fcl}%
  \BibitemOpen
  \bibfield  {author} {\bibinfo {author} {\bibfnamefont {A.}~\bibnamefont
  {Eichhorn}}, \bibinfo {author} {\bibfnamefont {R.}~\bibnamefont {Gold}},\
  and\ \bibinfo {author} {\bibfnamefont {A.}~\bibnamefont {Held}},\ }\bibfield
  {title} {\bibinfo {title} {{Horizonless Spacetimes As Seen by Present and
  Next-generation Event Horizon Telescope Arrays}},\ }\href
  {https://doi.org/10.3847/1538-4357/accced} {\bibfield  {journal} {\bibinfo
  {journal} {Astrophys. J.}\ }\textbf {\bibinfo {volume} {950}},\ \bibinfo
  {pages} {117} (\bibinfo {year} {2023})}\BibitemShut {NoStop}%
\bibitem [{\citenamefont {Ayzenberg}\ \emph {et~al.}(2023)\citenamefont
  {Ayzenberg} \emph {et~al.}}]{Ayzenberg:2023hfw}%
  \BibitemOpen
  \bibfield  {author} {\bibinfo {author} {\bibfnamefont {D.}~\bibnamefont
  {Ayzenberg}} \emph {et~al.},\ }\bibfield  {title} {\bibinfo {title}
  {{Fundamental Physics Opportunities with the Next-Generation Event Horizon
  Telescope}},\ }\href@noop {} {\  (\bibinfo {year} {2023})}\BibitemShut
  {NoStop}%
\bibitem [{\citenamefont {Eichhorn}\ and\ \citenamefont
  {Held}(2023)}]{Eichhorn:2022bbn}%
  \BibitemOpen
  \bibfield  {author} {\bibinfo {author} {\bibfnamefont {A.}~\bibnamefont
  {Eichhorn}}\ and\ \bibinfo {author} {\bibfnamefont {A.}~\bibnamefont
  {Held}},\ }\bibfield  {title} {\bibinfo {title} {{Quantum gravity lights up
  spinning black holes}},\ }\href
  {https://doi.org/10.1088/1475-7516/2023/01/032} {\bibfield  {journal}
  {\bibinfo  {journal} {JCAP}\ }\textbf {\bibinfo {volume} {01}},\ \bibinfo
  {pages} {032}}\BibitemShut {NoStop}%
\bibitem [{\citenamefont {Carballo-Rubio}\ \emph
  {et~al.}(2023{\natexlab{b}})\citenamefont {Carballo-Rubio}, \citenamefont
  {Di~Filippo}, \citenamefont {Liberati},\ and\ \citenamefont
  {Visser}}]{Carballo-Rubio:2022nuj}%
  \BibitemOpen
  \bibfield  {author} {\bibinfo {author} {\bibfnamefont {R.}~\bibnamefont
  {Carballo-Rubio}}, \bibinfo {author} {\bibfnamefont {F.}~\bibnamefont
  {Di~Filippo}}, \bibinfo {author} {\bibfnamefont {S.}~\bibnamefont
  {Liberati}},\ and\ \bibinfo {author} {\bibfnamefont {M.}~\bibnamefont
  {Visser}},\ }\bibfield  {title} {\bibinfo {title} {{A connection between
  regular black holes and horizonless ultracompact stars}},\ }\href
  {https://doi.org/10.1007/JHEP08(2023)046} {\bibfield  {journal} {\bibinfo
  {journal} {JHEP}\ }\textbf {\bibinfo {volume} {08}},\ \bibinfo {pages}
  {046}}\BibitemShut {NoStop}%
\bibitem [{\citenamefont {Arrechea}\ \emph {et~al.}(2024)\citenamefont
  {Arrechea}, \citenamefont {Barcel\'o}, \citenamefont {Carballo-Rubio},\ and\
  \citenamefont {Garay}}]{Arrechea:2023oax}%
  \BibitemOpen
  \bibfield  {author} {\bibinfo {author} {\bibfnamefont {J.}~\bibnamefont
  {Arrechea}}, \bibinfo {author} {\bibfnamefont {C.}~\bibnamefont {Barcel\'o}},
  \bibinfo {author} {\bibfnamefont {R.}~\bibnamefont {Carballo-Rubio}},\ and\
  \bibinfo {author} {\bibfnamefont {L.~J.}\ \bibnamefont {Garay}},\ }\bibfield
  {title} {\bibinfo {title} {{Ultracompact horizonless objects in order-reduced
  semiclassical gravity}},\ }\href
  {https://doi.org/10.1103/PhysRevD.109.104056} {\bibfield  {journal} {\bibinfo
   {journal} {Phys. Rev. D}\ }\textbf {\bibinfo {volume} {109}},\ \bibinfo
  {pages} {104056} (\bibinfo {year} {2024})}\BibitemShut {NoStop}%
\bibitem [{\citenamefont {Janis}\ \emph {et~al.}(1968)\citenamefont {Janis},
  \citenamefont {Newman},\ and\ \citenamefont {Winicour}}]{Janis:1968zz}%
  \BibitemOpen
  \bibfield  {author} {\bibinfo {author} {\bibfnamefont {A.~I.}\ \bibnamefont
  {Janis}}, \bibinfo {author} {\bibfnamefont {E.~T.}\ \bibnamefont {Newman}},\
  and\ \bibinfo {author} {\bibfnamefont {J.}~\bibnamefont {Winicour}},\
  }\bibfield  {title} {\bibinfo {title} {{Reality of the Schwarzschild
  Singularity}},\ }\href {https://doi.org/10.1103/PhysRevLett.20.878}
  {\bibfield  {journal} {\bibinfo  {journal} {Phys. Rev. Lett.}\ }\textbf
  {\bibinfo {volume} {20}},\ \bibinfo {pages} {878} (\bibinfo {year}
  {1968})}\BibitemShut {NoStop}%
\bibitem [{\citenamefont {Gyulchev}\ \emph {et~al.}(2020)\citenamefont
  {Gyulchev}, \citenamefont {Kunz}, \citenamefont {Nedkova}, \citenamefont
  {Vetsov},\ and\ \citenamefont {Yazadjiev}}]{Gyulchev:2020cvo}%
  \BibitemOpen
  \bibfield  {author} {\bibinfo {author} {\bibfnamefont {G.}~\bibnamefont
  {Gyulchev}}, \bibinfo {author} {\bibfnamefont {J.}~\bibnamefont {Kunz}},
  \bibinfo {author} {\bibfnamefont {P.}~\bibnamefont {Nedkova}}, \bibinfo
  {author} {\bibfnamefont {T.}~\bibnamefont {Vetsov}},\ and\ \bibinfo {author}
  {\bibfnamefont {S.}~\bibnamefont {Yazadjiev}},\ }\bibfield  {title} {\bibinfo
  {title} {{Observational signatures of strongly naked singularities: image of
  the thin accretion disk}},\ }\href
  {https://doi.org/10.1140/epjc/s10052-020-08575-7} {\bibfield  {journal}
  {\bibinfo  {journal} {Eur. Phys. J. C}\ }\textbf {\bibinfo {volume} {80}},\
  \bibinfo {pages} {1017} (\bibinfo {year} {2020})}\BibitemShut {NoStop}%
\bibitem [{\citenamefont {Gyulchev}\ \emph {et~al.}(2019)\citenamefont
  {Gyulchev}, \citenamefont {Nedkova}, \citenamefont {Vetsov},\ and\
  \citenamefont {Yazadjiev}}]{Gyulchev:2019tvk}%
  \BibitemOpen
  \bibfield  {author} {\bibinfo {author} {\bibfnamefont {G.}~\bibnamefont
  {Gyulchev}}, \bibinfo {author} {\bibfnamefont {P.}~\bibnamefont {Nedkova}},
  \bibinfo {author} {\bibfnamefont {T.}~\bibnamefont {Vetsov}},\ and\ \bibinfo
  {author} {\bibfnamefont {S.}~\bibnamefont {Yazadjiev}},\ }\bibfield  {title}
  {\bibinfo {title} {{Image of the Janis-Newman-Winicour naked singularity with
  a thin accretion disk}},\ }\href
  {https://doi.org/10.1103/PhysRevD.100.024055} {\bibfield  {journal} {\bibinfo
   {journal} {Phys. Rev. D}\ }\textbf {\bibinfo {volume} {100}},\ \bibinfo
  {pages} {024055} (\bibinfo {year} {2019})}\BibitemShut {NoStop}%
\bibitem [{\citenamefont {Deliyski}\ \emph {et~al.}(2024)\citenamefont
  {Deliyski}, \citenamefont {Gyulchev}, \citenamefont {Nedkova},\ and\
  \citenamefont {Yazadjiev}}]{Deliyski:2024wmt}%
  \BibitemOpen
  \bibfield  {author} {\bibinfo {author} {\bibfnamefont {V.}~\bibnamefont
  {Deliyski}}, \bibinfo {author} {\bibfnamefont {G.}~\bibnamefont {Gyulchev}},
  \bibinfo {author} {\bibfnamefont {P.}~\bibnamefont {Nedkova}},\ and\ \bibinfo
  {author} {\bibfnamefont {S.}~\bibnamefont {Yazadjiev}},\ }\bibfield  {title}
  {\bibinfo {title} {{Observing naked singularities by the present and
  next-generation Event Horizon Telescope}},\ }\href@noop {} {\  (\bibinfo
  {year} {2024})}\BibitemShut {NoStop}%
\bibitem [{\citenamefont {Gralla}\ \emph {et~al.}(2019)\citenamefont {Gralla},
  \citenamefont {Holz},\ and\ \citenamefont {Wald}}]{Gralla:2019xty}%
  \BibitemOpen
  \bibfield  {author} {\bibinfo {author} {\bibfnamefont {S.~E.}\ \bibnamefont
  {Gralla}}, \bibinfo {author} {\bibfnamefont {D.~E.}\ \bibnamefont {Holz}},\
  and\ \bibinfo {author} {\bibfnamefont {R.~M.}\ \bibnamefont {Wald}},\
  }\bibfield  {title} {\bibinfo {title} {{Black Hole Shadows, Photon Rings, and
  Lensing Rings}},\ }\href {https://doi.org/10.1103/PhysRevD.100.024018}
  {\bibfield  {journal} {\bibinfo  {journal} {Phys. Rev. D}\ }\textbf {\bibinfo
  {volume} {100}},\ \bibinfo {pages} {024018} (\bibinfo {year}
  {2019})}\BibitemShut {NoStop}%
\bibitem [{\citenamefont {Gralla}\ \emph {et~al.}(2020)\citenamefont {Gralla},
  \citenamefont {Lupsasca},\ and\ \citenamefont {Marrone}}]{Gralla:2020srx}%
  \BibitemOpen
  \bibfield  {author} {\bibinfo {author} {\bibfnamefont {S.~E.}\ \bibnamefont
  {Gralla}}, \bibinfo {author} {\bibfnamefont {A.}~\bibnamefont {Lupsasca}},\
  and\ \bibinfo {author} {\bibfnamefont {D.~P.}\ \bibnamefont {Marrone}},\
  }\bibfield  {title} {\bibinfo {title} {{The shape of the black hole photon
  ring: A precise test of strong-field general relativity}},\ }\href
  {https://doi.org/10.1103/PhysRevD.102.124004} {\bibfield  {journal} {\bibinfo
   {journal} {Phys. Rev. D}\ }\textbf {\bibinfo {volume} {102}},\ \bibinfo
  {pages} {124004} (\bibinfo {year} {2020})}\BibitemShut {NoStop}%
\bibitem [{\citenamefont {Johnson}\ \emph {et~al.}(2020)\citenamefont {Johnson}
  \emph {et~al.}}]{Johnson:2019ljv}%
  \BibitemOpen
  \bibfield  {author} {\bibinfo {author} {\bibfnamefont {M.~D.}\ \bibnamefont
  {Johnson}} \emph {et~al.},\ }\bibfield  {title} {\bibinfo {title} {{Universal
  interferometric signatures of a black hole\textquoteright{}s photon ring}},\
  }\href {https://doi.org/10.1126/sciadv.aaz1310} {\bibfield  {journal}
  {\bibinfo  {journal} {Sci. Adv.}\ }\textbf {\bibinfo {volume} {6}},\ \bibinfo
  {pages} {eaaz1310} (\bibinfo {year} {2020})}\BibitemShut {NoStop}%
\bibitem [{\citenamefont {Himwich}\ \emph {et~al.}(2020)\citenamefont
  {Himwich}, \citenamefont {Johnson}, \citenamefont {Lupsasca},\ and\
  \citenamefont {Strominger}}]{Himwich:2020msm}%
  \BibitemOpen
  \bibfield  {author} {\bibinfo {author} {\bibfnamefont {E.}~\bibnamefont
  {Himwich}}, \bibinfo {author} {\bibfnamefont {M.~D.}\ \bibnamefont
  {Johnson}}, \bibinfo {author} {\bibfnamefont {A.}~\bibnamefont {Lupsasca}},\
  and\ \bibinfo {author} {\bibfnamefont {A.}~\bibnamefont {Strominger}},\
  }\bibfield  {title} {\bibinfo {title} {{Universal polarimetric signatures of
  the black hole photon ring}},\ }\href
  {https://doi.org/10.1103/PhysRevD.101.084020} {\bibfield  {journal} {\bibinfo
   {journal} {Phys. Rev. D}\ }\textbf {\bibinfo {volume} {101}},\ \bibinfo
  {pages} {084020} (\bibinfo {year} {2020})}\BibitemShut {NoStop}%
\bibitem [{\citenamefont {Hadar}\ \emph {et~al.}(2021)\citenamefont {Hadar},
  \citenamefont {Johnson}, \citenamefont {Lupsasca},\ and\ \citenamefont
  {Wong}}]{Hadar:2020fda}%
  \BibitemOpen
  \bibfield  {author} {\bibinfo {author} {\bibfnamefont {S.}~\bibnamefont
  {Hadar}}, \bibinfo {author} {\bibfnamefont {M.~D.}\ \bibnamefont {Johnson}},
  \bibinfo {author} {\bibfnamefont {A.}~\bibnamefont {Lupsasca}},\ and\
  \bibinfo {author} {\bibfnamefont {G.~N.}\ \bibnamefont {Wong}},\ }\bibfield
  {title} {\bibinfo {title} {{Photon Ring Autocorrelations}},\ }\href
  {https://doi.org/10.1103/PhysRevD.103.104038} {\bibfield  {journal} {\bibinfo
   {journal} {Phys. Rev. D}\ }\textbf {\bibinfo {volume} {103}},\ \bibinfo
  {pages} {104038} (\bibinfo {year} {2021})}\BibitemShut {NoStop}%
\bibitem [{\citenamefont {Lupsasca}\ \emph {et~al.}(2024)\citenamefont
  {Lupsasca}, \citenamefont {C\'ardenas-Avenda\~no}, \citenamefont {Palumbo},
  \citenamefont {Johnson}, \citenamefont {Gralla}, \citenamefont {Marrone},
  \citenamefont {Galison}, \citenamefont {Tiede},\ and\ \citenamefont
  {Keeble}}]{Lupsasca:2024xhq}%
  \BibitemOpen
  \bibfield  {author} {\bibinfo {author} {\bibfnamefont {A.}~\bibnamefont
  {Lupsasca}}, \bibinfo {author} {\bibfnamefont {A.}~\bibnamefont
  {C\'ardenas-Avenda\~no}}, \bibinfo {author} {\bibfnamefont {D.~C.~M.}\
  \bibnamefont {Palumbo}}, \bibinfo {author} {\bibfnamefont {M.~D.}\
  \bibnamefont {Johnson}}, \bibinfo {author} {\bibfnamefont {S.~E.}\
  \bibnamefont {Gralla}}, \bibinfo {author} {\bibfnamefont {D.~P.}\
  \bibnamefont {Marrone}}, \bibinfo {author} {\bibfnamefont {P.}~\bibnamefont
  {Galison}}, \bibinfo {author} {\bibfnamefont {P.}~\bibnamefont {Tiede}},\
  and\ \bibinfo {author} {\bibfnamefont {L.}~\bibnamefont {Keeble}},\
  }\bibfield  {title} {\bibinfo {title} {{The Black Hole Explorer: Photon Ring
  Science, Detection and Shape Measurement}},\ }\href@noop {} {\  (\bibinfo
  {year} {2024})}\BibitemShut {NoStop}%
\bibitem [{\citenamefont {Cardoso}\ \emph {et~al.}(2009)\citenamefont
  {Cardoso}, \citenamefont {Miranda}, \citenamefont {Berti}, \citenamefont
  {Witek},\ and\ \citenamefont {Zanchin}}]{Cardoso:2008bp}%
  \BibitemOpen
  \bibfield  {author} {\bibinfo {author} {\bibfnamefont {V.}~\bibnamefont
  {Cardoso}}, \bibinfo {author} {\bibfnamefont {A.~S.}\ \bibnamefont
  {Miranda}}, \bibinfo {author} {\bibfnamefont {E.}~\bibnamefont {Berti}},
  \bibinfo {author} {\bibfnamefont {H.}~\bibnamefont {Witek}},\ and\ \bibinfo
  {author} {\bibfnamefont {V.~T.}\ \bibnamefont {Zanchin}},\ }\bibfield
  {title} {\bibinfo {title} {{Geodesic stability, Lyapunov exponents and
  quasinormal modes}},\ }\href {https://doi.org/10.1103/PhysRevD.79.064016}
  {\bibfield  {journal} {\bibinfo  {journal} {Phys. Rev. D}\ }\textbf {\bibinfo
  {volume} {79}},\ \bibinfo {pages} {064016} (\bibinfo {year}
  {2009})}\BibitemShut {NoStop}%
\bibitem [{\citenamefont {Bozza}\ and\ \citenamefont
  {Scarpetta}(2007)}]{Bozza:2007gt}%
  \BibitemOpen
  \bibfield  {author} {\bibinfo {author} {\bibfnamefont {V.}~\bibnamefont
  {Bozza}}\ and\ \bibinfo {author} {\bibfnamefont {G.}~\bibnamefont
  {Scarpetta}},\ }\bibfield  {title} {\bibinfo {title} {{Strong deflection
  limit of black hole gravitational lensing with arbitrary source distances}},\
  }\href {https://doi.org/10.1103/PhysRevD.76.083008} {\bibfield  {journal}
  {\bibinfo  {journal} {Phys. Rev. D}\ }\textbf {\bibinfo {volume} {76}},\
  \bibinfo {pages} {083008} (\bibinfo {year} {2007})}\BibitemShut {NoStop}%
\bibitem [{\citenamefont {Novello}\ \emph {et~al.}(2002)\citenamefont
  {Novello}, \citenamefont {Visser},\ and\ \citenamefont
  {Volovik}}]{Novello:2002qg}%
  \BibitemOpen
  \bibinfo {editor} {\bibfnamefont {M.}~\bibnamefont {Novello}}, \bibinfo
  {editor} {\bibfnamefont {M.}~\bibnamefont {Visser}},\ and\ \bibinfo {editor}
  {\bibfnamefont {G.}~\bibnamefont {Volovik}},\ eds.,\ \href
  {https://doi.org/10.1142/4861} {\emph {\bibinfo {title} {{Artificial black
  holes}}}}\ (\bibinfo  {publisher} {World Scientific},\ \bibinfo {year}
  {2002})\BibitemShut {NoStop}%
\bibitem [{\citenamefont {Novello}\ \emph {et~al.}(2001)\citenamefont
  {Novello}, \citenamefont {Salim}, \citenamefont {De~Lorenci},\ and\
  \citenamefont {Elbaz}}]{Novello:2001fv}%
  \BibitemOpen
  \bibfield  {author} {\bibinfo {author} {\bibfnamefont {M.}~\bibnamefont
  {Novello}}, \bibinfo {author} {\bibfnamefont {J.~M.}\ \bibnamefont {Salim}},
  \bibinfo {author} {\bibfnamefont {V.~A.}\ \bibnamefont {De~Lorenci}},\ and\
  \bibinfo {author} {\bibfnamefont {E.}~\bibnamefont {Elbaz}},\ }\bibfield
  {title} {\bibinfo {title} {{Nonlinear electrodynamics can generate a closed
  space - like path for photons}},\ }\href
  {https://doi.org/10.1103/PhysRevD.63.103516} {\bibfield  {journal} {\bibinfo
  {journal} {Phys. Rev. D}\ }\textbf {\bibinfo {volume} {63}},\ \bibinfo
  {pages} {103516} (\bibinfo {year} {2001})}\BibitemShut {NoStop}%
\bibitem [{\citenamefont {Allahyari}\ \emph {et~al.}(2020)\citenamefont
  {Allahyari}, \citenamefont {Khodadi}, \citenamefont {Vagnozzi},\ and\
  \citenamefont {Mota}}]{Allahyari:2019jqz}%
  \BibitemOpen
  \bibfield  {author} {\bibinfo {author} {\bibfnamefont {A.}~\bibnamefont
  {Allahyari}}, \bibinfo {author} {\bibfnamefont {M.}~\bibnamefont {Khodadi}},
  \bibinfo {author} {\bibfnamefont {S.}~\bibnamefont {Vagnozzi}},\ and\
  \bibinfo {author} {\bibfnamefont {D.~F.}\ \bibnamefont {Mota}},\ }\bibfield
  {title} {\bibinfo {title} {{Magnetically charged black holes from non-linear
  electrodynamics and the Event Horizon Telescope}},\ }\href
  {https://doi.org/10.1088/1475-7516/2020/02/003} {\bibfield  {journal}
  {\bibinfo  {journal} {JCAP}\ }\textbf {\bibinfo {volume} {02}},\ \bibinfo
  {pages} {003}}\BibitemShut {NoStop}%
\bibitem [{\citenamefont {Toshmatov}\ \emph {et~al.}(2019)\citenamefont
  {Toshmatov}, \citenamefont {Stuchl\'\i{}k}, \citenamefont {Ahmedov},\ and\
  \citenamefont {Malafarina}}]{Toshmatov:2019gxg}%
  \BibitemOpen
  \bibfield  {author} {\bibinfo {author} {\bibfnamefont {B.}~\bibnamefont
  {Toshmatov}}, \bibinfo {author} {\bibfnamefont {Z.}~\bibnamefont
  {Stuchl\'\i{}k}}, \bibinfo {author} {\bibfnamefont {B.}~\bibnamefont
  {Ahmedov}},\ and\ \bibinfo {author} {\bibfnamefont {D.}~\bibnamefont
  {Malafarina}},\ }\bibfield  {title} {\bibinfo {title} {{Relaxations of
  perturbations of spacetimes in general relativity coupled to nonlinear
  electrodynamics}},\ }\href {https://doi.org/10.1103/PhysRevD.99.064043}
  {\bibfield  {journal} {\bibinfo  {journal} {Phys. Rev. D}\ }\textbf {\bibinfo
  {volume} {99}},\ \bibinfo {pages} {064043} (\bibinfo {year}
  {2019})}\BibitemShut {NoStop}%
\bibitem [{\citenamefont {Cardoso}\ \emph {et~al.}(2014)\citenamefont
  {Cardoso}, \citenamefont {Crispino}, \citenamefont {Macedo}, \citenamefont
  {Okawa},\ and\ \citenamefont {Pani}}]{Cardoso:2014sna}%
  \BibitemOpen
  \bibfield  {author} {\bibinfo {author} {\bibfnamefont {V.}~\bibnamefont
  {Cardoso}}, \bibinfo {author} {\bibfnamefont {L.~C.~B.}\ \bibnamefont
  {Crispino}}, \bibinfo {author} {\bibfnamefont {C.~F.~B.}\ \bibnamefont
  {Macedo}}, \bibinfo {author} {\bibfnamefont {H.}~\bibnamefont {Okawa}},\ and\
  \bibinfo {author} {\bibfnamefont {P.}~\bibnamefont {Pani}},\ }\bibfield
  {title} {\bibinfo {title} {{Light rings as observational evidence for event
  horizons: long-lived modes, ergoregions and nonlinear instabilities of
  ultracompact objects}},\ }\href {https://doi.org/10.1103/PhysRevD.90.044069}
  {\bibfield  {journal} {\bibinfo  {journal} {Phys. Rev. D}\ }\textbf {\bibinfo
  {volume} {90}},\ \bibinfo {pages} {044069} (\bibinfo {year}
  {2014})}\BibitemShut {NoStop}%
\bibitem [{\citenamefont {Cunha}\ \emph {et~al.}(2017)\citenamefont {Cunha},
  \citenamefont {Berti},\ and\ \citenamefont {Herdeiro}}]{Cunha:2017qtt}%
  \BibitemOpen
  \bibfield  {author} {\bibinfo {author} {\bibfnamefont {P.~V.~P.}\
  \bibnamefont {Cunha}}, \bibinfo {author} {\bibfnamefont {E.}~\bibnamefont
  {Berti}},\ and\ \bibinfo {author} {\bibfnamefont {C.~A.~R.}\ \bibnamefont
  {Herdeiro}},\ }\bibfield  {title} {\bibinfo {title} {{Light-Ring Stability
  for Ultracompact Objects}},\ }\href
  {https://doi.org/10.1103/PhysRevLett.119.251102} {\bibfield  {journal}
  {\bibinfo  {journal} {Phys. Rev. Lett.}\ }\textbf {\bibinfo {volume} {119}},\
  \bibinfo {pages} {251102} (\bibinfo {year} {2017})}\BibitemShut {NoStop}%
\bibitem [{\citenamefont {Cardoso}\ and\ \citenamefont
  {Pani}(2019)}]{Cardoso:2019rvt}%
  \BibitemOpen
  \bibfield  {author} {\bibinfo {author} {\bibfnamefont {V.}~\bibnamefont
  {Cardoso}}\ and\ \bibinfo {author} {\bibfnamefont {P.}~\bibnamefont {Pani}},\
  }\bibfield  {title} {\bibinfo {title} {{Testing the nature of dark compact
  objects: a status report}},\ }\href
  {https://doi.org/10.1007/s41114-019-0020-4} {\bibfield  {journal} {\bibinfo
  {journal} {Living Rev. Rel.}\ }\textbf {\bibinfo {volume} {22}},\ \bibinfo
  {pages} {4} (\bibinfo {year} {2019})}\BibitemShut {NoStop}%
\bibitem [{\citenamefont {Keir}(2016)}]{Keir:2014oka}%
  \BibitemOpen
  \bibfield  {author} {\bibinfo {author} {\bibfnamefont {J.}~\bibnamefont
  {Keir}},\ }\bibfield  {title} {\bibinfo {title} {{Slowly decaying waves on
  spherically symmetric spacetimes and ultracompact neutron stars}},\ }\href
  {https://doi.org/10.1088/0264-9381/33/13/135009} {\bibfield  {journal}
  {\bibinfo  {journal} {Class. Quant. Grav.}\ }\textbf {\bibinfo {volume}
  {33}},\ \bibinfo {pages} {135009} (\bibinfo {year} {2016})}\BibitemShut
  {NoStop}%
\bibitem [{\citenamefont {Murk}\ and\ \citenamefont
  {Soranidis}(2024)}]{Murk:2024nod}%
  \BibitemOpen
  \bibfield  {author} {\bibinfo {author} {\bibfnamefont {S.}~\bibnamefont
  {Murk}}\ and\ \bibinfo {author} {\bibfnamefont {I.}~\bibnamefont
  {Soranidis}},\ }\bibfield  {title} {\bibinfo {title} {{Light rings and
  causality for nonsingular ultracompact objects sourced by nonlinear
  electrodynamics}},\ }\href@noop {} {\  (\bibinfo {year} {2024})}\BibitemShut
  {NoStop}%
\bibitem [{\citenamefont {Shaikh}\ \emph {et~al.}(2019)\citenamefont {Shaikh},
  \citenamefont {Kocherlakota}, \citenamefont {Narayan},\ and\ \citenamefont
  {Joshi}}]{Shaikh:2018lcc}%
  \BibitemOpen
  \bibfield  {author} {\bibinfo {author} {\bibfnamefont {R.}~\bibnamefont
  {Shaikh}}, \bibinfo {author} {\bibfnamefont {P.}~\bibnamefont
  {Kocherlakota}}, \bibinfo {author} {\bibfnamefont {R.}~\bibnamefont
  {Narayan}},\ and\ \bibinfo {author} {\bibfnamefont {P.~S.}\ \bibnamefont
  {Joshi}},\ }\bibfield  {title} {\bibinfo {title} {{Shadows of spherically
  symmetric black holes and naked singularities}},\ }\href
  {https://doi.org/10.1093/mnras/sty2624} {\bibfield  {journal} {\bibinfo
  {journal} {Mon. Not. Roy. Astron. Soc.}\ }\textbf {\bibinfo {volume} {482}},\
  \bibinfo {pages} {52} (\bibinfo {year} {2019})}\BibitemShut {NoStop}%
\bibitem [{\citenamefont {Schee}\ and\ \citenamefont
  {Stuchlik}(2015)}]{Schee:2015nua}%
  \BibitemOpen
  \bibfield  {author} {\bibinfo {author} {\bibfnamefont {J.}~\bibnamefont
  {Schee}}\ and\ \bibinfo {author} {\bibfnamefont {Z.}~\bibnamefont
  {Stuchlik}},\ }\bibfield  {title} {\bibinfo {title} {{Gravitational lensing
  and ghost images in the regular Bardeen no-horizon spacetimes}},\ }\href
  {https://doi.org/10.1088/1475-7516/2015/06/048} {\bibfield  {journal}
  {\bibinfo  {journal} {JCAP}\ }\textbf {\bibinfo {volume} {06}},\ \bibinfo
  {pages} {048}}\BibitemShut {NoStop}%
\end{thebibliography}%

\end{document}